\documentclass{aa}  

\usepackage{txfonts}

\usepackage{newtxtext,newtxmath,upgreek}

\usepackage{graphicx}   
\usepackage{soul}
\usepackage[dvipsnames]{xcolor}
\usepackage[breaklinks=true,colorlinks=true,allcolors=blue]{hyperref}
\usepackage{stfloats}
\usepackage{amsmath}    
\usepackage{amssymb}    
\usepackage{natbib, booktabs, multirow, tabularx} 
\usepackage{hhline}
\usepackage{placeins}
\usepackage{threeparttable}

\usepackage{float}

\restylefloat{figure}
\begin{document}

\raggedbottom
\title{Investigating the detection rates and inference of gravitational-wave and radio emission from black hole neutron star mergers\thanks{Data used to plot the images have been uploaded at: \url{http://doi.org/10.5281/zenodo.6573093}}}
\titlerunning{Investigating the gravitational-wave and radio emission from black hole neutron star mergers}
\author{
     Oliver~M.~Boersma      \inst{\ref{uva} , \ref{astron}}, 
     Joeri~van~Leeuwen    \inst{\ref{astron}, \ref{uva}}}
\institute{
Anton Pannekoek Institute, University of Amsterdam, Postbus 94249, 1090 GE Amsterdam, The Netherlands\label{uva}
  \and
ASTRON, the Netherlands Institute for Radio Astronomy, Oude Hoogeveensedijk 4,7991 PD Dwingeloo, The Netherlands\label{astron}}

% \abstract{}{}{}{}{} 
% 5 {} token are mandatory
 
  \abstract
   {Black hole neutron star (BHNS) mergers have recently been detected through their gravitational-wave (GW) emission. While no electromagnetic emission has yet been confidently associated with these systems, observing any such emission could provide information on, for example, the neutron star equation of state. 
   Black hole neutron star mergers could produce electromagnetic emission as a short gamma-ray burst (sGRB) and/or an sGRB afterglow upon interaction with the circum-merger medium.}
   {We make predictions for the expected  detection rates with the Square Kilometre Array Phase 1 (SKA1) of sGRB radio afterglows associated with BHNS mergers. We also investigate the benefits of a multi-messenger analysis in inferring the properties of the merging binary.}
   {We simulated a population of BHNS mergers, making use of recent stellar population synthesis results, and estimated their sGRB afterglow flux to obtain the detection rates with SKA1. 
   We investigate how this rate depends on the GW detector sensitivity, the primary black hole spin, and the neutron star equation of state. We then performed a multi-messenger Bayesian inference study on a fiducial BHNS merger. We simulated its sGRB afterglow and GW emission as input to this study, using recent models for both, and take systematic errors into account. } 
   {The expected rates of a combined GW and radio detection with the current-generation GW detectors are likely low. Due to the much increased sensitivity of future GW detectors such as the Einstein Telescope, the chances of an sGRB localisation and radio detection increase substantially. The unknown distribution of the black hole spin has a big influence on the detection rates, however, and it is a large source of uncertainty. Furthermore, when placing our fiducial BHNS merger at 50 and 100 Mpc, we are able to infer both the binary source parameters and the parameters of the sGRB afterglow simultaneously if we combine the GW and radio data. The radio data provide useful extra information on the binary parameters, such as the mass ratio, but this is limited by the systematic errors involved. For our fiducial binary at 200 Mpc, it is considerably more difficult to adequately infer the parameters of the system.}
   {The probability of finding an sGRB afterglow of a BHNS merger is low in the near future but will rise significantly when the next-generation GW detectors come online. 
   Combining information from GW data with radio data is crucial for characterising the jet properties. 
   A better understanding of the systematics will further increase the amount of information on the binary parameters that can be extracted from this radio data.}


   \maketitle
%
%________________________________________________________________

\section{Introduction}
\begin{figure*}[h]
    \centering
    \includegraphics[width=0.9\textwidth]{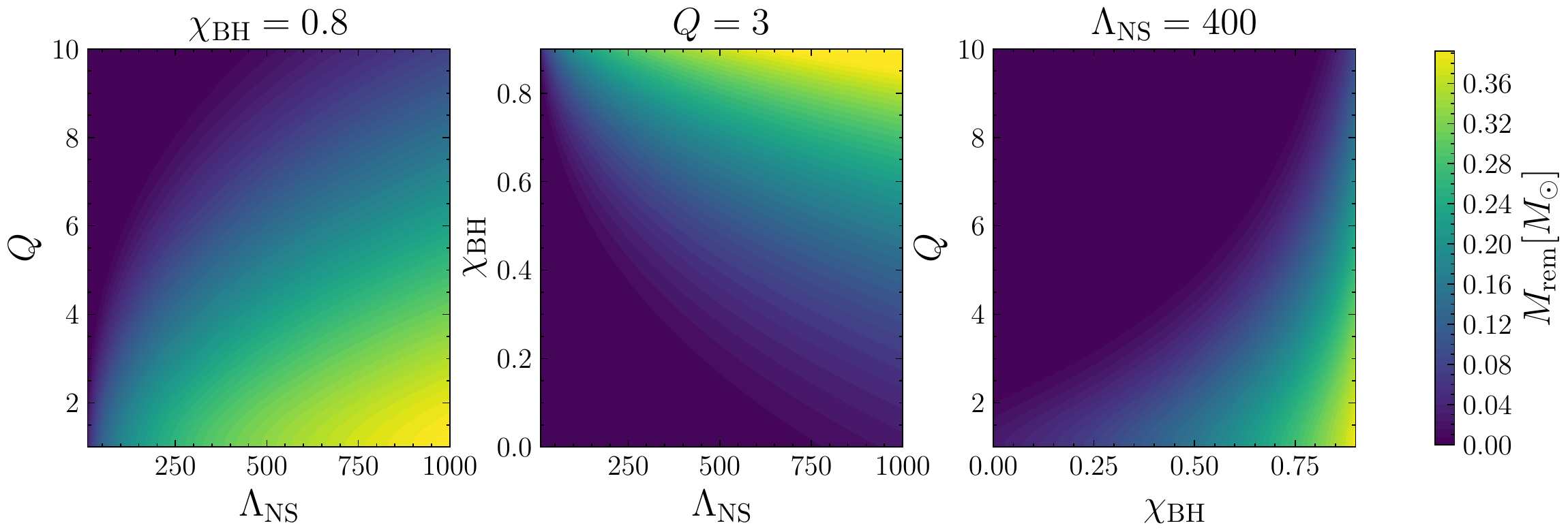}
    \caption{Visualisation of the fitting formula for the remnant mass in~\citet{foucart_remnant_2018}, showing the dependence on the BH spin ($\chi_\mathrm{BH}$), mass ratio ($Q$), and NS tidal deformability ($\Lambda_\mathrm{NS}$). Best-fit values for the free parameters are assumed. In each of the three panels, $\chi_\mathrm{BH}, \ Q \  \mathrm{, or} \ \Lambda_\mathrm{NS}$ is kept constant, as shown in the title of the panel, and the other two parameters are varied. The coloured regions indicate the amount of remnant mass ($M_\mathrm{rem}$) in solar masses.}
    \label{fig:ej_viz}
\end{figure*}
The detection of the first two black hole neutron star (BHNS) mergers, GW200105 and GW200115, by the advanced Laser Interferometer Gravitational-Wave Observatory (aLIGO) and Virgo detector network~\citep{aasi_advanced_2015,acernese_advanced_2014} completed the trifecta of compact binary coalescence gravitational-wave (GW) observations~\citep{abbott_observation_2021}. The third GW catalogue (GWTC-3) by the GW detector network contains a total of 90 significant detections from binary black hole (BBH), BHNS, and binary neutron star (BNS) mergers~\citep{the_ligo_scientific_collaboration_the_virgo_collaboration_the_kagra_collaboration_gwtc-3_2021}. The Kamioka Gravitational Wave Detector (KAGRA)~\citep{somiya_detector_2012,the_kagra_collaboration_interferometer_2013} came online and was included in the network in early 2020. This rich source of data has brought forth numerous exciting results on its own in areas of research such as tests of general relativity \citep{the_ligo_scientific_collaboration_the_virgo_collaboration_the_kagra_collaboration_tests_2021} and the cosmic expansion history~\citep{the_ligo_scientific_collaboration_the_virgo_collaboration_the_kagra_collaboration_constraints_2021}. 

Of particular interest to astronomers are the GW observations of 
BNS and BHNS systems because of the possibility of a complementary electromagnetic (EM) signature\footnote{It has been hypothesised that BBH mergers can also be a source of EM emission (see e.g.~\citealt{liebling_electromagnetic_2016})}. This possibility was first confirmed with the BNS source GW170817~\citep{ligo_scientific_collaboration_and_virgo_collaboration_gw170817_2017}, which was accompanied by EM radiation from the dynamical ejecta of a kilonova (see e.g.~\citealt{chornock_electromagnetic_2017,coulter_swope_2017}), 
the short gamma-ray burst (sGRB) GRB170817A~\citep[e.g.][]{abbott_gravitational_2017}, and the afterglow of the jet interacting with the interstellar environment~\citep[e.g.][]{alexander_electromagnetic_2017,haggard_deep_2017,hallinan_radio_2017}. No such emission was detected from either BHNS mergers GW200105 or GW200115 despite a multitude of follow-up campaigns being performed after their detections~\citep[e.g.][]{antier_grandma_2020,abbasi_probing_2021,abe_search_2021,paterson_searches_2021,ridnaia_search_2020,kasliwal_kilonova_2020,ashkar_hess_2021,anand_optical_2021}\footnote{See also~\citealt{httpsgcn1:online,httpsgcn2:online}.}. 

The probability of detecting any EM emission of a BHNS merger is tightly connected to the ejecta that might be produced during and after the merger. In BHNS mergers, such ejecta originates from the tidal disruption of the neutron star (NS) if the NS does not directly plunge into the black hole (BH). A small part of the tidally disrupted NS, which is not accreted onto the BH merger remnant, may become bound to a disk around the merger remnant or be ejected as unbounded material~\citep{foucart_remnant_2018}. The rotational energy of the BH merger remnant in conjunction with the magnetic field of the accretion disk could power an sGRB~\citep[e.g.][]{paschalidis_relativistic_2015}, which would produce a radio afterglow upon interacting with the circum-merger medium~\citep[e.g.][]{metzger_what_2012}. It has been suggested that some of the sGRBs detected to date originate from BHNS mergers~\citep{gompertz_search_2020}. Other sources of radio emission, such as the afterglow of the unbound dynamical ejecta, have also been proposed~\citep[e.g.][]{nakar_detectable_2011}.

The fraction of systems in which tidal disruption occurs is still uncertain as it depends on unknown distributions of the binary mass ratio, NS compactness, and BH spin. Still, this fraction is assumed to be small~\citep{zappa_black-hole_2019,fragione_black-holeneutron-star_2021}. For example, while BHNS mergers with large BH spins aligned with the system angular momentum produce the most ejecta mass, high spins do not seem to be consistent with current GW detections~\citep{abbott_population_2021}. Furthermore, a hard NS equation of state (EOS) also increases the amount of ejecta, but this is again disfavoured by current GW detections~\citep{the_ligo_scientific_collaboration_and_the_virgo_collaboration_gw170817_2018,fragione_black-holeneutron-star_2021}.

While the rate of BHNS mergers with any EM emission is likely low, this is offset by the high potential science impact of a detection. An EM observation associated with a BHNS merger could, for example, provide information on the NS EOS~\citep{pannarale_prospects_2014,ascenzi_constraining_2019,fragione_constraining_2021}, constrain the BH spin~\citep{barbieri_light-curve_2019}, or help determine the progenitor class if it is unclear from the GW detection alone~\citep{hinderer_distinguishing_2019}. Such types of multi-messenger analyses have been performed extensively on GW170817 (e.g.~\citealt{radice_gw170817_2018,coughlin_constraints_2018,radice_multimessenger_2019,coughlin_multimessenger_2019,raaijmakers_constraining_2020,capano_stringent_2020,dietrich_multimessenger_2020,breschi_at2017gfo_2021}), and various general Bayesian multi-messenger frameworks have been developed with this goal in mind (e.g.~\citealt{breschi_bayesian_2021,raaijmakers_challenges_2021,nicholl_tight_2021}).

In this paper we investigate both the expected rates and the parameter inference of radio observations of sGRB afterglows associated with BHNS mergers. We build on recent population synthesis results~\citep{broekgaarden_impact_2021} and connect them to an analytical estimate of the sGRB afterglow flux~\citep{nakar_detectability_2002,duque_radio_2019}. This allows us to make predictions for the sGRB afterglow radio detection rates with the Square Kilometre Array Phase 1~\citep[SKA1;][]{braun_anticipated_2019}. We then perform a comprehensive joint analysis of simulated GW and radio data in a Bayesian framework, taking systematic errors into account. Here, we make use of a recent model of the full afterglow light curve~\citep{ryan_gamma-ray_2020} and explore the benefits of a multi-messenger analysis in the inference of the source properties. Notably, we do the parameter inference of the GW data and the radio data simultaneously instead of using a sequential approach~\citep[see e.g.][]{barbieri_light-curve_2019}. 

The rest of the paper is structured as follows. In Sect.~\ref{sec:ejoutflows} we give an overview of the connection between the binary properties and the ejecta mass and describe the assumed jet launch mechanism and jet energy. We derive the expected radio afterglow detection rates in Sect.~\ref{sec:aglowdetrates}. We then describe our Bayesian multi-messenger framework setup in Sect.~\ref{sec:MMPI} and show the results for a fiducial BHNS merger in Sect.~\ref{sec:PEresults}. We discuss our findings in Sect.~\ref{sec:discussion} and end the paper with a summary and conclusion in Sect.~\ref{sec:conclusion}.
\section{Ejecta outflows of BHNS mergers}
\label{sec:ejoutflows}
\subsection{Computation of the ejecta mass}
\label{sec:ejmasscomp}
The amount of bound disk and unbound dynamical ejecta in BHNS mergers is heavily dependent on the binary source parameters of the mass ratio $Q = M_{\mathrm{BH}} / M_{\mathrm{NS}}$, with $M_{\mathrm{BH}}$ the mass of the BH and $M_{\mathrm{NS}}$ the mass of the NS, the tidal deformability of the NS $\Lambda_{\mathrm{NS}}$, and the dimensionless BH spin $\chi_{\mathrm{BH}}$. In general, comparable mass binaries with high $\chi_{\mathrm{BH}}$ and a stiff assumed EOS (high $\Lambda_{\mathrm{NS}}$) produce the most ejecta~\citep{foucart_remnant_2018}. The analyses in this paper are limited to non-precessing systems so $\chi_{\mathrm{BH}}$ and the dimensionless NS spin $\chi_{\mathrm{NS}}$ refer to the component of the spin parallel to the orbital angular momentum.

To fully capture the relevant physics and predict the ejecta properties, numerical relativity (NR) simulations are necessary, but they bring a large computational burden. As an approximation, we used analytical formulae from the literature which are fits to such simulations, covering a range of binary source parameters. These fits incorporate various free parameters, which can change depending on the precise form of the fitting formula and the types of NR simulations used to constrain the free parameters. To avoid any ambiguity on how we implemented these fits, we list the formulae together with the best-fit values of the free parameters below.

~\citet{foucart_remnant_2018} provide a formula, fit to 75 NR simulations, for calculating the remnant mass normalised to the baryonic mass of the NS ($M^b_{\mathrm{NS}}$):
\begin{equation}
\label{eq:mrem}
    {M}_{\mathrm{rem}} /M^b_{\mathrm{NS}} = \Big[\mathrm{Max}\Big(\alpha\frac{1-2C_{\mathrm{NS}}}{\eta^{1/3}} - \beta \hat{R}_{\mathrm{ISCO}}\frac{C_{\mathrm{NS}}}{\eta} + \gamma,0\Big)\Big]^{\delta},
\end{equation}
with best-fit values $\alpha \ = \ 0.406$, $\beta \ = \ 0.139$, $\gamma \ = \ 0.255,$ and $\delta \ = \ 1.761$. Here, $\eta=Q/(1+Q)^2$ and $\hat{R}_{\mathrm{ISCO}} = {R}_{\mathrm{ISCO}} / M_\mathrm{BH}$ is the innermost stable circular orbit radius normalised by the BH mass:
\begin{align}
    \hat{R}_{\mathrm{ISCO}} &= 3 + Z_2 -\mathrm{sgn}(\chi_\mathrm{BH})\sqrt{(3-Z_1)(3 + Z_1 + 2Z_2)}\\
    Z_1 &= 1 + (1-\chi^2_\mathrm{BH})^{1/3}[(1+\chi_{\mathrm{BH}})^{1/3} + (1-\chi_{\mathrm{BH}})^{1/3}]\\
    Z_2 &= \sqrt{3\chi^2_{\mathrm{BH}} + Z^2_1}.
\end{align}
To compute the compactness of the NS $C_\mathrm{NS}$, we used the approximately universal C-Love relation for NSs of Eq. 78 in~\citet{yagi_approximate_2017}:
\begin{equation}
    C_\mathrm{NS} = \sum_{\text{k}=0}^{{2}} a_k(\log \Lambda_\mathrm{NS})^k,
\end{equation}
with best-fit coefficients $a_0 = 0.360$, $a_1 = -0.0355$ and $a_2 = 0.000705$. While this relation is not a perfect substitute for integrating a NS EOS to obtain $C_\mathrm{NS}$, it performs well across a wide range of EOSs and has much less computational overhead. From $C_\mathrm{NS}$ we could, again approximately, compute $M^b_{\mathrm{NS}}$~\citep{lattimer_neutron_2001}:
\begin{equation}
    M^b_{\mathrm{NS}} = M_{\mathrm{NS}}\Big(1 + \frac{0.6C_\mathrm{NS}}{1-0.5C_\mathrm{NS}}\Big).
\end{equation}
The dependence of ${M}_{\mathrm{rem}}$ on $\chi_\mathrm{BH}$, $Q$, and $\Lambda_\mathrm{NS}$ is visualised in Fig.~\ref{fig:ej_viz}. 

Given ${M}_{\mathrm{rem}}$, we can calculate the amount of disk mass ($M_\mathrm{disk}$) if we have an estimate of the amount of unbound dynamical ejecta mass ($M_\mathrm{dyn}$):
\begin{equation}
    M_\mathrm{disk} = {M}_{\mathrm{rem}} - {M}_{\mathrm{dyn}}.
\end{equation}
We again turned to fits of NR simulations~\citep{kruger_estimates_2020} for ${M}_{\mathrm{dyn}}$:
\begin{equation}
\label{eq:mdyn}
    {M}_{\mathrm{dyn}} /M^b_{\mathrm{NS}} = a_1Q^{n_1}\frac{1-2C_\mathrm{NS}}{C_\mathrm{NS}} - a_2Q^{n_2}\hat{R}_\mathrm{ISCO} + a_4,
\end{equation}
with best-fit coefficients $a_1 = 0.007116$, $a_2 = 0.001436$, $a_4 =
-0.02762$, $n_1 = 0.8636$, and $n_2 = 1.6840$.

Parts of the disk itself can also become unbounded through disk wind outflows that are either thermally or magnetically driven.~\citet{raaijmakers_challenges_2021} derive a simple formula that broadly captures the dependence found in NR simulations of the disk wind ejecta mass $M_\mathrm{ej,disk}$ on $Q$~\citep{fernandez_landscape_2020}:
\begin{equation}
    M_\mathrm{ej,disk}/M_\mathrm{disk} = \xi_1 + \frac{\xi_2 - \xi_1}{1 + e^{1.5(Q-3)}}.
\end{equation}
Here we assumed average values for the free parameters of $\xi_1 = 0.18$ and $\xi_2 = 0.29$.
\subsection{GRB jet}
\label{sec:grbjet}
\begin{figure*}[h]
    \centering
    \includegraphics[width=0.9\textwidth]{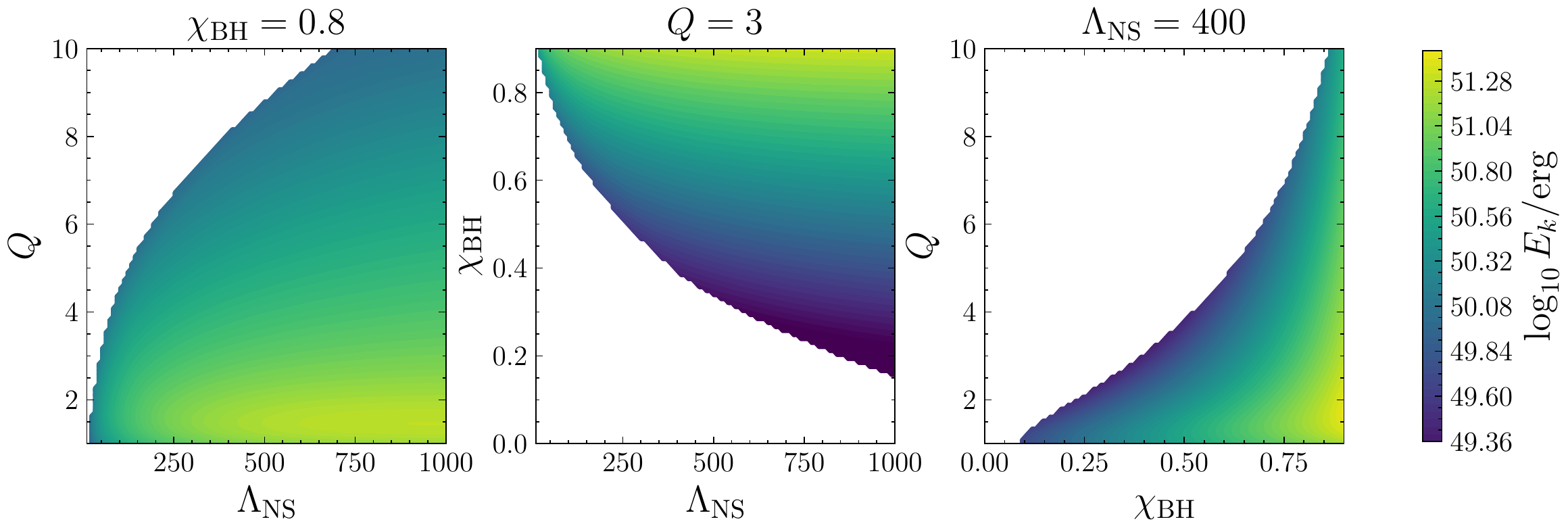}
    \caption{Same as Fig.~\ref{fig:ej_viz} except now the final kinetic energy of the jet in Eq.~\ref{eq:Ekjet} is visualised. The regions of the parameter space with no colour indicate either that $M_\mathrm{disk}=0$ or that the minimal accreted disk mass for sGRB creation, $M_\mathrm{acc} = 0.03 \ M_{\odot}$, is not reached. }
    \label{fig:Ek_viz}
\end{figure*}
GRB170817A firmly established BNS mergers as a source of sGRBs~\citep[e.g.][]{abbott_gravitational_2017}. Whether BHNS mergers are also able to produce such ultra-relativistic highly collimated outflows, or jets, is not yet fully understood (see~\citealt{kyutoku_coalescence_2021} for a recent review of BHNS mergers and the expected EM emission). If the NS is tidally disrupted, more mass may be accreted in BHNS mergers than in BNS mergers, which could increase the energy budget of the jet~\citep[see][and references therein]{gompertz_search_2020}. Furthermore, the dynamical part of the ejecta is only present close to the equatorial plane in BHNS mergers~\citep{kyutoku_anisotropic_2013} so it cannot choke a possible jet. Some gas pressure from surrounding ejecta might be necessary, however, to create a highly collimated, and thus jet, outflow (see e.g.~\citealt{nagakura_jet_2014}). Unlike the dynamical ejecta, the disk wind ejecta are present in the polar regions. It is not clear whether an ultra-relativistic outflow can overcome such ejecta with sufficient collimation.~\citet{just_neutron-star_2016} still find that BHNS mergers are able to harbour ultra-relativistic jets because of the lack of polar dynamical ejecta.

A further topic of debate is the mechanism through which the jet is launched. Two often proposed candidates are the Blandford-Znajek (BZ) mechanism~\citep{blandford_electromagnetic_1977} and neutrino pair annihilation~\citep{eichler_nucleosynthesis_1989,meszaros_high-entropy_1992,just_neutron-star_2016}.
\defcitealias{salafia_accretion--jet_2021}{SG21}
\citet[][hereafter referred to as  \citetalias{salafia_accretion--jet_2021}]{salafia_accretion--jet_2021}  
discuss both mechanisms in detail and derive disk accretion-to-jet energy conversion efficiencies. For GW170817, they calculate that both mechanisms have efficiencies that are consistent with GRB170817, and they can therefore not distinguish between the two.~\citet{kyutoku_coalescence_2021} argue that neutrino pair annihilation does not occur on a sufficiently long timescale to explain the observed sGRB duration. Here we thus assumed that a jet gets launched through the BZ mechanism and followed \citetalias{salafia_accretion--jet_2021} for the accretion-to-jet energy conversion efficiency (see~\citealt{barbieri_light-curve_2019} for a similar derivation). For a total accreted disk mass $M_\mathrm{acc} = M_\mathrm{disk} - M_\mathrm{ej,disk}$, we took $M_\mathrm{acc} \geq 0.03 \ M_{\odot}$ as a necessary condition to create an sGRB of $\sim$ 1s duration~\citep{stone_pulsations_2013,pannarale_prospects_2014}.

Taking a fairly typical gamma-ray burst prompt emission efficiency $f_{\gamma}$ of 10\%~\citep{beniamini_revised_2016}, the final kinetic energy of the jet responsible for the relativistic shock producing the afterglow becomes:
\begin{equation}
\label{eq:Ekjet}
    E_k = \frac{1}{2}(1-f_{\gamma})\eta_\mathrm{BZ}M_\mathrm{acc}c^2,
\end{equation}
where the factor one-half accounts for an identical counter-jet.~\citet{mckinney_measurement_2004} describe general-relativistic magneto-hydrodynamical simulations of spinning Kerr BHs with thick accretion disks. \citetalias{salafia_accretion--jet_2021} fit the following functions, extending the fit of~\citet{mckinney_total_2005}, to those simulations for the BZ accretion-to-jet energy conversion efficiency:
\begin{equation}
\label{eq:BZeff}
    \eta_\mathrm{BZ}=\begin{cases}
               1.52 \times 10^{-6} e^{a^\mathrm{f}_\mathrm{BH}/0.06}  &a^\mathrm{f}_\mathrm{BH}\quad \leq 0.25\\
               10^{-4}   \qquad \qquad 0.25 < &a^\mathrm{f}_\mathrm{BH}\quad \leq 0.505 \\
               0.068 \Omega^5_\mathrm{H}  &a^\mathrm{f}_\mathrm{BH}\quad > 0.505,
            \end{cases}
\end{equation}
where $a^\mathrm{f}_\mathrm{BH}$ is the spin of the final remnant BH and $\Omega_\mathrm{H}$ is the dimensionless angular frequency at the BH horizon:
\begin{equation}
\label{eq:omegah}
    \Omega_\mathrm{H} =\frac{a^\mathrm{f}_\mathrm{BH}} {1+\sqrt{1-(a^\mathrm{f}_\mathrm{BH})^2}}.
\end{equation}
To capture the dependence of $a^\mathrm{f}_\mathrm{BH}$ on the binary source parameters, we used the fits to NR simulations in~\citet{zappa_black-hole_2019}, which are extensions of fits to BBH mergers~\citep{jimenez-forteza_hierarchical_2017}. 
Compared to the formulae stated in \citetalias{salafia_accretion--jet_2021}, in the regime $a^\mathrm{f}_\mathrm{BH} \leq 0.25$
an additional factor of $10^{-2}$ is necessary~\citep{salafia_accretion--jet_2022}.

Analogous to Fig.~\ref{fig:ej_viz}, the dependence of $E_k$ on $\chi_\mathrm{BH}$, $Q$, and $\Lambda_\mathrm{NS}$ is visualised in Fig.~\ref{fig:Ek_viz}. Considerable parts of the parameter space do not produce any disk mass or not enough disk mass to satisfy the imposed threshold for sGRB creation. Similar to $M_\mathrm{rem}$, comparable mass binaries with high $\Lambda_\mathrm{NS}$ produce the highest jet energies. There is a strong dependence of $E_k$ on $\chi_\mathrm{BH}$ as well. This is partly inherited from $M_\mathrm{rem}$ but further strengthened by the dependence of $\eta_\mathrm{BZ}$ on $\chi_\mathrm{BH}$~\citep{zappa_black-hole_2019}. The dependence on $\chi_\mathrm{BH}$ has a big influence on the expected sGRB afterglow detection rates as we show in Sect.~\ref{sec:aglowdetrates}.  

To compute the afterglow emission of the jet when it shocks the interstellar medium, we took two approaches. For our parameter inference, full light curve models are required, which we detail in Sect.~\ref{sec:EML}. In the next section we look at the detection rates for the jet afterglow of BHNS mergers. Here we can limit ourselves to an analytical approximation for the peak flux.

\section{Afterglow detection rates with SKA1}
\label{sec:aglowdetrates}
In this section we derive quantitative estimates for the detection rate of sGRB afterglows associated with GW events of BHNS mergers. We rely on a recent population synthesis study done in\defcitealias{broekgaarden_impact_2021}{B21}
\citet[][hereafter referred to as \citetalias{broekgaarden_impact_2021}]{broekgaarden_impact_2021}. A detailed study on the influence of various input parameters such as the star formation rate density is beyond the scope of this work. Instead, we use their fiducial model `A000' (Sect. 3 of \citetalias{broekgaarden_impact_2021}) to obtain a baseline estimate of the detection rate given the GW detector type, the BH spin ($\chi_\mathrm{BH}$), and the NS EOS. 
\subsection{Population synthesis of BHNS mergers}
\citetalias{broekgaarden_impact_2021} provide both the final properties of the BHNS mergers after star formation and the merger rate per redshift ($z$). We followed their methods, which are derived from~\citet{neijssel_effect_2019}, and integrated over 250 redshift shells from $z=0$ to $z=0.5$ to get the total amount of BHNS mergers per year in this volume of space-time. In each redshift shell, we created BHNS mergers according to the merger rate (which is a function of e.g. the assumed star formation rate density) with various $M_\mathrm{BH}$ and $M_\mathrm{NS}$ given by the mass distribution for model A000. We distributed the mergers uniformly in sky position and orientation in order to calculate the GW signal-to-noise ratio (S/N). For a given GW detector sensitivity and S/N threshold, we then obtained the rate of detected GWs from BHNS mergers.

\begin{figure*}[h]
    \centering
    \includegraphics[width=0.8\textwidth]{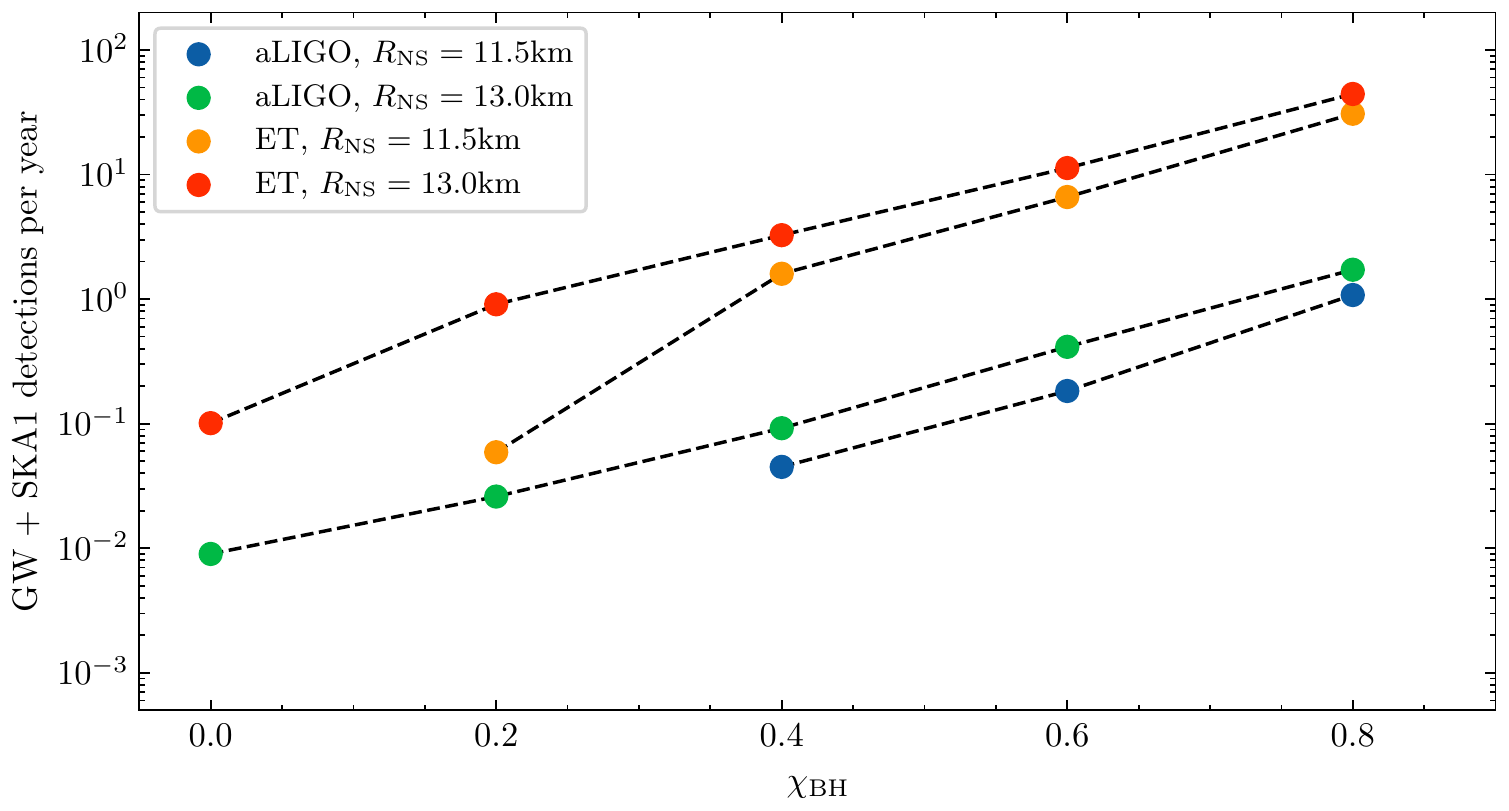}
    \caption{Combined detection rate of BHNS mergers in our simulated population observed through their GW emission and their associated sGRB afterglow emission. The radio emission is observed with SKA1-Mid at nominal frequency and sensitivity. The GW emission is observed with either an aLIGO detector network (blue and green circles) or an ET detector network (orange and red circles). Either a soft NS EOS ($R_\mathrm{NS}=11.5$ km) or a hard NS EOS ($R_\mathrm{NS}=13.0$ km) is assumed. The detection rate is shown as a function of the BH spin ($\chi_\mathrm{BH}$). The dashed black lines are not computed but connect the points to guide the eye.}
    \label{fig:det_rates}
\end{figure*}

To calculate the GW S/N, we proceeded similarly to~\citet{barrett_accuracy_2018} and \citetalias{broekgaarden_impact_2021}.~\citet{barrett_accuracy_2018} computed an interpolated grid of S/Ns for different sets of component masses calculated with GW waveforms IMRPhenomPv2~\citep{hannam_simple_2014,husa_frequency-domain_2016,khan_frequency-domain_2016} and SEOBNRv3~\citep{pan_inspiral-merger-ringdown_2014,babak_validating_2017}. They marginalised over the external parameters~\citep{finn_observing_1993} to compute an averaged detection probability instead of a single S/N value. We omitted the marginalisation and calculated the S/N value instead because we needed the inclination as input for the magnitude of the afterglow flux. We have disregarded the effects of spin and NS tidal disruption on the GW waveform here but take them into account in Sect.~\ref{sec:MMPI}. We looked at both a network of second-generation (2G) GW detectors at design sensitivity and a future network of third-generation (3G) detectors including the Einstein Telescope (ET) with the ET-D sensitivity curve~\citep{hild_sensitivity_2011}. For both networks, we calculated the S/N in a single detector, either aLIGO or ET, and took a detection threshold $\mathrm{S/N}_\mathrm{thresh}\geq 8$ as a standard proxy for a detection, and sufficient localisation for follow-up, with the entire network.

For all BHNS mergers we created, we computed the amount, if any, of disk mass and the resulting kinetic energy of the gamma-ray-burst jet $E_k$ using Eqs.~\ref{eq:mrem} through~\ref{eq:omegah}. For the ejecta formulae, we also needed to specify an EOS and a BH spin. We set the EOS by fixing the radius of the NS, in line with \citetalias{broekgaarden_impact_2021}, to either $R_\mathrm{NS}=11.5$ km (consistent with GW observations; see e.g,~\citealt{the_ligo_scientific_collaboration_and_the_virgo_collaboration_gw170817_2018}) or $R_\mathrm{NS}=13.0$ km (consistent with NICER observations; see~\citealt{miller_psr_2019,raaijmakers_constraints_2021,riley_nicer_2021}). As the distribution of BH spins in BHNS mergers is still highly uncertain, we used an equal (average) value for all the mergers in a single population and varied this value as $\chi_\mathrm{BH}=\{0,0.2,0.4,0.6,0.8\}$ to study its influence. In the next section we use $E_k$ to obtain the peak flux magnitude of the possible sGRB afterglow.
\subsection{sGRB afterglow}
\label{sec:sgrbafterglow}
Similar to previous work~\citep{boersma_search_2021}, we followed~\citet{duque_radio_2019} for an estimate of the sGRB afterglow peak flux based on the theory in~\citet{nakar_detectability_2002}:
\begin{equation}
\label{eq:peakflux}
F_{\mathrm{p},\nu} \propto E_{\mathrm{0}} \ \theta_{c}^{2} \ n_0^{\frac{p+1}{4}} \ \epsilon_{\mathrm{e}}^{p-1} \ \epsilon_{\mathrm{B}}^{\frac{p+1}{4}} \ \nu^{\frac{1-p}{2}} \ d_L^{-2} \ (1+z)^{ \frac{3-p}{2}} \ \mathrm{max}\Big(\theta_c,\theta_\mathrm{obs} \Big)^{-2p},    
\end{equation}
where $\theta_{c}$ is the opening angle of the jet core, $n_0$ is the circumburst density, $\epsilon_{\mathrm{e}}$, $\epsilon_{\mathrm{B}}$ and $p$ are shock microphysics parameters, $\nu$ is the observing frequency, $d_L$ is the luminosity distance and $\theta_
\mathrm{obs}$ is the observing angle. We assumed the jet is produced orthogonal to the orbital plane so that $\theta_{\mathrm{obs}}$ is equal to the inclination angle $\iota$ of the binary merger used in the GW analysis. 

As shown in Eq.~\ref{eq:peakflux}, the observed flux does not directly depend on $E_k$ but on the on-axis isotropic-equivalent jet energy $E_0$. We used a standard formula for top-hat jets to convert between the two: $E_0 = E_k/(1-\cos(\theta_c))$, where the opening angle is fixed to a representative value of $\theta_c=0.1 \ \mathrm{rad}$ ($\approx 5.7$ deg)~\citep{beniamini_electrons_2017}. The literature on gamma-ray-burst  afterglow modelling often uses such a simplified top-hat jet approximation instead of a structured jet model (see e.g.~\citealt{aksulu_exploring_2022} for a recent study). The top-hat jet, in contrast to the structured jet models, assumes no angular dependence of the jet energy, which is a good approximation at small inclination angles. This belief breaks down for afterglows associated with GW detections as these will most likely be viewed off-axis. This was first demonstrated by GW170817 (see e.g.~\citealt{ryan_gamma-ray_2020} and references therein) and is expected to largely hold true for future detections as well, as we show in the next section. While the difference in the full afterglow light curve is substantial between top-hat and structured jets, the discrepancy for the peak flux is not as large~\citep{duque_radio_2019}. Because we focus on detectability only as a function of the peak flux in this section, we stick to Eq.~\ref{eq:peakflux} and employ a structured jet model later in Sect.~\ref{sec:MMPI}.

Similar to~\citet{hotokezaka_hubble_2019}, we fixed $p=2.2$ and $\epsilon_e = 0.1$ to fiducial values. These values are consistent with the observed gamma-ray-burst  population~\citep{beniamini_electrons_2017,aksulu_exploring_2022}. Both $n_0$ and $\epsilon_B$ have a much broader population distributions than $\epsilon_e$ or $p$ so we did not fix these parameters. We took the same approach as~\citet{duque_radio_2019} and considered a log-normal distribution for both parameters with mean $\mu=10^{-3}$ and standard deviation $\sigma=0.75$. Furthermore, $\epsilon_B$ was constrained to the range $[10^{-4},10^{-2}]$. We assumed a flat $\Lambda$ cold dark matter cosmology with parameters given by the WMAP9 dataset~\citep{hinshaw_nine-year_2013}. We observed the afterglows with the SKA1-Mid telescope of SKA, at a nominal frequency of $\nu=1.43 \ $GHz, nominal rms continuum noise of $\sigma_\mathrm{RMS}=2 \ \mu$Jy~\citep{braun_anticipated_2019} and claimed a detection when $F_{p,1.43}$ passes the $5\sigma_\mathrm{RMS}$ threshold. 

We have used 20 combinations of detector type, $R_\mathrm{NS}$ and $\chi_\mathrm{BH}$ in total. We ran each combination for 1000 realisations of one year. We show the averaged detection rates in Fig.~\ref{fig:det_rates}.

\subsection{Rates}
\begin{figure*}[h]
    \centering
    \includegraphics[width=0.8\textwidth]{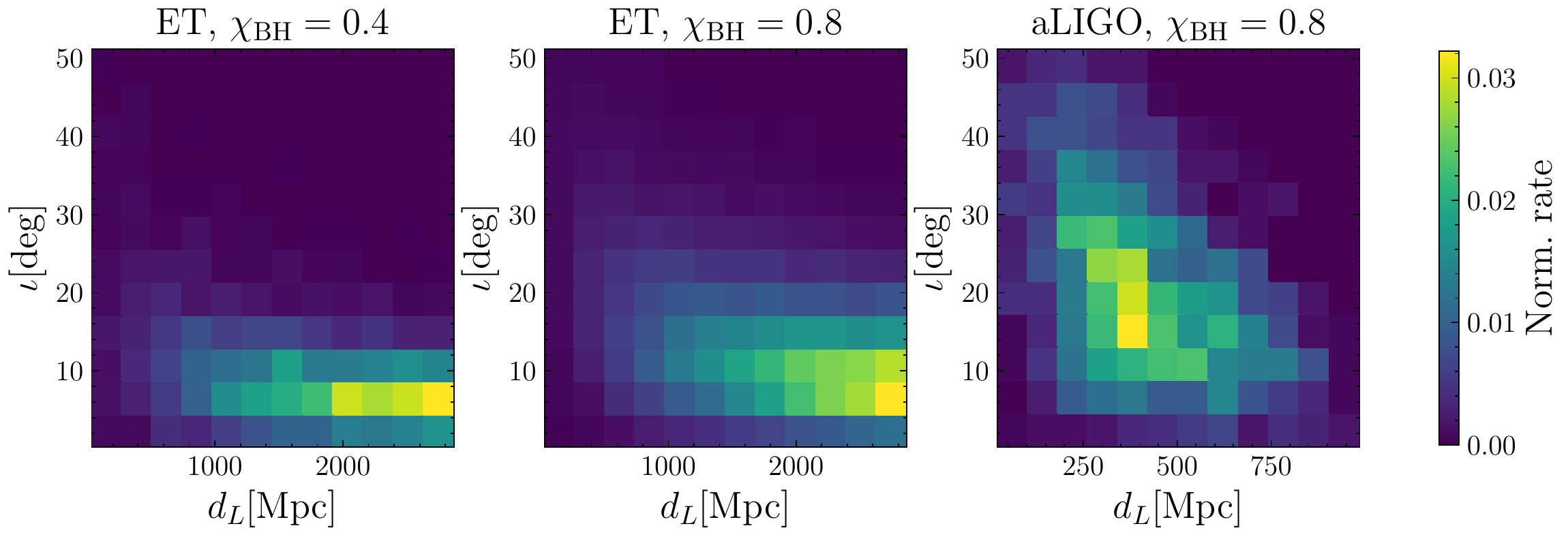}
    \caption{Normalised GW and radio detection rates for the ET and $\chi_\mathrm{BH} = \{0.4,0.8\}$ combinations (left and middle panel) and for the aLIGO and $\chi_\mathrm{BH} = 0.8$ combination (right panel). The rates are normalised to the combined detection rates of Fig.~\ref{fig:det_rates}, given by (from left to right): 3.3 $\mathrm{yr}^{-1}$, 44 $\mathrm{yr}^{-1}$, and 1.7 $\mathrm{yr}^{-1}$. A hard NS EOS ($R_\mathrm{NS} = 13.0$ km) is used in all cases. The rates are shown as a function of the distance and inclination angle of the detected sources.}
    \label{fig:inc_dist_rates}
\end{figure*}
For the aLIGO network, we obtain the same averaged 11 detected GW mergers per year as \citetalias{broekgaarden_impact_2021} verifying our methods. We also calculated the percentage of those GW detections with some form of NS tidal disruption. For $\chi_\mathrm{BH}=0$,  we again recover similar percentages as \citetalias{broekgaarden_impact_2021} of 1.3\% and 3.9\% with $R_\mathrm{NS}=11.5$ km and $R_\mathrm{NS}=13.0$ km, respectively. We provide some additional figures of the simulated populations in Appendix~\ref{app:bhnspop}. We focus on Fig.~\ref{fig:det_rates} and Fig.~\ref{fig:inc_dist_rates} below.

For a network of 2G detectors, we find that the SKA1-Mid detection rates of sGRB afterglows associated with BHNS GW events are likely low. Only for a population of BHNSs with probably unrealistically high $\chi_\mathrm{BH} \sim 0.8$~\citep{fragione_black-holeneutron-star_2021}, the combined detection rate per year nears unity. Transitioning from a soft ($R_\mathrm{NS}=11.5 \ \mathrm{km}$) to a hard ($R_\mathrm{NS}=13.0 \ \mathrm{km}$) NS EOS does increase the rates two- to three-fold but a combined detection remains rare for all but the highest $\chi_\mathrm{BH}$. 

For the soft EOS, we do not observe any combined events with the 2G network in our 1000 realisations for $\chi_\mathrm{BH}=\{0,0.2\}$. Similarly for the soft EOS with the 3G network, we do not observe any combined events for $\chi_\mathrm{BH}=0$. In those cases, that would imply an upper limit, assuming a Poisson distribution, at a 95\% confidence level of less than one combined detection per $\sim$ 300 years.

The situation changes considerably for a network of 3G detectors and non-zero BH spin. Now, one combined detection per year is not unlikely for $\chi_\mathrm{BH} \sim 0.2$, assuming a hard NS EOS. Many combined detections are expected per year for $\chi_\mathrm{BH} \sim 0.4 - 0.8$ but we stress this is most certainly an overestimate of the detection rates.

The stark differences between the rates for the 2G network and the 3G network are a result of the much increased sensitivity of the latter. We find, like~\citet{dobie_radio_2021} (see~\citealt{hotokezaka_radio_2016} for an earlier study on the radio detectability of GW mergers), that SKA1-Mid is able to observe afterglows out to gigaparsec distances, which is beyond the range of aLIGO. Our 3G network, however, is able to observe most of the mergers occurring in our chosen space-time volume, which enables many more combined detections with SKA1-Mid. 
This is visualised in Fig.~\ref{fig:inc_dist_rates}, where the combined GW and radio detection rates are plotted as a function of the distance and inclination angle. This is done for the ET and $\chi_\mathrm{BH} = \{0.4,0.8\}$ combinations and for the aLIGO and $\chi_\mathrm{BH} = 0.8$ combination. The hard NS EOS is used and all rates are normalised to their respective total rate, given in Fig.~\ref{fig:det_rates},  for ease of comparison. Clearly most of the detected sources with the 3G network and SKA1-Mid are located at much greater distances than those detected with the 2G network. The inclination angle distribution of the observed sources also varies with GW detector type and BH spin. In general, because of the strong dependence of the peak flux on the inclination angle, most sources are detected at relatively small $\iota$. Nearby sources (i.e. those detected with aLIGO) with comparatively higher fluxes can still be observed at larger $\iota \gtrsim 20$ deg as well. Similarly, sources with high $\chi_\mathrm{BH} \sim 0.8$ and subsequently larger jet energies are also more readily detected at larger $\iota$. For the ET and $\chi_\mathrm{BH} = 0.4$ combination, we predict about $~30$\% of the afterglows to be observed on-axis ($\iota \leq \theta_c$).

A 3G network is expected to detect mergers at significantly larger redshifts than our maximum of $z = 0.5$~\citep{maggiore_science_2020}. We also find that $\sim 25-35\%$ (depending on the chosen $\chi_\mathrm{BH}$) of our detected afterglows in the radio are between $0.4<z<0.5$. This implies SKA1-Mid will be sensitive enough to detect afterglows with $z>0.5$ as well. While these relatively distant mergers are not taken into account here, some of these will thus be observable both through GWs and their afterglows. Such mergers will increase the rates for the 3G network and SKA1-Mid combination further but we note that a sufficient localisation for follow-up is not guaranteed at larger distances~\citep{maggiore_science_2020}. 

The combined detection rates are strongly dependent on the assumed $\chi_\mathrm{BH}$ and an approximate log-linear relation is visible in Fig.~\ref{fig:det_rates}. We attribute this partly to the formulation of the BZ accretion-to-jet energy conversion efficiency in Eq.~\ref{eq:BZeff}. $E_0$ is directly proportional to this efficiency and by extension $F_{p,1.43}$ as well. For our set of $\chi_\mathrm{BH}$, we are usually in the regime where $a^\mathrm{f}_\mathrm{BH} > 0.505$~\citep{zappa_black-hole_2019}. As the final BH spin $a^\mathrm{f}_\mathrm{BH}$ is approximately linear in $\chi_\mathrm{BH}$~\citep{zappa_black-hole_2019}, the log-linear dependence on $\chi_\mathrm{BH}$ follows naturally from Eq.~\ref{eq:BZeff} in this regime. The dependence of $M_\mathrm{acc}$ on $\chi_\mathrm{BH}$ is less intuitively understood, but we observe a roughly log-linear slope there too. 

In conclusion, the chances of finding any sGRB afterglow of a BHNS merger in the near future are slim.~\citet{fragione_black-holeneutron-star_2021} even argue that it is unlikely to detect most forms of EM emission associated with BHNS mergers when looking at the small fraction of mergers with ejecta mass. We remain cautiously optimistic on the basis of our simulations that the low probability of ejecta mass might be offset by the sheer amount of mergers a 3G detector network will be able to observe. Other forthcoming GW detectors built in the more immediate future, such as the LIGO Voyager~\citep{adhikari_cryogenic_2020}, also promise significantly increased rates of detection.

We end this section with a few caveats to our results. Any radio source count with a negative slope will find most objects just above the instrument flux density limit (see e.g.~\citealt{mooley_sensitive_2013}). Any changes in the final sensitivity of SKA1-Mid will thus have a big impact on the detection numbers. If we would set the detection threshold at a more conservative $10\sigma_\mathrm{RMS}$, for example, we would lose $\sim$ 40\% - 50\% of the previously detected afterglows. A further source of uncertainty is the disk mass required for sGRB creation. We set the threshold at the lower end of the uncertainty range~\citep{stone_pulsations_2013,pannarale_prospects_2014}. Increasing the mass required will consequently lower the rates. Furthermore, we would like to iterate that we have used only one population synthesis model of the sizeable 420 model variations in \citetalias{broekgaarden_impact_2021}. They find a large range in both the predicted GW detections per year as well as the percentage of those detections with any remnant mass over their model variations. We provide point estimates here for the fiducial model but remind the reader that such numbers are inherently surrounded by large population uncertainties given the few BHNS GW detections to date. In the future, even if no EM emission is detected, more GW observations will allow the population of BHNS mergers to be better understood. 

\section{Multi-messenger parameter inference}
\label{sec:MMPI}
In Sect.~\ref{sec:aglowdetrates} we consider single flux measurements of sGRB afterglows associated with BHNS mergers. A sole measurement above the detection threshold could already give some hints on the binary source properties by confirming the presence of ejecta mass. Much more physics, however, is encompassed in a light curve consisting of multiple data points. In the remainder of this paper we investigate, given a GW observation, what the added benefit of a radio light curve could be in inferring the binary source properties. Conversely, we examine the ability of a GW observation to help constrain the parameters of the sGRB afterglow too.  
\subsection{Bayesian framework}
We took a Bayesian approach to incorporate both GW and EM radiation in our parameter inference. This method has been used extensively in analysing GW data (see e.g.~\citealt{romero-shaw_bayesian_2020}) and we proceeded similarly to~\citet{raaijmakers_challenges_2021} when incorporating the EM data. 

To compute the posterior density function of a certain set of signal parameters $\vec{\theta}$ we made use of Bayes' theorem. For a given signal $d$ and a model of the signal $h(\vec{\theta})$, the posterior density function is proportional to:
\begin{equation} \label{eq:Bayes}
    p(\vec{\theta}|d,h(\vec{\theta})) \propto \mathcal{L}(d|\vec{\theta},h(\vec{\theta}))p(\vec{\theta}|h(\vec{\theta})),
\end{equation}
where $\mathcal{L}(d|\vec{\theta},h(\vec{\theta}))$ is the likelihood function and $p(\vec{\theta}|h(\vec{\theta}))$ is the prior information on $\vec{\theta}$ given the model. For multiple signals, either gravitational or EM in nature, the joint likelihood becomes the product of the $N$ individual likelihoods assuming the noise streams in the various detectors are uncorrelated:
\begin{equation}
    \mathcal{L}_{\textrm{joint}} = \prod_{\text{i}=1}^{{N}} \mathcal{L}_i(d_i|\vec{\theta},h_i(\vec{\theta})).
\end{equation}
In practice, we sampled from the joint log-likelihood where the product over the individual likelihoods is replaced with a sum over their natural log counterpart. To sample from $\log \mathcal{L}_{\textrm{joint}}$, we employed the \texttt{MultiNest} nested sampler~\citep{feroz_multinest_2009} in its Python implementation incorporated into the \texttt{bilby} Python package~\citep{ashton_bilby_2019}. While \texttt{bilby} is primarily developed for pure GW data analysis, its flexibility and extensive documentation make it well suited for our purposes. In the next sections we discuss the specific forms of the GW and EM likelihoods.
\subsection{GW likelihood}
\label{sec:GWL}
We used a network of three GW detectors consisting of the two aLIGO detectors (in Hanford and Livingston) and the Virgo detector at their design sensitivities, hereafter referred to as the LHV network. The log-likelihood of this network takes the standard form (see e.g.~\citealt{cutler_gravitational_1994}):
\begin{equation}
\label{eq:loglikeGW}
    \log \mathcal{L}_{\mathrm{GW}}  \propto 
    -\frac{1}{2}\sum_{\text{i}=1}^{{N_d = 3}} \bigg \langle d^{\mathrm{GW}}_i- h^{\mathrm{GW}}_i(\vec{\theta}_{\mathrm{GW}}) \ | \ d^{\mathrm{GW}}_i- h^{\mathrm{GW}}_i(\vec{\theta}_{\mathrm{GW}}) \bigg \rangle \,,
\end{equation}
where the inner product for two GW strains $a(t)$ and $b(t)$ is defined as
\begin{equation}
\label{eq:overlap}
\langle a|b \rangle = 4\Re \int \frac{\tilde{a}^*(f)\tilde{b}(f)}{S_n(f)} df.
\end{equation}
Here the tilde superscript implies a Fourier transform, and $S_n(f)$ is the 
power spectral density of the detector's noise. In \texttt{bilby}, the likelihood of Eq.~\eqref{eq:loglikeGW} is calculated using the \textit{GravitationalWaveTransient} class (see~\citealt{ashton_bilby_2019} for more details).

For our GW waveform model $h^{\mathrm{GW}}(\vec{\theta}_{\mathrm{GW}}),$ we chose a recent model, `$\mathrm{SEOBNRv4\_ROM\_NRTidalv2\_NSBH}$'~\citep{matas_aligned-spin_2020}, which is specifically tailored to aligned-spin BHNS mergers. It includes tidal effects and also accounts for NS tidal disruption in the late inspiral, merger and ring-down part of the waveform. The data $d^{\mathrm{GW}}$ in each detector was generated by projecting the same waveform model onto the detector frame and adding coloured Gaussian noise scaled by the power spectral density.   
The set of signal parameters $\vec{\theta}_{\mathrm{GW}}$ used to generate the waveform consists of various parameters intrinsic and extrinsic to the source. While we have defined most of these parameters in previous sections already, we list them here for clarity:
(i) the mass of the BH, $M_{\mathrm{BH}}$, and the mass of the NS, $M_{\mathrm{NS}}$. 
    As it is relatively inefficient to sample in the two component masses because of strong correlations between them, we transformed them to the chirp mass, $\mathcal{M}_c$, and the mass ratio, $q$\footnote{Not to be confused with $Q=1/q$.}. The chirp mass is defined as 
    \begin{equation}\mathcal{M}_c = \frac{(M_{\mathrm{NS}} \cdot M_{\mathrm{BH}})^{(3/5)}}{(M_{\mathrm{NS}} + M_{\mathrm{BH}})^{(1/5)}}, \ \mathrm{and} \ q = M_{\mathrm{NS}}/M_{\mathrm{BH}};
    \end{equation}
    (ii) the spin of the BH, $\chi_{\mathrm{BH}}$, and the spin of the NS, $\chi_{\mathrm{NS}}$;  (iii) the tidal deformability of the NS, $\Lambda_\mathrm{NS}$. We set the tidal deformability of the BH to zero; (iv) the inclination angle, $\iota$, and the polarisation, $\psi$; and (v) the sky position in right ascension, $\alpha$, and declination, $\delta$, and the luminosity distance, $d_{L}$. By assuming an EM counterpart was detected and the source was thus localised to a host galaxy, we did not need to sample in these parameters and we could set them to their true injected values.

To fully specify the waveform, the time of coalescence $t_c$ and the phase of coalescence $\phi_c$ need to be given as well. Because these parameters are not of interest for our analysis but can drastically increase sampling time, we analytically marginalised over them using built-in functions in \texttt{bilby}.

\subsection{EM likelihood}
\label{sec:EML}
\begin{figure*}[h]
    \centering
    \includegraphics[width=0.8\textwidth]{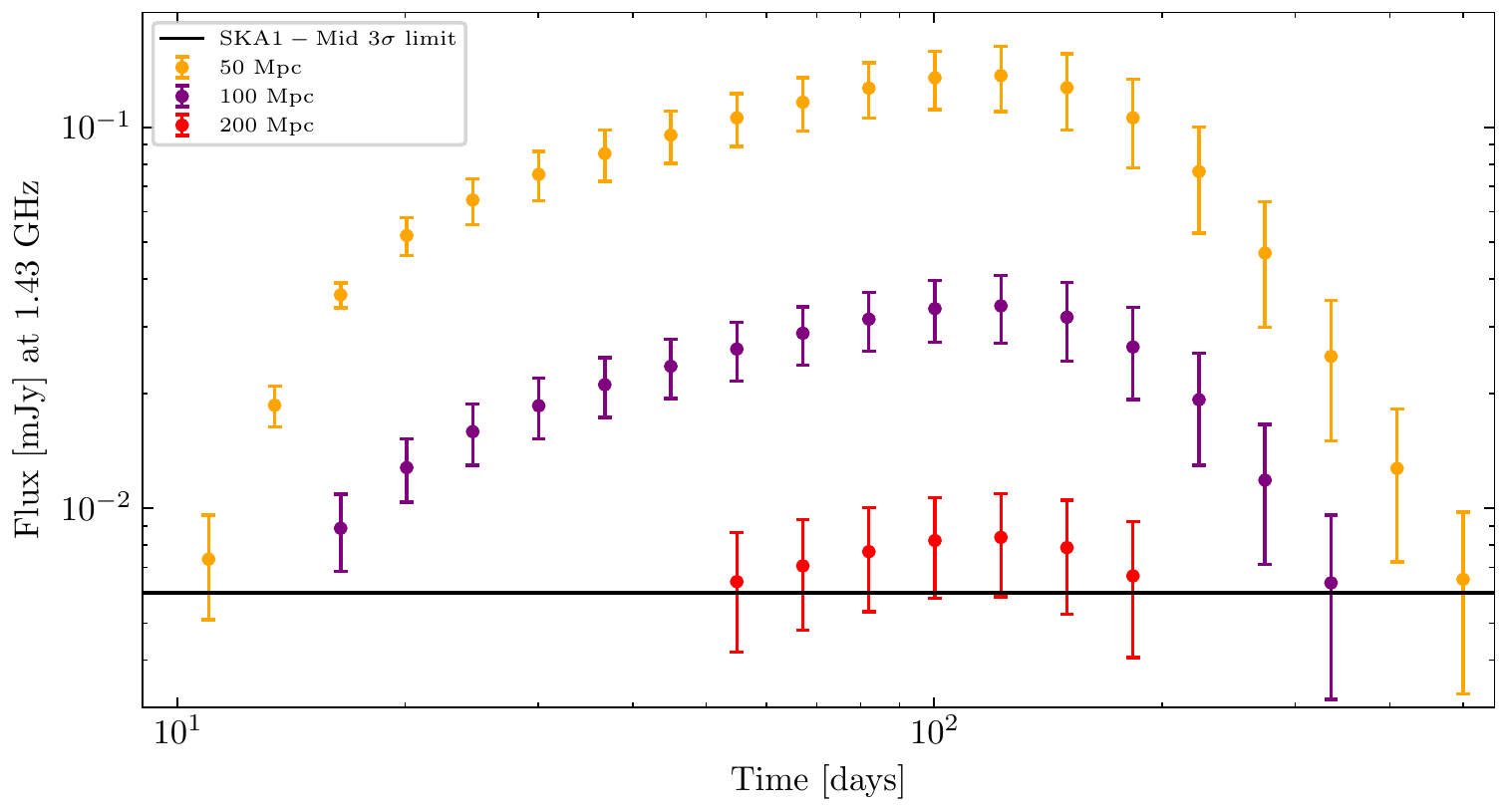}
    \caption{sGRB afterglow radio light curves, computed with \texttt{afterglowpy}, of our fiducial binary at the nominal SKA1-Mid frequency, $\nu=1.43$ GHz. The curves are plotted on a log-log scale. Twenty simulated observations were taken between 11 and 500 days post-merger. The orange circles indicate the 20 observations above the 3$\sigma_\mathrm{RMS}$ detection threshold when the binary is placed at $d_L=50$ Mpc. The purple circles indicate the 16 observations above the 3$\sigma_\mathrm{RMS}$ detection threshold when the binary is placed at $d_L=100$ Mpc. The red circles indicate the seven observations above the 3$\sigma_\mathrm{RMS}$ detection threshold when the binary is placed at $d_L=200$ Mpc. The solid black line shows the SKA1-Mid $3\sigma_\mathrm{RMS}$ detection threshold with $\sigma_\mathrm{RMS}=2 \ \mu$Jy. The error bars give the total error as a combination of the measurement error and the systematic error.}
    \label{fig:lightcurves}
\end{figure*}
For our model $h^{\mathrm{EM}}(\vec{\theta}_{\mathrm{EM}})$ of the jet synchrotron afterglow at radio frequencies, we turned to the \texttt{afterglowpy} Python package~\citep{ryan_gamma-ray_2020}. It provides functionality to quickly calculate synthetic light curves at various frequencies for afterglows arising from structured jets. It produces reasonably accurate light curves compared to codes using relativistic numerical hydrodynamic simulations such as \texttt{BoxFit}~\citep{eerten_gamma-ray_2012} at a small fraction of the compute time~\citep{ryan_gamma-ray_2020}. This makes it ideal for parameter estimation studies where the model needs to be calculated many times. We refer to~\citet{ryan_gamma-ray_2020} for further details on \texttt{afterglowpy}. We used the `power law jet' model light curves from \texttt{afterglowpy} in this work.  In general, these light curves are a function of the same parameters as the top-hat jet peak flux in Eq.~\ref{eq:peakflux}. Some additional parameters are necessary for structured jets, however. We summarise here all the parameters $\vec{\theta}_\mathrm{EM}$ needed to generate the light curves:
 (i) the observing frequency, $\nu$, and observing times, $t_\mathrm{obs}$; (ii) the core opening angle of the jet, $\theta_c$, which determines the effective width of the jet core; (iii) the on-axis isotropic-equivalent jet energy, $E_0$; (iv) the observing angle, $\theta_{\mathrm{obs}}$. As in Sect.~\ref{sec:sgrbafterglow}, we assumed $\theta_{\mathrm{obs}} = \iota$; (v) the truncation angle, $\theta_w$, which determines how far the wings of the jet extend; (vi) the power law index, $b$, which determines how energetic the wings of the jet are; (vii) the circumburst density, $n_0$. We treated $n_0$ as being uniform, though not fixed, in value; (viii) the index of the Lorentz factor distribution, $p$, of the accelerated electrons and the fraction, $\epsilon_e$, of thermal energy in those electrons. As in Sect.~\ref{sec:sgrbafterglow}, we set $p=2.2$ and $\epsilon_e = 0.1$; (ix) the fraction, $\epsilon_B$, of magnetic energy relative to thermal energy; (x) the fraction of accelerated electrons, $\xi_N$. We fixed $\xi_N = 1$, in line with~\citet{ryan_gamma-ray_2020}; and (xi) the luminosity distance, $d_L$, which we set to the true injected value, as in Sect. \ref{sec:GWL}.

We assumed both Gaussian measurement errors $\sigma_\mathrm{RMS}$ and systematic errors $\sigma_{\mathrm{sys}}$ on $k$ data points $d^{\mathrm{EM}}$ generated using \texttt{afterglowpy}. This leads to the following likelihood for our EM data:
\begin{equation}
\label{eq:loglikeEM}
    \log \mathcal{L}_\mathrm{EM} = -\frac{1}{2}\sum_{\text{i}=1}^{{k}}\Bigg[ \frac{\Big(d^{\mathrm{EM}}_i - h_i^{\mathrm{EM}}(\vec{\theta}_{\mathrm{EM}})\Big)^2}{\sigma_\mathrm{RMS}^2 + \sigma_{\mathrm{sys},i}^2} - \log(2\pi (\sigma_\mathrm{RMS}^2 + \sigma_{\mathrm{sys},i}^2))\Bigg]
.\end{equation}

\subsection{Connecting the EM and GW likelihoods}
\label{sec:conEMGW}
Similar to Sect.~\ref{sec:aglowdetrates}, we expressed the kinetic jet energy, $E_k$, in terms of the binary source parameters. In this way, we connect our EM measurements to our GW measurements. We thus did not sample in $E_k$, or more precisely $E_0$, directly but used Eqs.~\ref{eq:mrem} through~\ref{eq:omegah} to calculate $E_k$ from $\mathcal{M}_c$, $q$, $\chi_\mathrm{BH}$ and $\Lambda_\mathrm{NS}$. After converting the resulting $E_k$ to $E_0$ using the specific formula for structured jets in the \texttt{afterglowpy} code, the EM data points, $d^\mathrm{EM}$, could be computed from the rest of the parameters in $\vec{\theta}_\mathrm{EM}$. Our complete set of sampling parameters is thus
\begin{equation}
    \vec{\theta} = \vec{\theta}_\mathrm{GW} \cup \vec{\theta}_\mathrm{EM} = \{\mathcal{M}_c,q,\chi_\mathrm{BH},\chi_\mathrm{NS},\Lambda_\mathrm{NS},\iota,\psi,\theta_c,\theta_w,b,n_0,\epsilon_B\}.
\end{equation}
Our joint log likelihood is simply the sum of Eqs.~\ref{eq:loglikeGW} and~\ref{eq:loglikeEM}:$~\log \mathcal{L}_\mathrm{joint} = \log \mathcal{L}_\mathrm{GW} + \log \mathcal{L}_\mathrm{EM}$.

We did not assume any other dependence of the parameters in $\vec{\theta}_\mathrm{EM}$ on $\vec{\theta}_\mathrm{GW}$. In reality, these dependences likely do exist to some degree. We return to this point in Sect.~\ref{sec:discussion}.
\subsection{Systematics}
\label{sec:systematics}
The error $\sigma_\mathrm{sys}$ represents the uncertainty that arises when converting the binary source parameters into EM data points using fits to NR data. These fits are not exact and we used their residual errors as input for $\sigma_\mathrm{sys}$.To incorporate the errors in the remnant mass and the dynamical ejecta, we generated 1000 samples from two Gaussian distributions in $M_\mathrm{rem}$ and $M_\mathrm{dyn}$ each with means given by Eq.~\ref{eq:mrem} and Eq.~\ref{eq:mdyn}, respectively. The standard deviation for each distribution is assumed to be a relative error of 15\% in $M_\mathrm{rem}$~\citep{foucart_remnant_2018} and an absolute error of $\sigma_\mathrm{dyn}= 0.0047 \ M_{\odot}$ in $M_\mathrm{dyn}$~\citep{kruger_estimates_2020}. For each generated combination of samples, we calculated the $k$ data points of our light curve as detailed in Sect.~\ref{sec:conEMGW}. We set $\sigma_{\mathrm{sys},i}$ equal to the standard deviation of the resulting flux distribution for each data point $d^{\mathrm{EM}}_i$. We discuss more potential sources of systematic errors in Sect.~\ref{sec:discussion}.
  
Having specified our Bayesian framework, we apply it to a fiducial BHNS merger.

\subsection{Fiducial BHNS merger}
\begin{table}[h]
    %\centering
    \caption{Parameters of our fiducial BHNS merger and the chosen prior type and range for the parameter inference.}
    \begin{tabular}{lllll}
    %\hhline{========}
       Parameter  & Fiducial value & Prior type & Range 
       \\
       \hline
       $M_\mathrm{BH}$ & $9.0 \ M_{\odot}$ & - & -   \\
       $M_\mathrm{NS}$ & $1.7 \ M_{\odot}$ & - & -   \\
       $\mathcal{M}_c$ & $3.198 \ M_{\odot}$ & Uniform & $(2.7,3.7)$   \\
       $q$ & $0.189$ & Uniform & $(0.05,1.0)$  \\
       $\chi_\mathrm{BH}$ & $0.7$ & Uniform & $(-0.9,0.9)$ \\
       $\chi_\mathrm{NS}$ & $0.02$ & Uniform & $(-0.05,0.05)$ \\
       $\Lambda_\mathrm{NS}$ & $400$ & Uniform & $(10,3000)$ \\
       $\iota$ & $0.4 \ \mathrm{rad}$ & Sine & $(0,\pi/2)$ \\
       $\psi$ & $2.659 \ \mathrm{rad}$ & Uniform & $(0,\pi)$ \\
       $\theta_c$ & $0.05 \ \mathrm{rad}$ & Uniform & $(0,\pi/2)$ \\
       $\theta_w$ & $0.2 \ \mathrm{rad}$ & Uniform & $(0,12\theta_c)$\\
       $b$ & 6.0 & Uniform & $(0,10)$ \\
       $n_0$ & $10^{-3} \ \mathrm{cm}^{-3}$ & Log-Uniform & $(10^{-5},10^3)$ \\
       $\epsilon_B$ & $10^{-3.7}$ & Log-Uniform & $(10^{-5},1)$ \\
       $d_L$ & $50,100, 200 \ \mathrm{Mpc}$ & - & -   
       
    \end{tabular}
    \label{tab:paramsandpriors}
\end{table}
For our fiducial BHNS merger, we chose binary source parameters comparable to  GW200105~\citep{abbott_observation_2021} but adjusted to generate enough ejecta mass for an observable sGRB afterglow. Particularly, we set a high $\chi_\mathrm{BH}=0.7,$ which is beneficial for detectability as shown in Fig.~\ref{fig:det_rates}. We set $\Lambda_\mathrm{NS} = 400,$ which implies a hard EOS for the corresponding NS mass. We chose a small $\chi_\mathrm{NS}=0.02$, consistent with the spins observed in galactic BNSs~\citep{burgay_increased_2003}. We picked a relatively large inclination angle, $\iota=0.4 \ \mathrm{rad,}$ consistent with GW170817 (see e.g.~\citealt{ryan_gamma-ray_2020}), and a random value for $\psi$ in $(0,\pi)$. The other parameters, $\theta_c,\theta_w,b,n_0$, and $\epsilon_B$, all have fiducial values similar to the inferred values of GW170817 for the power law jet model in~\citet{ryan_gamma-ray_2020}. While we did not sample in $d_L$, we varied the injected value to generate GW and sGRB afterglow detections with different S/Ns. All fiducial parameters are listed in Table~\ref{tab:paramsandpriors}.
\subsection{Setup}
\begin{figure*}[h]
    \centering
    \mbox{\includegraphics[width=0.48\textwidth]{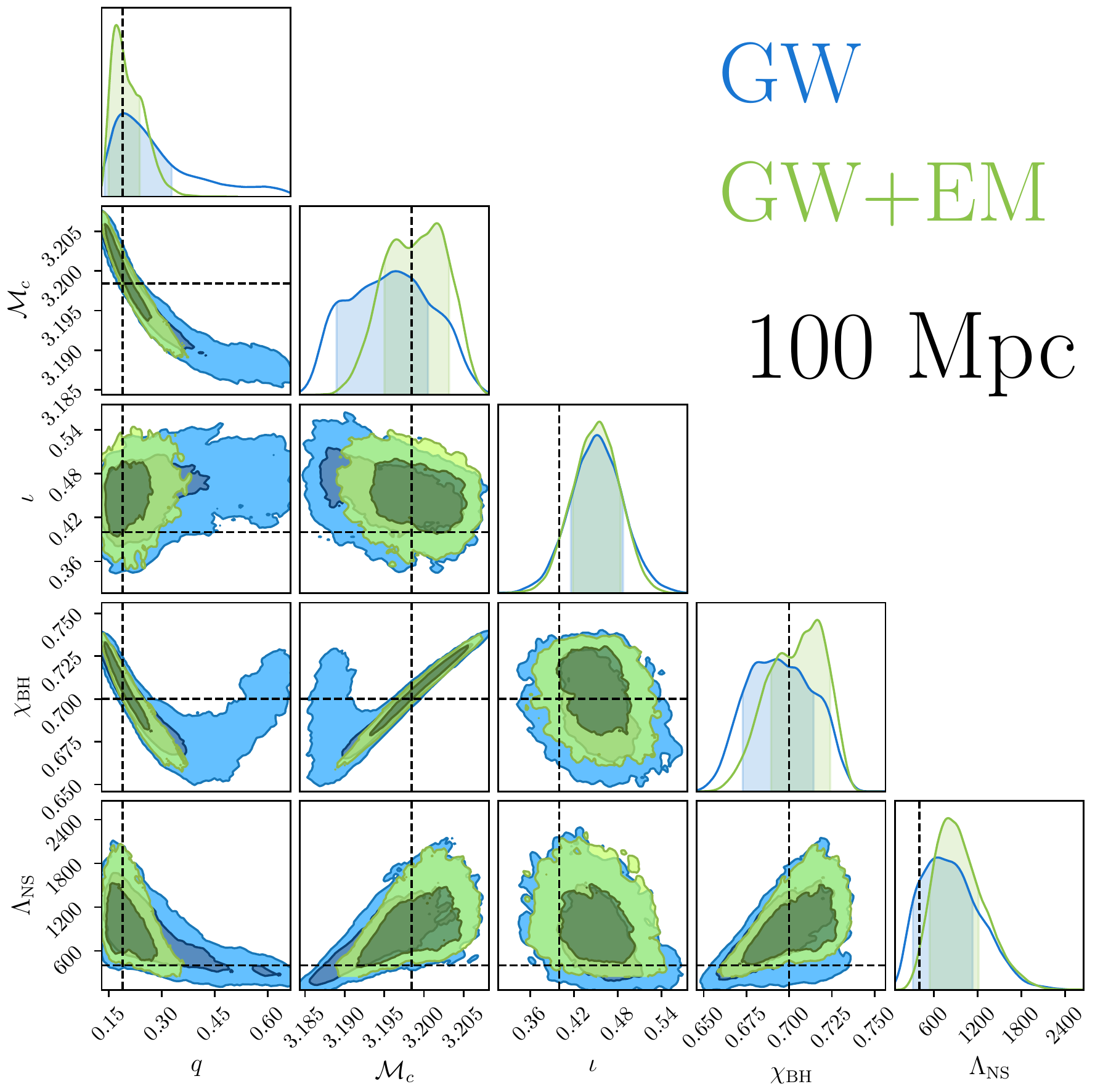}}
    \mbox{\includegraphics[width=0.48\textwidth]{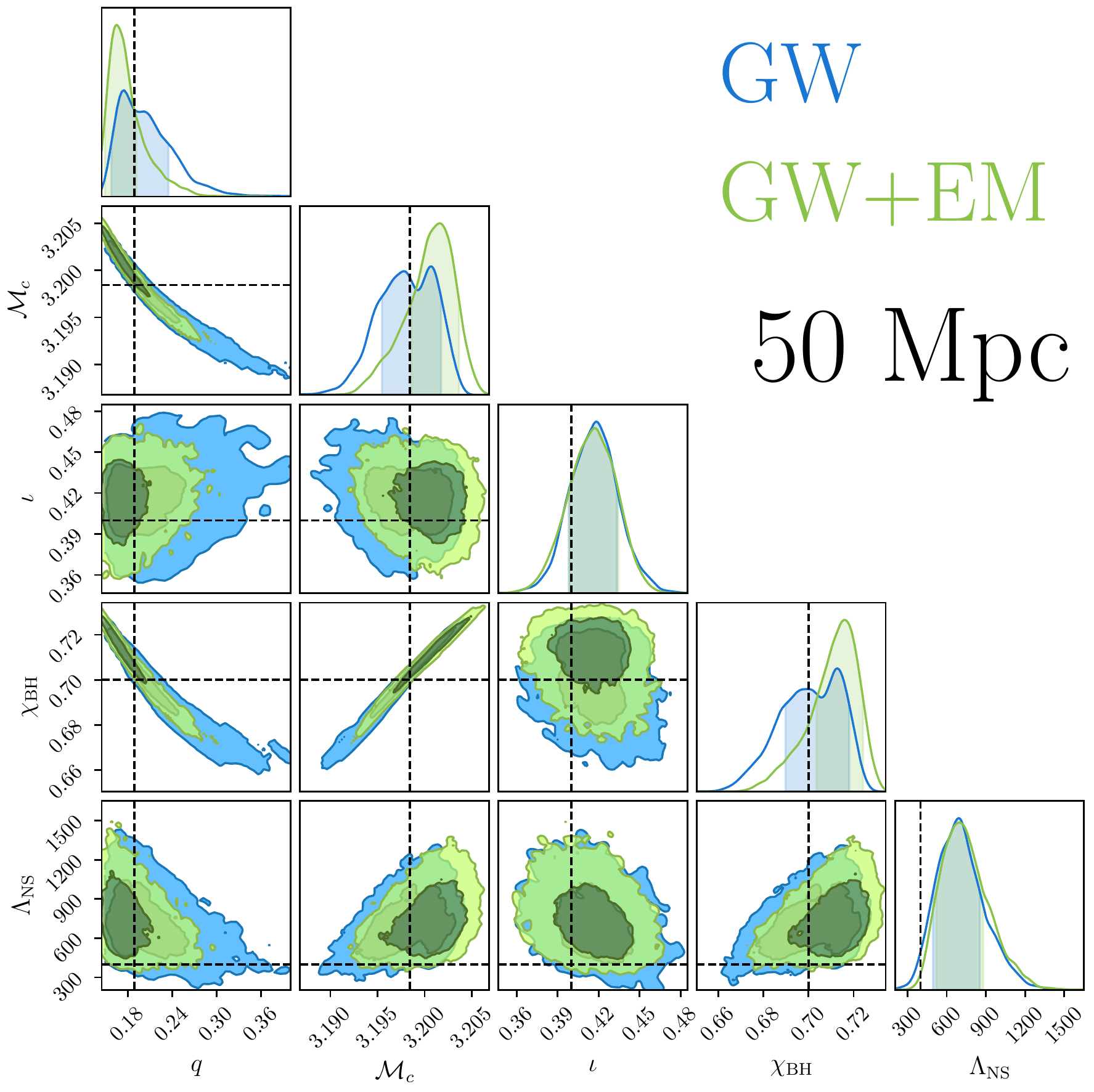}}
    \caption{Posterior distributions of the parameters $\vec{\theta}_\mathrm{GW}$ as inferred for our fiducial binary at $d_L=100$ Mpc (left panel) and $d_L=50$ Mpc (right panel). Blue contours give the posterior distribution when only using the GW data, whereas the green contours give the posterior distribution when using both the GW and radio data. The dark and light shaded regions give the 68\% and 95\% confidence levels, respectively, for all posteriors shown. The fiducial injected values are indicated by the dashed black lines.}
    \label{fig:PEGW}
\end{figure*}
We injected an 8s BHNS merger signal, using the methods in Sect.~\ref{sec:GWL}, with our fiducial parameters into the LHV GW detector network starting at a frequency of 40 Hz and we used a cutoff frequency of 4096 Hz. We supposed the merger was sufficiently localised when the targeted radio follow-up was commenced. We then started observing with SKA1-Mid at $\nu = \textrm{1.43 GHz}$ 11 days after the merger and took 20 observations geometrically spaced until 500 days post-merger. We set a constant SKA1-Mid measurement error at $\sigma_\mathrm{RMS}=2 \ \mu\mathrm{Jy}$ and assumed a data point was detected if it passed the $3\sigma_\mathrm{RMS}$ threshold. Because we took multiple measurements of the same source instead of just a single observation, the detection threshold could be set lower for each measurement than in Sect.~\ref{sec:aglowdetrates}. All 20 observations at $d_L=50$ Mpc give a flux measurement above the threshold. At $d_L=100$ Mpc, the first and last two observations have flux measurements below this value and were not taken into account. At $d_L=200$ Mpc, only seven observations between approximately 55 and 183 days post-merger are above the detection threshold. We calculated the systematic errors as described in Sect.~\ref{sec:systematics} for our fiducial parameter set. We show the resulting light curves in Fig.~\ref{fig:lightcurves}. 

To perform the parameter estimation in $\vec{\theta}$, we chose the broad uninformative priors listed in Table~\ref{tab:paramsandpriors}. Specifically, we took uniform priors for $\mathcal{M}_c,q,\chi_\mathrm{BH},\chi_\mathrm{NS},\Lambda_\mathrm{NS},\psi,\theta_c,\theta_w, $ and $b$, while for $n_0$ and $\epsilon_B$ we chose log-uniform priors. We assumed an isotropic prior for the orientation of the binary, $\iota$. We initialised \texttt{MultiNest} with standard parameters through \texttt{bilby} and sampled using 1008 live points\footnote{For efficient parallelisation, the amount of live points should be a multiple of the number of cores used, 48 in our case, to perform the inference on.}. For each $d_L$, we did two runs: one incorporating both GW and EM data ($\log \mathcal{L}_\mathrm{joint}$) and one using only GW data ($\log \mathcal{L}_\mathrm{GW}$) for comparison.

For the two values of $d_L = 50$ Mpc and $d_L = 100$ Mpc, the GW signal network S/N is 117 and 59, respectively. These are well above the canonical threshold of $\mathrm{S/N}_\mathrm{thresh} \geq 8$. In those cases, we are focusing on the effects of an sGRB afterglow detection in the presence of an already loud GW signal. This is interesting for a few reasons. In the 3G detector network era, high GW S/N detections will become commonplace because of the increased sensitivity. There are other benefits to 3G detectors, such as their broader frequency range, that are not considered here. Still, a loud GW detection in our LHV network will serve as a reasonable testing ground for the extra information an sGRB afterglow detection with SKA1-Mid might give in this scenario. Furthermore, placing the merger relatively nearby allows all parts of the light curve to be well sampled at SKA1-Mid sensitivity. This enables a robust study of the radio afterglow in the limit of small relative measurement errors. Even so, as we describe in Sect.~\ref{sec:aglowdetrates}, most radio afterglows will be detected close to the detection limit. Here, the measurement errors become substantial relative to the observed flux. To investigate this scenario too, we also placed our fiducial binary at $d_L = 200$ Mpc. At this distance, the light curve consists of fewer measurements, as shown in Fig.~\ref{fig:lightcurves}, and the relative error on those measurements is larger as well. The GW signal is also weaker but the network S/N of 29 is still above the detection threshold. 

We show the results in the next section.
\section{Results}
\begin{figure}[h]
    \centering
    \includegraphics[width=0.45\textwidth]{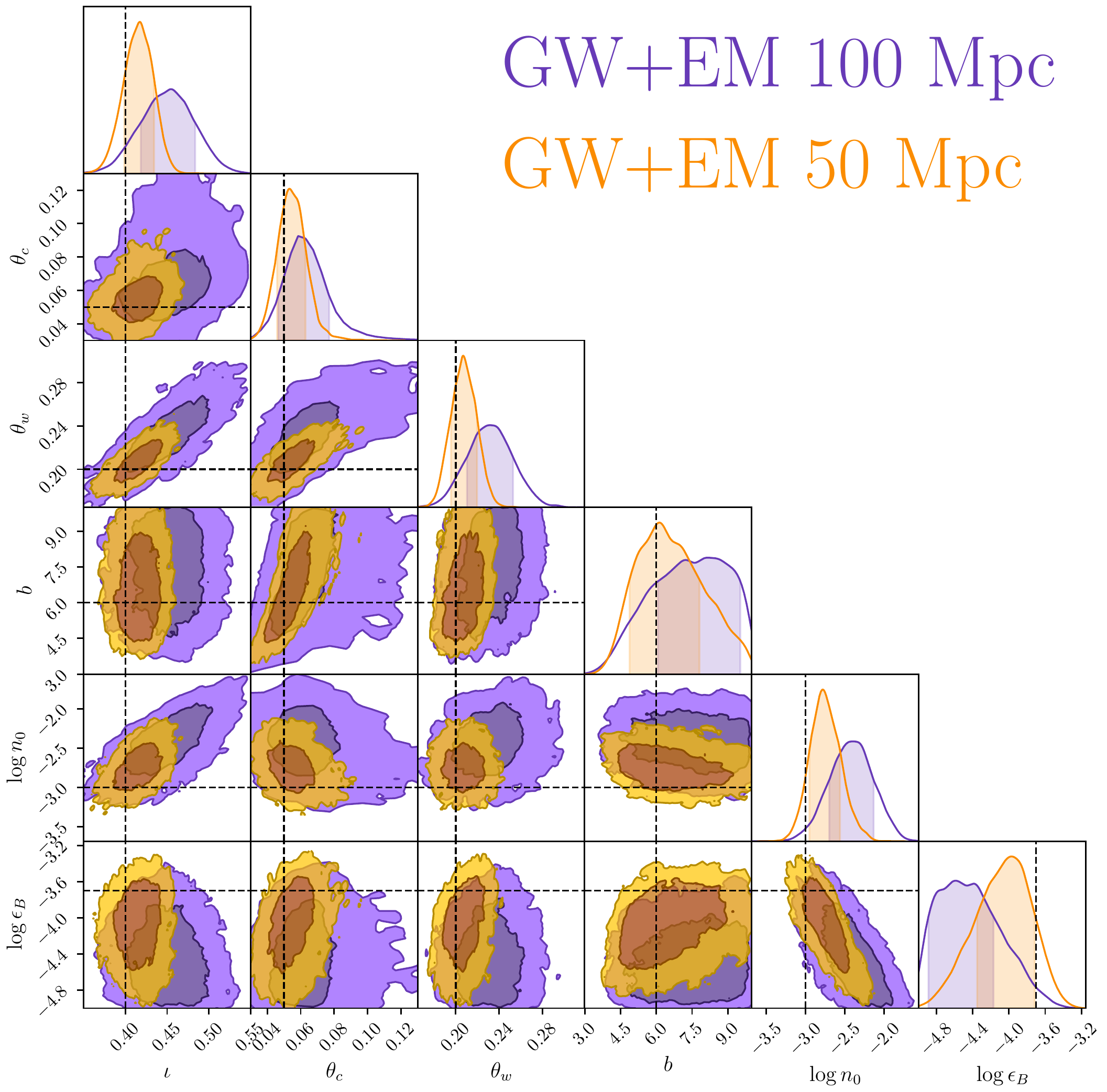}
    \caption{Posterior distributions of the parameters $\vec{\theta}_\mathrm{EM}$. Purple contours give the inferred posterior distribution from the combination of the GW and radio data when the binary is placed at $d_L=100$ Mpc. Orange contours give the inferred posterior distribution from the combination of the GW and radio data when the binary is placed at $d_L=50$ Mpc. The fiducial injected values are indicated by the dashed black lines.}
    \label{fig:PEEM}
\end{figure}
\label{sec:PEresults}
\subsection{50 Mpc and 100 Mpc}
In Fig.~\ref{fig:PEGW} we show the posteriors for $\vec{\theta}_\mathrm{GW}$ for $d_L= 100$ Mpc (left panel) and $d_L= 50$ Mpc (right panel) when sampling from $\log \mathcal{L}_\mathrm{GW}$ and when sampling from $\log \mathcal{L}_\mathrm{joint}$. The parameters $\vec{\theta}_\mathrm{EM}$ are only inferred in the latter case. The posteriors of these parameters are shown in Fig.~\ref{fig:PEEM} for $d_L= 100$ Mpc and for $d_L= 50$ Mpc.

At 100 Mpc, including EM data of the afterglow provides a range of improvements in the estimates of the binary source parameters over using the GW data alone. It is most notable in the marginalised posterior of $q$, where the long tail towards more equal mass ratios can be ruled out using the SKA1-Mid observations. At high $q$, corresponding to more equal $M_\mathrm{BH}$ and $M_\mathrm{NS}$, the amount of disk mass likely becomes too large to still be consistent with the observed light curve. Both $\mathcal{M}_c$ and $\chi_\mathrm{BH}$ are well constrained from the GW data alone, and their estimates are slightly improved with the EM data added. Because the afterglow does not depend on either $\chi_\mathrm{NS}$ or $\psi$ in our model, the EM data provides no additional information on their already weak constraints by the GW data and we do not show their posteriors in Fig.~\ref{fig:PEGW}.

Despite the high GW S/N at $d_L= 50$ Mpc, the inclusion of EM data still gives notable improvements in the binary source parameter estimates at this distance too. Similar to the results at $d_L= 100$ Mpc, the run with the EM data has less support for more equal mass ratios and gives some additional constraining power on $\mathcal{M}_c$ and $\chi_\mathrm{BH}$ compared to the run with only the GW data.

At $d_L= 100$ Mpc, the posterior range of $\Lambda_\mathrm{NS}$ is somewhat reduced when including the EM data while at $d_L = 50$ Mpc, the EM data gives no additional information on $\Lambda_\mathrm{NS}$. In both cases, the peak of the posteriors are shifted to higher values compared to the injected value. From our testing, we attribute this bias to the specific random noise realisation used for all inference runs (this is a known effect; see~\citealt{nissanke_exploring_2010}).

\begin{figure*}[h]
    \centering
    \mbox{\includegraphics[width=0.48\textwidth]{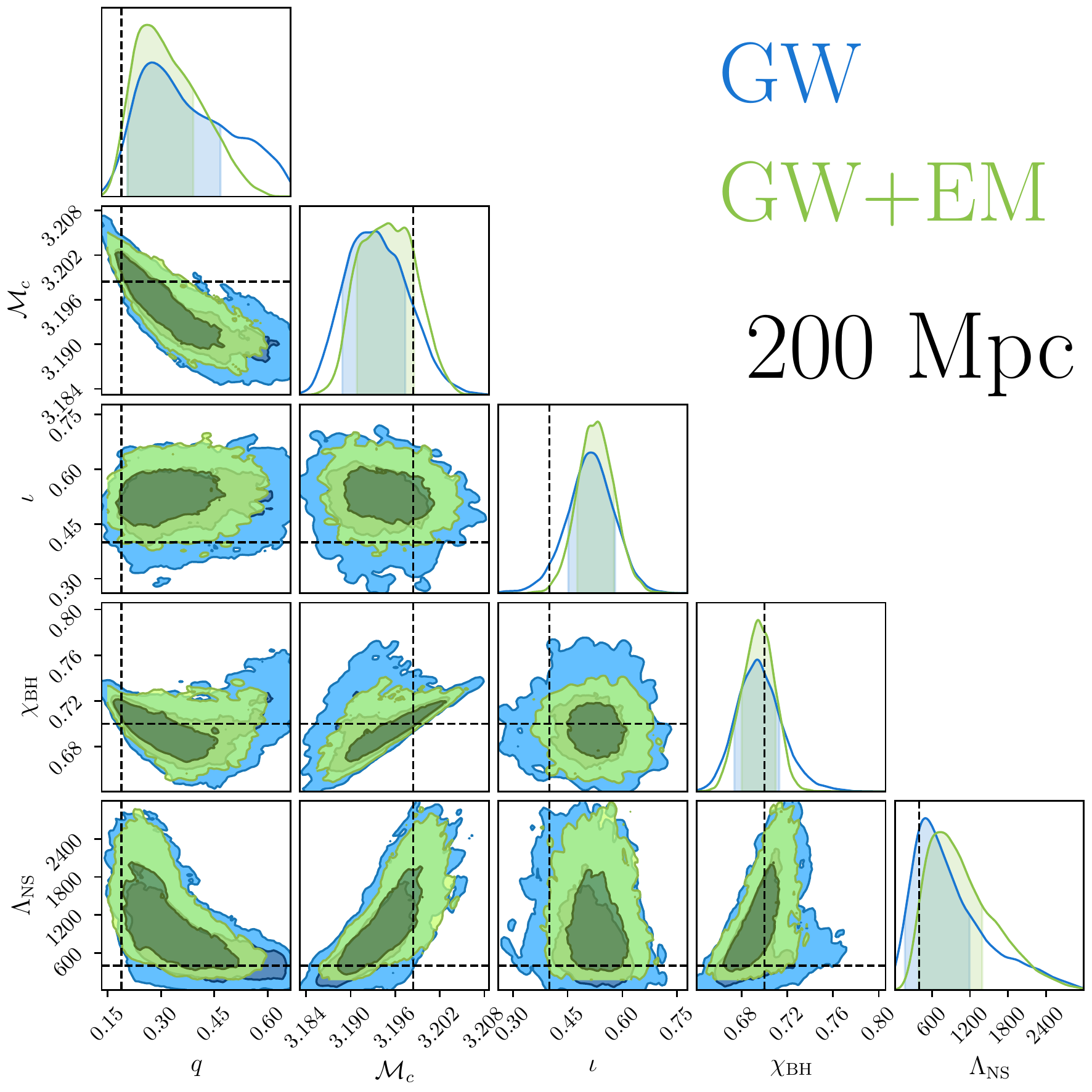}}
    \mbox{\includegraphics[width=0.48\textwidth]{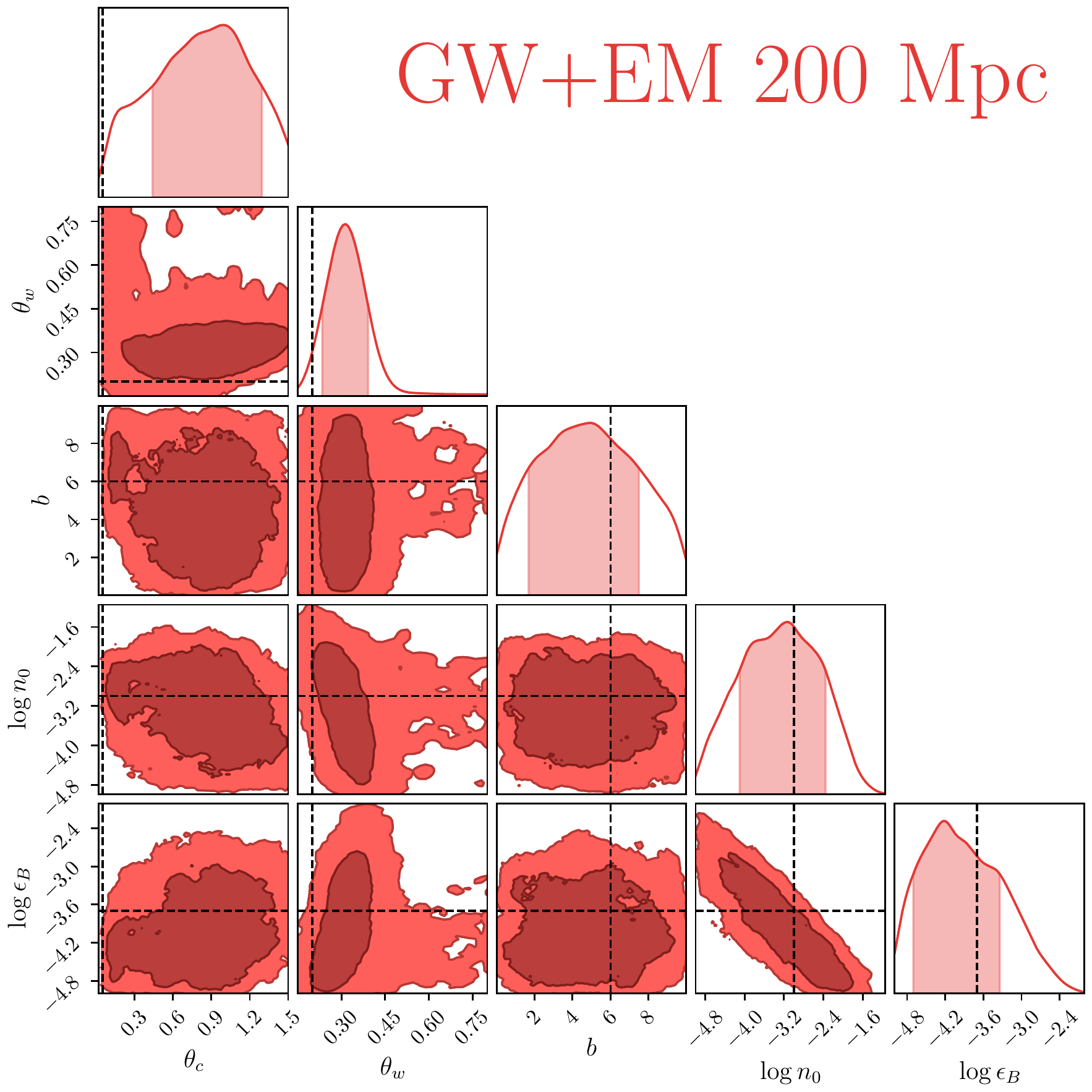}}
    \caption{Posterior distributions of the parameters $\vec{\theta}_\mathrm{GW}$ (left panel) and $\vec{\theta}_\mathrm{EM}$ (right panel) as inferred for our fiducial binary at $d_L=200$ Mpc. In the left panel, blue contours give the posterior distribution when only using the GW data, whereas the green contours give the posterior distribution when using both the GW and radio data. In the right panel, red contours give the inferred posterior distribution from the combination of the GW and radio data. The fiducial injected values are indicated by the dashed black lines.}
    \label{fig:PE200mpc}
\end{figure*}

An immediate advantage of knowing $d_L$ beforehand through a host galaxy association, is the resolved degeneracy between $d_L$ and $\iota$ in the GW inference~\citep{abbott_gravitational_2017,the_ligo_scientific_collaboration_and_the_virgo_collaboration_gravitational-wave_2017}. The resulting GW data estimate on $\iota$ is essential to break various degeneracies in $\vec{\theta}_\mathrm{EM}$, specifically between $\iota$ and $\theta_c$~\citep{nakar_afterglow_2021}. Turning our attention to the posteriors of $\vec{\theta}_\mathrm{EM}$ in Fig.~\ref{fig:PEEM}, it is clear that the angles $\iota$, $\theta_c$ and $\theta_w$ are quite well constrained at both distances. Still, at $d_L= 50$ Mpc, the improved estimate of $\iota$ from the GW data helps improve the estimates on $\theta_c$ and $\theta_w$ further. The EM data gives almost no additional information on $\iota$ compared to only the GW data in Fig.~\ref{fig:PEGW}. For the marginalised posterior of $b$ in Fig.~\ref{fig:PEEM}, only the lower half of the prior range can be confidently ruled out at either distance. We are able to infer $n_0$ and $\epsilon_B$ to uncertainties of $0.5 - 2$ dex at both distances. 
\subsection{200 Mpc}
In Fig.~\ref{fig:PE200mpc} we show the posteriors at $d_L = 200$ Mpc for $\vec{\theta}_\mathrm{GW}$ (left panel) and $\vec{\theta}_\mathrm{EM}$ (right panel). At this distance, it becomes considerably more difficult to fully characterise the binary. In regards to the binary source parameters, both $\mathcal{M}_c$ and $\chi_\mathrm{BH}$ are still inferred relatively well from the GW data, while the lower GW S/N impedes a precise estimate of $q$, $\iota$ and $\Lambda_\mathrm{NS}$. Including the EM data gives some improvement in the inference of these parameters with the biggest improvement in $q$ akin to the results at $d_L=50, 100$ Mpc. Still, the large measurement and systematic errors on the radio observations allow the light curve to be only loosely constrained. This is clearly visible in the inference of the $\vec{\theta}_\mathrm{EM}$ parameters. Most information on the geometry of the jet is now lost as the posterior of $\theta_c$ spans almost the entire prior range; for $b$ the estimate is similarly weak. The estimate on $\theta_w$ is weakened as well compared to the previous cases. From the posteriors of $n_0$ and $\epsilon_B$, we can only rule out the upper end of their respective prior ranges.

\section{Discussion}
\label{sec:discussion}
In summary, we are able to recover most of the injected parameters of our fiducial binary with reasonable confidence at both 50 Mpc and 100 Mpc. The observed sGRB afterglow clearly provides additional constraints on the binary source parameters despite the high GW S/Ns. Conversely, crucial degeneracies are broken in the parameters of the afterglow light curve using the GW data. The GW observations thus play an essential role in fitting the corresponding EM data. At 200 Mpc, both the GW and EM data provide less information on the binary and the parameters of the sGRB afterglow in particular become fairly undetermined. 

Inferring the parameters of both the GW data and the EM data simultaneously is not trivial. Various degeneracies come into play, arising from, for example, the conversion of four binary parameters to one jet energy parameter. These and other aspects warrant further discussion, below.

From the parameters in $\vec{\theta}_\mathrm{EM}$, $\epsilon_B$ and $n_0$ are the most strongly correlated with the binary source parameters. We show the combined posterior of the binary source parameters together with $n_0$ and $\epsilon_B$ in Appendix~\ref{app:PE} for $d_L = 50, 100$ Mpc. To improve our constraints on the binary parameters further, it would be helpful to thus constrain $n_0$ and $\epsilon_B$ better too. This might be achieved, for $\epsilon_B$ in particular, by using additional EM data at different frequencies (e.g.~\citealt{beniamini_energies_2015}). If the observations at other frequencies are in the same cooling regime, however, such homothetic light curves give no additional information (e.g.~\citealt{ryan_gamma-ray_2020}). Better constraints on the binary parameters could in turn improve the estimates of $n_0$ and $\epsilon_B$. 

\begin{figure}[h]
    \centering
   \includegraphics[width=0.48\textwidth]{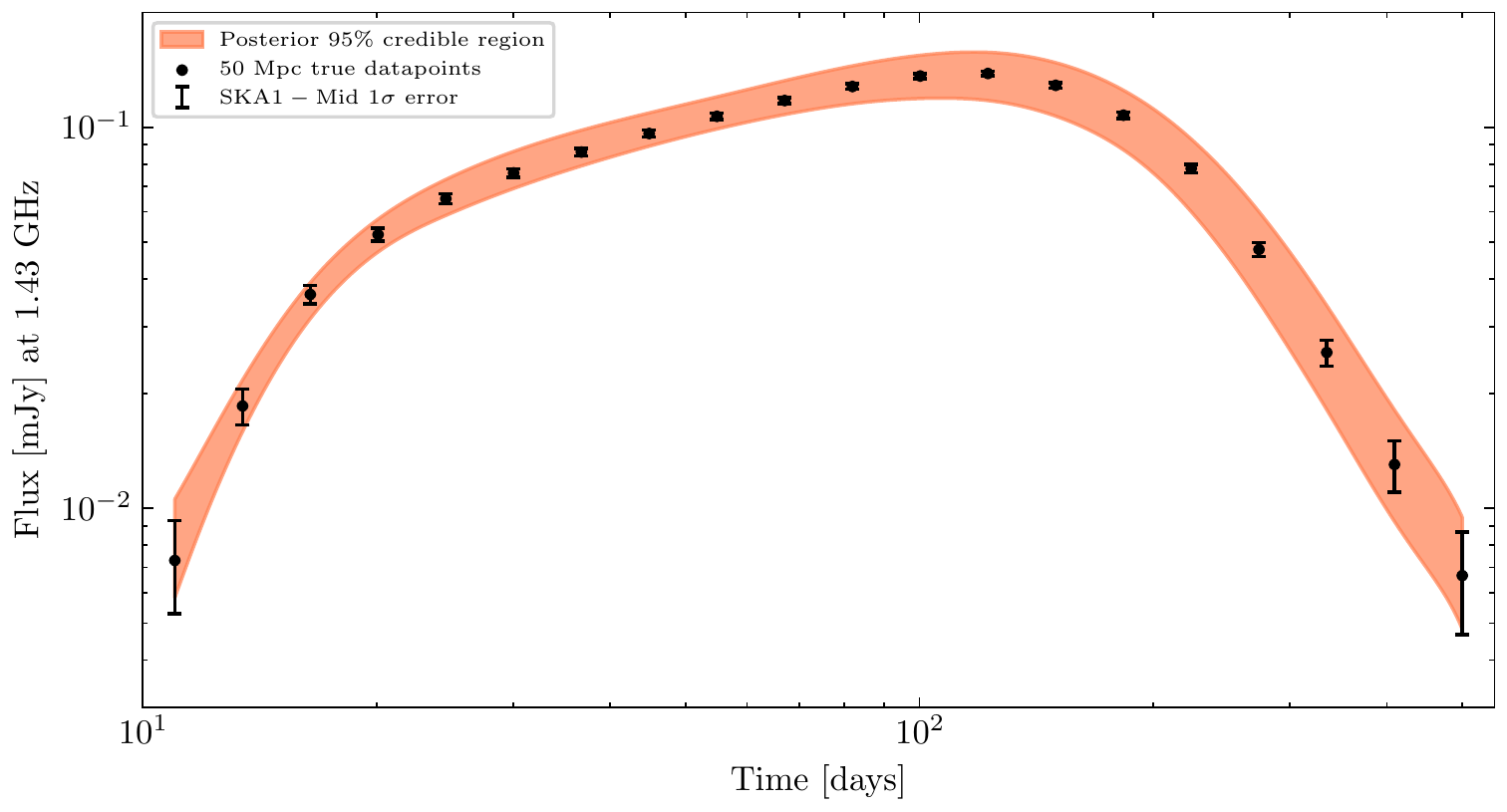}
    \caption{Resulting fit to the light curve of our fiducial binary at $d_L=50$ Mpc. The 95\% credible region of the fit, indicated by the shaded region, is computed from the posterior samples of the combined GW and EM data inference in Sect.~\ref{sec:PEresults}. The black circles give the 20 simulated observations, and the error bars correspond to the SKA1-Mid measurement errors.}
    \label{fig:postdraws}
\end{figure}

For most of the data points on our light curves at $d_L= 50, 100$ Mpc, the systematic errors described in Sect.~\ref{sec:systematics} are substantially larger than the measurement errors. These systematics currently thus have a big effect on the inference power of well-observed EM data, where most observations have fluxes significantly above the sensitivity of a next-generation radio telescope such as SKA1-Mid. This is visible in Fig.~\ref{fig:postdraws}, where we have computed the uncertainty in the fit of the light curve from the posterior samples of the combined GW and EM data inference at $d_L=50$ Mpc. The shaded area gives the 95\% credible region of the resulting fit. The black error bars, corresponding to the SKA1-Mid measurement error on our 20 simulated observations, are much smaller than this uncertainty region for most data points. Thus, the systematic errors, and not the measurement errors, govern how well the light curve can be constrained at these distances. At 200 Mpc, the situation changes. While the systematic errors are not insignificant, the measurement errors are now the dominant source of uncertainty for all seven data points. Not only does this limit our ability to gain additional constraints on the binary parameters, it also hinders any attempt to resolve the structure of the jet. Besides, for example, a good estimate of the inclination angle, a well-measured afterglow peak flux and time are necessary to determine the jet opening angle~\citep{nakar_afterglow_2021}. We conclude that this is not the case for our light curve at 200 Mpc and that this might not hold true for most afterglows that will be detected close the telescope sensitivity limit either. 

For a low GW S/N, a well-measured afterglow light curve might provide better constraints on the binary source parameters than what we found with the poorly sampled light curve at 200 Mpc discussed in this work. However, the aforementioned systematics will still limit how much information can be gained. Even if those errors are brought down in the future, other systematics in the sGRB afterglow, such as interstellar scintillation at higher frequencies~\citep{aksulu_new_2020}, can also have an effect on the EM data inference. A proper understanding of these and other systematic uncertainties and how they propagate through the Bayesian inference framework (see e.g.~\citealt{carson_constraining_2019} for a discussion on the systematics from different EOSs) will be essential to extract as much information as possible out of sGRB afterglow observations. Additional information on the inclination angle through, for example, Very Long Baseline Interferometry measurements~\citep{mooley_superluminal_2018,hotokezaka_hubble_2019} will be required as well if this angle is not sufficiently inferred from the low S/N GW data. Furthermore, to fully characterise the sGRB afterglow, it might be necessary to infer $\epsilon_e$,~$p$ and $\xi_N$ too instead of setting them to often used canonical values (e.g.~\citealt{aksulu_exploring_2022}). In this work, we have also ignored the uncertainties in the jet launching mechanism and focused on the systematics of Sect.~\ref{sec:systematics}. We leave a full discussion of the uncertainties in the launching mechanism for future work but give some brief remarks here.

\citet[][\citetalias{salafia_accretion--jet_2021}]{salafia_accretion--jet_2021}  note that the fit of $\eta_\mathrm{BZ}$ in Eq.~\ref{eq:BZeff} pertains to simulations of BH-accretion disk systems hosted by active galactic nuclei. Such systems differ in terms of, for example, the accretion rate from BH merger remnants of compact object mergers. The estimated posterior of the accretion-to-jet energy conversion efficiency $\eta$ of GW170817 in \citetalias{salafia_accretion--jet_2021} is still consistent with $\eta_\mathrm{BZ}$. The spread in this posterior covers multiple decades in $\eta$, however, and it is thus hard to make definitive conclusions. More EM detections accompanying BNS mergers in the future will bring down the uncertainty in $\eta$ for this class of compact object mergers. It remains to be seen whether this efficiency is similar for BHNS mergers given, for example, the differences in ejecta properties. Updated NR simulations will help in answering this question, even if an EM detection of a BHNS sGRB afterglow does not occur soon.

Given the large uncertainties in deriving the energy of the jet from the binary source parameters, it will also be necessary to better understand the dependence of jet features, such as the opening angle on the source properties.~\citet{lazzati_two_2021}, for instance, perform a hydrodynamic simulation of a relativistic jet propagating in BNS merger ejecta and characterise the jet properties. A (semi-)analytical description derived from the mapping of the output of many such simulations to a variety of associated input source parameters will be a valuable tool in the inference of a single observed source. It is worth bearing in mind that the influence of the central BH engine and surrounding ejecta on the jet propagation and opening angle is still under much debate, certainly in BHNS mergers~\citep{kyutoku_coalescence_2021}. Systematics will be a limiting factor when incorporating such descriptions too.

We have only focused on sGRB afterglow observations in this work. Combining multiple sources of EM radiation (see e.g.~\citealt{dietrich_multimessenger_2020}) will be key to exploit the full potential of multi-messenger astronomy. Incorporating kilonova physics and emission~\citep{raaijmakers_challenges_2021,nicholl_tight_2021} could improve the estimates on the binary source parameters further, which we will explore in future work.

We end this discussion with some technical considerations. Perhaps the most straightforward and often utilised way of combining multi-messenger emission is taking the marginalised posteriors of the analysis of one emission type as prior input for the analysis of another type of emission.~\citet{nicholl_tight_2021}, for example, took 1D GW parameter estimates of GW170817 as input for some of the priors of their kilonova model. In this way, a (re-)analysis of the GW data is avoided, making the inference clearly more efficient computationally. Furthermore, it is not always necessary to do a full re-analysis of the data if information on only one parameter (e.g. the inclination angle) is needed, which is readily available from the marginalised posterior. If there are strong correlations in the parameter space, however, such a sequential approach comes at the cost of losing that valuable information between sets of parameters. This information is preserved when the full likelihood of the GW data is used as in our work. To consistently combine not only GW and afterglow emission but also kilonova emission moving forwards, we argue that a joint likelihood Bayesian analysis is preferred (and perhaps necessary) over a sequential approach because of the many correlations between the parameters.

A method that sits somewhere in between a sequential and a joint approach is making a multi-dimensional density estimate of the sampled GW posterior to preserve the correlations. This estimate can then be used in place of the full likelihood. We have experimented with kernel density estimates, generated by software packages such as \texttt{kalepy}~\citep{kelley_kalepy_2021}, but did not find these approximations to always converge to the same results as the full likelihood. We suspect this to be a result of insufficient accuracy in the estimates of the tails of the distributions. Other density estimates, such as methods based on Gaussian processes~\citep{demilio_density_2021}, could fair better in this regard but exploring such alternatives was beyond the scope of this work.

\section{Summary and conclusion}
\label{sec:conclusion}
In this paper we have looked at the prospects for detecting and analysing BHNS mergers using both GW and radio emission. We modelled the radio emission from an accompanying sGRB jet afterglow launched from the ejecta disk surrounding the merger remnant.  We used fits from the literature, connecting the binary source properties to the accreted disk mass and to the accretion-to-jet-energy conversion efficiency of the jet. 

We first performed a population synthesis study to derive the expected rates of GW BHNS observations with an sGRB afterglow detection at SKA1-Mid radio frequency and sensitivity. We explored the impact of the current generation of GW detectors (2G) versus the future generation (3G), the NS EOS, and the BH spin of the binary, $\chi_\mathrm{BH}$. We find that rates of around one combined detection per year with a 2G detector network and SKA1-Mid are only possible for an unrealistically high $\chi_\mathrm{BH}$ of $\sim 0.8$. This is a result of the strong dependence of the jet energy, and thus radio flux, on $\chi_\mathrm{BH}$. The increased sensitivity of a 3G network will allow for similar rates near one combined detection per year for lower $\chi_\mathrm{BH}$  of $\sim  0.2$. Such values of the BH spin are more in line with the expected values in BHNS mergers based on current GW detections. The probability of finding a combined GW and radio detection of a BHNS merger in the near future is thus low. This will increase substantially with forthcoming GW detectors that will be able to detect many more BHNS mergers, increasing the chances of an sGRB localisation and radio detection. Furthermore, transitioning from a soft to a hard NS EOS increases the expected rates two- to three-fold. 

We then examined our ability to infer the properties of a fiducial BHNS merger using simulated aLIGO and Virgo GW observations in conjunction with SKA1-Mid radio observations of its sGRB afterglow. We employed a recent waveform model for the GW signal and modelled the full afterglow light curve using \texttt{afterglowpy}. We performed a joint Bayesian analysis, combining the likelihood of the GW data with the likelihood of the radio data. We also incorporated systematic errors from the conversion of the binary source parameters to the disk mass. We placed our fiducial binary at three distances to study the effect of different GW and sGRB afterglow detection S/Ns. 

We find that it is possible to simultaneously recover both the binary source parameters, such as the chirp mass, and parameters of the jet afterglow, such as the opening angle, with reasonable confidence for nearby binaries. When placing our fiducial binary at 100 Mpc, including the radio data provided various improvements in the inference of the binary source parameters compared to just using the GW data. This was most noticeable in the well-improved estimate of the mass ratio. At 50 Mpc, despite the already high GW S/N, the estimates of the binary source parameters were also improved with the EM data used in the inference. At these distances, we find that the systematic errors were likely the limiting factor in how much information could be extracted from the radio light curve. On the other hand, the inclination angle estimate of the GW data broke various degeneracies in the afterglow jet parameters of the radio data. Specifically, parameters pertaining to the geometry of the jet were well determined from the combination of GW and radio data at both distances. Other parameters, such as the circumburst density, were recovered with larger relative uncertainties. Thus, in a close-by real life detection, we can be reasonably confident that we can determine the properties of the system. If the binary is farther away, it becomes much harder to get a complete picture of the system. For our fiducial binary at 200 Mpc, only the chirp mass and BH spin were easily inferred, while most of the information on the jet parameters was lost. If an afterglow is observed close to the telescope detection threshold, as for our binary at 200 Mpc, we conclude that the benefits of including radio data in the analysis of the binary are limited.

In the future, it will be necessary to get a more complete understanding of the connection between the binary source parameters and the resulting properties of the jet afterglow. Lowering the associated systematic uncertainties is important for really leveraging the depth of information encompassed in the afterglow signal. Numerical relativity simulations will continue to be instrumental in this regard, notably when combined with future GW detections of BHNS mergers to constrain the physical processes further. 

To summarise, we conclude that the upcoming generation of GW detectors will provide a unique opportunity to study the population of BHNS mergers, even in the absence of an sGRB afterglow. For high S/N GW signals, which will be routine occurrences with future GW detectors, tight constraints on the inferred binary parameters are possible. If an associated afterglow is observed, which is, as we have argued, not unlikely in the 3G detector era, a combined analysis of the GW and radio data will be essential for characterising the jet afterglow properties. 
For a sufficiently bright afterglow, this joint analysis can also provide additional information on the binary parameters, but only if the modelling systematics are well understood. 

\begin{acknowledgements}
We thank Pikky Atri, Geert Raaijmakers and Samaya Nissanke for the helpful discussions and suggestions on this work. We thank the anonymous referee for their thoughtful comments which have improved this work. This research was supported by Vici research program 'ARGO' with project number 639.043.815, financed by the Dutch Research Council (NWO).
JVL further acknowledges funding through  CORTEX (NWA.1160.18.316), under the research programme NWA-ORC, financed by NWO;
and from the European Research Council under the 
European Union's Seventh Framework Programme (FP/2007-2013)/ERC Grant Agreement No. 617199 (`ALERT').
\end{acknowledgements}

%-------------------------------------------------------------------

\bibliographystyle{yahapj}
\bibliography{references}

\begin{thebibliography}{}
\providecommand\natexlab[1]{#1}
\providecommand\JournalTitle[1]{#1}

\bibitem[{Aasi {et~al.}(2015)Aasi, Abbott, Abbott, Abbott, Abernathy, Ackley,
  Adams, Adams, Addesso, Adhikari, Adya, Affeldt, Aggarwal, Aguiar, Ain, Ajith,
  Alemic, Allen, Amariutei, Anderson, Anderson, Arai, Araya, Arceneaux, Areeda,
  Ashton, Ast, Aston, Aufmuth, Aulbert, Aylott, Babak, Baker, Ballmer,
  Barayoga, Barbet, Barclay, Barish, Barker, Barr, Barsotti, Bartlett, Barton,
  Bartos, Bassiri, Batch, Baune, Behnke, Bell, Bell, Benacquista, Bergman,
  Bergmann, Berry, Betzwieser, Bhagwat, Bhandare, Bilenko, Billingsley, Birch,
  Biscans, Biwer, Blackburn, Blackburn, Blair, Blair, Bock, Bodiya, Bojtos,
  Bond, Bork, Born, Bose, Brady, Braginsky, Brau, Bridges, Brinkmann, Brooks,
  Brown, Brown, Brown, Buchman, Buikema, Buonanno, Cadonati, Bustillo, Camp,
  Cannon, Cao, Capano, Caride, Caudill, Cavaglià, Cepeda, Chakraborty,
  Chalermsongsak, Chamberlin, Chao, Charlton, Chen, Cho, Cho, Chow,
  Christensen, Chu, Chung, Ciani, Clara, Clark, Collette, Cominsky, Constancio,
  Cook, Corbitt, Cornish, Corsi, Costa, Coughlin, Countryman, Couvares, Coward,
  Cowart, Coyne, Coyne, Craig, Creighton, Creighton, Cripe, Crowder, Cumming,
  Cunningham, Cutler, Dahl, Canton, Damjanic, Danilishin, Danzmann, Dartez,
  Dave, Daveloza, Davies, Daw, DeBra, Pozzo, Denker, Dent, Dergachev, DeRosa,
  DeSalvo, Dhurandhar, D´ıaz, Palma, Dojcinoski, Dominguez, Donovan, Dooley,
  Doravari, Douglas, Downes, Driggers, Du, Dwyer, Eberle, Edo, Edwards,
  Edwards, Effler, Eggenstein, Ehrens, Eichholz, Eikenberry, Essick, Etzel,
  Evans, Evans, Factourovich, Fairhurst, Fan, Fang, Farr, Farr, Favata, Fays,
  Fehrmann, Fejer, Feldbaum, Ferreira, Fisher, Frei, Freise, Frey, Fricke,
  Fritschel, Frolov, Fuentes-Tapia, Fulda, Fyffe, Gair, Gaonkar, Gehrels,
  Gergely´, Giaime, Giardina, Gleason, Goetz, Goetz, Gondan, González,
  Gordon, Gorodetsky, Gossan, Goßler, Gräf, Graff, Grant, Gras, Gray,
  Greenhalgh, Gretarsson, Grote, Grunewald, Guido, Guo, Gushwa, Gustafson,
  Gustafson, Hacker, Hall, Hammond, Hanke, Hanks, Hanna, Hannam, Hanson,
  Hardwick, Harry, Harry, Hart, Hartman, Haster, Haughian, Hee, Heintze,
  Heinzel, Hendry, Heng, Heptonstall, Heurs, Hewitson, Hild, Hoak, Hodge,
  Hollitt, Holt, Hopkins, Hosken, Hough, Houston, Howell, Hu, Huerta, Hughey,
  Husa, Huttner, Huynh, Huynh-Dinh, Idrisy, Indik, Ingram, Inta, Islas, Isler,
  Isogai, Iyer, Izumi, Jacobson, Jang, Jawahar, Ji, Jiménez-Forteza, Johnson,
  Jones, Jones, Ju, Haris, Kalogera, Kandhasamy, Kang, Kanner, Katsavounidis,
  Katzman, Kaufer, Kaufer, Kaur, Kawabe, Kawazoe, Keiser, Keitel, Kelley,
  Kells, Keppel, Key, Khalaidovski, Khalili, Khazanov, Kim, Kim, Kim, Kim, Kim,
  King, King, Kinzel, Kissel, Klimenko, Kline, Koehlenbeck, Kokeyama,
  Kondrashov, Korobko, Korth, Kozak, Kringel, Krishnan, Krueger, Kuehn, Kumar,
  Kumar, Kuo, Landry, Lantz, Larson, Lasky, Lazzarini, Lazzaro, Le, Leaci,
  Leavey, Lebigot, Lee, Lee, Lee, Leong, Levin, Levine, Lewis, Li, Libbrecht,
  Libson, Lin, Littenberg, Lockerbie, Lockett, Logue, Lombardi, Lormand, Lough,
  Lubinski, Lück, Lundgren, Lynch, Ma, Macarthur, MacDonald, Machenschalk,
  MacInnis, Macleod, Magaña-Sandoval, Magee, Mageswaran, Maglione, Mailand,
  Mandel, Mandic, Mangano, Mansell, Márka, Márka, Markosyan, Maros, Martin,
  Martin, Martynov, Marx, Mason, Massinger, Matichard, Matone, Mavalvala,
  Mazumder, Mazzolo, McCarthy, McClelland, McCormick, McGuire, McIntyre,
  McIver, McLin, McWilliams, Meadors, Meinders, Melatos, Mendell, Mercer,
  Meshkov, Messenger, Meyers, Miao, Middleton, Mikhailov, Miller, Miller,
  Millhouse, Ming, Mirshekari, Mishra, Mitra, Mitrofanov, Mitselmakher,
  Mittleman, Moe, Mohanty, Mohapatra, Moore, Moraru, Moreno, Morriss, Mossavi,
  Mow-Lowry, Mueller, Mueller, Mukherjee, Mullavey, Munch, Murphy, Murray,
  Mytidis, Nash, Nayak, Necula, Nedkova, Newton, Nguyen, Nielsen, Nissanke,
  Nitz, Nolting, Normandin, Nuttall, Ochsner, O'Dell, Oelker, Ogin, Oh, Oh,
  Ohme, Oppermann, Oram, O'Reilly, Ortega, O'Shaughnessy, Osthelder, Ott,
  Ottaway, Ottens, Overmier, Owen, Padilla, Pai, Pai, Palashov, Pal-Singh, Pan,
  Pankow, Pannarale, Pant, Papa, Paris, Patrick, Pedraza, Pekowsky, Pele, Penn,
  Perreca, Phelps, Pierro, Pinto, Pitkin, Poeld, Post, Poteomkin, Powell,
  Prasad, Predoi, Premachandra, Prestegard, Price, Principe, Privitera, Prix,
  Prokhorov, Puncken, Pürrer, Qin, Quetschke, Quintero, Quiroga,
  Quitzow-James, Raab, Rabeling, Radkins, Raffai, Raja, Rajalakshmi, Rakhmanov,
  Ramirez, Raymond, Reed, Reid, Reitze, Reula, Riles, Robertson, Robie,
  Rollins, Roma, Romano, Romanov, Romie, Rowan, Rüdiger, Ryan, Sachdev,
  Sadecki, Sadeghian, Saleem, Salemi, Sammut, Sandberg, Sanders, Sannibale,
  Santiago-Prieto, Sathyaprakash, Saulson, Savage, Sawadsky, Scheuer,
  Schilling, Schmidt, Schnabel, Schofield, Schreiber, Schuette, Schutz, Scott,
  Scott, Sellers, Sengupta, Sergeev, Serna, Sevigny, Shaddock, Shahriar,
  Shaltev, Shao, Shapiro, Shawhan, Shoemaker, Sidery, Siemens, Sigg, Silva,
  Simakov, Singer, Singer, Singh, Sintes, Slagmolen, Smith, Smith, Smith,
  Smith-Lefebvre, Son, Sorazu, Souradeep, Staley, Stebbins, Steinke,
  Steinlechner, Steinlechner, Steinmeyer, Stephens, Steplewski, Stevenson,
  Stone, Strain, Strigin, Sturani, Stuver, Summerscales, Sutton, Szczepanczyk,
  Szeifert, Talukder, Tanner, Tápai, Tarabrin, Taracchini, Taylor, Tellez,
  Theeg, Thirugnanasambandam, Thomas, Thomas, Thorne, Thorne, Thrane, Tiwari,
  Tomlinson, Torres, Torrie, Traylor, Tse, Tshilumba, Ugolini, Unnikrishnan,
  Urban, Usman, Vahlbruch, Vajente, Valdes, Vallisneri, Veggel, Vass, Vaulin,
  Vecchio, Veitch, Veitch, Venkateswara, Vincent-Finley, Vitale, Vo, Vorvick,
  Vousden, Vyatchanin, Wade, Wade, Wade, Walker, Wallace, Walsh, Wang, Wang,
  Wang, Ward, Warner, Was, Weaver, Weinert, Weinstein, Weiss, Welborn, Wen,
  Wessels, Westphal, Wette, Whelan, Whitcomb, White, Whiting, Wilkinson,
  Williams, Williams, Williamson, Willis, Willke, Wimmer, Winkler, Wipf,
  Wittel, Woan, Worden, Xie, Yablon, Yakushin, Yam, Yamamoto, Yancey, Yang,
  Zanolin, Zhang, Zhang, Zhang, Zhang, Zhao, Zhou, Zhu, Zucker, Zuraw, \&
  Zweizig}]{aasi_advanced_2015}
Aasi, a.~J., Abbott, B.~P., Abbott, R., {et~al.} 2015,
  \href{http://dx.doi.org/10.1088/0264-9381/32/7/074001}{\JournalTitle{Classical
  and Quantum Gravity}, 32, 074001}, publisher: IOP Publishing

\bibitem[{Abbasi {et~al.}(2021)Abbasi, Ackermann, Adams, Aguilar, Ahlers,
  Ahrens, Alispach, Alves~Jr., Amin, An, Andeen, Anderson, Ansseau, Anton,
  Argüelles, Ashida, Axani, Bai, V., Barbano, Barwick, Bastian, Basu, Baur,
  Bay, Beatty, Becker, Tjus, Bellenghi, BenZvi, Berley, Bernardini, Besson,
  Binder, Bindig, Blaufuss, Blot, Bontempo, Borowka, Böser, Botner, Böttcher,
  Bourbeau, Bradascio, Braun, Bron, Brostean-Kaiser, Browne, Burgman, Busse,
  Campana, Chen, Chirkin, Choi, Clark, Clark, Classen, Coleman, Collin, Conrad,
  Coppin, Correa, Cowen, Cross, Dave, De~Clercq, DeLaunay, Dembinski, Deoskar,
  De~Ridder, Desai, Desiati, de~Vries, de~Wasseige, de~With, DeYoung, Dharani,
  Diaz, Díaz-Vélez, Dujmovic, Dunkman, DuVernois, Dvorak, Ehrhardt, Eller,
  Engel, Erpenbeck, Evans, Evenson, Fazely, Fiedlschuster, Fienberg, Filimonov,
  Finley, Fischer, Fox, Franckowiak, Friedman, Fritz, Fürst, Gaisser,
  Gallagher, Ganster, Garcia, Garrappa, Gerhardt, Ghadimi, Glaser, Glauch,
  Glüsenkamp, Goldschmidt, Gonzalez, Goswami, Grant, Grégoire, Griswold,
  Gündüz, Günther, Haack, Hallgren, Halliday, Halve, Halzen, Minh, Hanson,
  Hardin, Harnisch, Haungs, Hauser, Hebecker, Helbing, Henningsen, Hettinger,
  Hickford, Hignight, Hill, Hill, Hoffman, Hoffmann, Hoinka, Hokanson-Fasig,
  Hoshina, Huang, Huber, Huber, Hultqvist, Hünnefeld, Hussain, In, Iovine,
  Ishihara, Jansson, Japaridze, Jeong, Jones, Joppe, Kang, Kang, Kang, Kappes,
  Kappesser, Karg, Karl, Karle, Katz, Kauer, Kellermann, Kelley, Kheirandish,
  Kin, Kintscher, Kiryluk, Klein, Koirala, Kolanoski, Kontrimas, Köpke,
  Kopper, Kopper, Koskinen, Koundal, Kovacevich, Kowalski, Kurahashi, Kyriacou,
  Lad, Gualda, Lanfranchi, Larson, Lauber, Lazar, Lee, Leonard, Leszczyńska,
  Li, Liu, Liubarska, Lohfink, Mariscal, Lu, Lucarelli, Ludwig, Luszczak, Lyu,
  Ma, Madsen, Mahn, Makino, Mancina, Mari\{ş\}, Maruyama, Mase, McElroy,
  McNally, Meagher, Medina, Meier, Meighen-Berger, Merz, Micallef, Mockler,
  Montaruli, Moore, Morse, Moulai, Naab, Nagai, Naumann, Necker,
  Nguy\{{\textbackslash}{\textasciitilde}\{ê\}\}n, Niederhausen, Nisa,
  Nowicki, Nygren, Pollmann, Oehler, Olivas, O'Sullivan, Pandya, Pankova, Park,
  Parker, Paudel, Paul, Heros, Philippen, Pieloth, Pieper, Pittermann, Pizzuto,
  Plum, Popovych, Porcelli, Rodriguez, Price, Pries, Przybylski, Raab, Raissi,
  Rameez, Rawlins, Rea, Rehman, Reimann, Renzi, Resconi, Reusch, Rhode,
  Richman, Riedel, Robertson, Roellinghoff, Rongen, Rott, Ruhe, Ryckbosch,
  Cantu, Safa, Saffer, Herrera, Sandrock, Sandroos, Santander, Sarkar, Sarkar,
  Satalecka, Scharf, Schaufel, Schieler, Schlunder, Schmidt, Schneider,
  Schneider, Schröder, Schumacher, Sclafani, Seckel, Seunarine, Sharma,
  Shefali, Silva, Skrzypek, Smithers, Snihur, Soedingrekso, Soldin,
  Spannfellner, Spiczak, Spiering, Stachurska, Stamatikos, Stanev, Stein,
  Stettner, Steuer, Stezelberger, Stürwald, Stuttard, Sullivan, Taboada,
  Tenholt, Ter-Antonyan, Tilav, Tischbein, Tollefson, Tomankova, Tönnis,
  Toscano, Tosi, Trettin, Tselengidou, Tung, Turcati, Turcotte, Turley,
  Twagirayezu, Ty, Elorrieta, Valtonen-Mattila, Vandenbroucke, van Eijndhoven,
  Vannerom, van Santen, Verpoest, Vraeghe, Walck, Wallace, Watson, Weaver,
  Weigel, Weindl, Weiss, Weldert, Wendt, Werthebach, Weyrauch, Whelan,
  Whitehorn, Wiebusch, Williams, Wolf, Woschnagg, Wrede, Wulff, Xu, Xu, Yanez,
  Yoshida, Yu, Yuan, \& Zhang}]{abbasi_probing_2021}
Abbasi, R., Ackermann, M., Adams, J., {et~al.} 2021,
  \href{http://arxiv.org/abs/2105.13160}{\JournalTitle{arXiv:2105.13160
  [astro-ph]}}, arXiv: 2105.13160

\bibitem[{Abbott {et~al.}(2017)Abbott, Abbott, Abbott, Acernese, Ackley, Adams,
  Adams, Addesso, Adhikari, Adya, Affeldt, Afrough, Agarwal, Agathos, Agatsuma,
  Aggarwal, Aguiar, Aiello, Ain, Ajith, Allen, Allen, Allocca, Aloy, Altin,
  Amato, Ananyeva, Anderson, Anderson, Angelova, Antier, Appert, Arai, Araya,
  Areeda, Arnaud, Arun, Ascenzi, Ashton, Ast, Aston, Astone, Atallah, Aufmuth,
  Aulbert, AultONeal, Austin, Avila-Alvarez, Babak, Bacon, Bader, Bae, Baker,
  Baldaccini, Ballardin, Ballmer, Banagiri, Barayoga, Barclay, Barish, Barker,
  Barkett, Barone, Barr, Barsotti, Barsuglia, Barta, Bartlett, Bartos, Bassiri,
  Basti, Batch, Bawaj, Bayley, Bazzan, Bécsy, Beer, Bejger, Belahcene, Bell,
  Berger, Bergmann, Bero, Berry, Bersanetti, Bertolini, Betzwieser, Bhagwat,
  Bhandare, Bilenko, Billingsley, Billman, Birch, Birney, Birnholtz, Biscans,
  Biscoveanu, Bisht, Bitossi, Biwer, Bizouard, Blackburn, Blackman, Blair,
  Blair, Blair, Bloemen, Bock, Bode, Boer, Bogaert, Bohe, Bondu, Bonilla,
  Bonnand, Boom, Bork, Boschi, Bose, Bossie, Bouffanais, Bozzi, Bradaschia,
  Brady, Branchesi, Brau, Briant, Brillet, Brinkmann, Brisson, Brockill,
  Broida, Brooks, Brown, Brown, Brunett, Buchanan, Buikema, Bulik, Bulten,
  Buonanno, Buskulic, Buy, Byer, Cabero, Cadonati, Cagnoli, Cahillane,
  Bustillo, Callister, Calloni, Camp, Canepa, Canizares, Cannon, Cao, Cao,
  Capano, Capocasa, Carbognani, Caride, Carney, Diaz, Casentini, Caudill,
  Cavaglià, Cavalier, Cavalieri, Cella, Cepeda, Cerdá-Durán, Cerretani,
  Cesarini, Chamberlin, Chan, Chao, Charlton, Chase, Chassande-Mottin,
  Chatterjee, Chatziioannou, Cheeseboro, Chen, Chen, Chen, Cheng, Chia,
  Chincarini, Chiummo, Chmiel, Cho, Cho, Chow, Christensen, Chu, Chua, Chua,
  Chung, Chung, Ciani, Ciolfi, Cirelli, Cirone, Clara, Clark, Clearwater,
  Cleva, Cocchieri, Coccia, Cohadon, Cohen, Colla, Collette, Cominsky, Jr,
  Conti, Cooper, Corban, Corbitt, Cordero-Carrión, Corley, Cornish, Corsi,
  Cortese, Costa, Coughlin, Coughlin, Coulon, Countryman, Couvares, Covas,
  Cowan, Coward, Cowart, Coyne, Coyne, Creighton, Creighton, Cripe, Crowder,
  Cullen, Cumming, Cunningham, Cuoco, Canton, Dálya, Danilishin, D'Antonio,
  Danzmann, Dasgupta, Costa, Dattilo, Dave, Davier, Davis, Daw, Day, De, DeBra,
  Degallaix, Laurentis, Deléglise, Pozzo, Demos, Denker, Dent, Pietri,
  Dergachev, Rosa, DeRosa, Rossi, DeSalvo, Varona, Devenson, Dhurandhar, Díaz,
  Fiore, Giovanni, Girolamo, Lieto, Pace, Palma, Renzo, Doctor, Dolique,
  Donovan, Dooley, Doravari, Dorrington, Douglas, Álvarez, Downes, Drago,
  Dreissigacker, Driggers, Du, Ducrot, Dupej, Dwyer, Edo, Edwards, Effler,
  Eggenstein, Ehrens, Eichholz, Eikenberry, Eisenstein, Essick, Estevez,
  Etienne, Etzel, Evans, Evans, Factourovich, Fafone, Fair, Fairhurst, Fan,
  Farinon, Farr, Farr, Fauchon-Jones, Favata, Fays, Fee, Fehrmann, Feicht,
  Fejer, Fernandez-Galiana, Ferrante, Ferreira, Ferrini, Fidecaro, Finstad,
  Fiori, Fiorucci, Fishbach, Fisher, Fitz-Axen, Flaminio, Fletcher, Fong, Font,
  Forsyth, Forsyth, Fournier, Frasca, Frasconi, Frei, Freise, Frey, Frey,
  Fries, Fritschel, Frolov, Fulda, Fyffe, Gabbard, Gadre, Gaebel, Gair,
  Gammaitoni, Ganija, Gaonkar, Garcia-Quiros, Garufi, Gateley, Gaudio, Gaur,
  Gayathri, Gehrels, Gemme, Genin, Gennai, George, George, Gergely, Germain,
  Ghonge, Ghosh, Ghosh, Ghosh, Giaime, Giardina, Giazotto, Gill, Glover, Goetz,
  Goetz, Gomes, Goncharov, González, Castro, Gopakumar, Gorodetsky, Gossan,
  Gosselin, Gouaty, Grado, Graef, Granata, Grant, Gras, Gray, Greco, Green,
  Gretarsson, Groot, Grote, Grunewald, Gruning, Guidi, Guo, Gupta, Gupta,
  Gushwa, Gustafson, Gustafson, Halim, Hall, Hall, Hamilton, Hammond, Haney,
  Hanke, Hanks, Hanna, Hannam, Hannuksela, Hanson, Hardwick, Harms, Harry,
  Harry, Hart, Haster, Haughian, Healy, Heidmann, Heintze, Heitmann, Hello,
  Hemming, Hendry, Heng, Hennig, Heptonstall, Heurs, Hild, Hinderer, Hoak,
  Hofman, Holt, Holz, Hopkins, Horst, Hough, Houston, Howell, Hreibi, Hu,
  Huerta, Huet, Hughey, Husa, Huttner, Huynh-Dinh, Indik, Inta, Intini, Isa,
  Isac, Isi, Iyer, Izumi, Jacqmin, Jani, Jaranowski, Jawahar, Jiménez-Forteza,
  Johnson, Johnson-McDaniel, Jones, Jones, Jonker, Ju, Junker, Kalaghatgi,
  Kalogera, Kamai, Kandhasamy, Kang, Kanner, Kapadia, Karki, Karvinen,
  Kasprzack, Kastaun, Katolik, Katsavounidis, Katzman, Kaufer, Kawabe,
  Kéfélian, Keitel, Kemball, Kennedy, Kent, Key, Khalili, Khan, Khan, Khan,
  Khazanov, Kijbunchoo, Kim, Kim, Kim, Kim, Kim, Kim, Kimbrell, King, King,
  Kinley-Hanlon, Kirchhoff, Kissel, Kleybolte, Klimenko, Knowles, Koch,
  Koehlenbeck, Koley, Kondrashov, Kontos, Korobko, Korth, Kowalska, Kozak,
  Krämer, Kringel, Krishnan, Królak, Kuehn, Kumar, Kumar, Kumar, Kuo,
  Kutynia, Kwang, Lackey, Lai, Landry, Lang, Lange, Lantz, Lanza,
  Lartaux-Vollard, Lasky, Laxen, Lazzarini, Lazzaro, Leaci, Leavey, Lee, Lee,
  Lee, Lee, Lee, Lehmann, Lenon, Leonardi, Leroy, Letendre, Levin, Li, Linker,
  Littenberg, Liu, Lo, Lockerbie, London, Lord, Lorenzini, Loriette, Lormand,
  Losurdo, Lough, Lousto, Lovelace, Lück, Lumaca, Lundgren, Lynch, Ma, Macas,
  Macfoy, Machenschalk, MacInnis, Macleod, Hernandez, Magaña-Sandoval,
  Zertuche, Magee, Majorana, Maksimovic, Man, Mandic, Mangano, Mansell, Manske,
  Mantovani, Marchesoni, Marion, Márka, Márka, Markakis, Markosyan,
  Markowitz, Maros, Marquina, Martelli, Martellini, Martin, Martin, Martynov,
  Mason, Massera, Masserot, Massinger, Masso-Reid, Mastrogiovanni, Matas,
  Matichard, Matone, Mavalvala, Mazumder, McCarthy, McClelland, McCormick,
  McCuller, McGuire, McIntyre, McIver, McManus, McNeill, McRae, McWilliams,
  Meacher, Meadors, Mehmet, Meidam, Mejuto-Villa, Melatos, Mendell, Mercer,
  Merilh, Merzougui, Meshkov, Messenger, Messick, Metzdorff, Meyers, Miao,
  Michel, Middleton, Mikhailov, Milano, Miller, Miller, Miller, Millhouse,
  Milovich-Goff, Minazzoli, Minenkov, Ming, Mishra, Mitra, Mitrofanov,
  Mitselmakher, Mittleman, Moffa, Moggi, Mogushi, Mohan, Mohapatra, Montani,
  Moore, Moraru, Moreno, Morriss, Mours, Mow-Lowry, Mueller, Muir, Mukherjee,
  Mukherjee, Mukherjee, Mukund, Mullavey, Munch, Muñiz, Muratore, Murray,
  Napier, Nardecchia, Naticchioni, Nayak, Neilson, Nelemans, Nelson, Nery,
  Neunzert, Nevin, Newport, Newton, Ng, Nguyen, Nichols, Nielsen, Nissanke,
  Nitz, Noack, Nocera, Nolting, North, Nuttall, Oberling, O'Dea, Ogin, Oh, Oh,
  Ohme, Okada, Oliver, Oppermann, Oram, O'Reilly, Ormiston, Ortega,
  O'Shaughnessy, Ossokine, Ottaway, Overmier, Owen, Pace, Page, Page, Pai, Pai,
  Palamos, Palashov, Palomba, Pal-Singh, Pan, Pan, Pang, Pang, Pankow,
  Pannarale, Pant, Paoletti, Paoli, Papa, Parida, Parker, Pascucci,
  Pasqualetti, Passaquieti, Passuello, Patil, Patricelli, Pearlstone, Pedraza,
  Pedurand, Pekowsky, Pele, Penn, Perez, Perreca, Perri, Pfeiffer, Phelps,
  Piccinni, Pichot, Piergiovanni, Pierro, Pillant, Pinard, Pinto, Pirello,
  Pitkin, Poe, Poggiani, Popolizio, Porter, Post, Powell, Prasad, Pratt,
  Pratten, Predoi, Prestegard, Prijatelj, Principe, Privitera, Prodi,
  Prokhorov, Puncken, Punturo, Puppo, Pürrer, Qi, Quetschke, Quintero,
  Quitzow-James, Raab, Rabeling, Radkins, Raffai, Raja, Rajan, Rajbhandari,
  Rakhmanov, Ramirez, Ramos-Buades, Rapagnani, Raymond, Razzano, Read,
  Regimbau, Rei, Reid, Reitze, Ren, Reyes, Ricci, Ricker, Rieger, Riles, Rizzo,
  Robertson, Robie, Robinet, Rocchi, Rolland, Rollins, Roma, Romano, Romel,
  Romie, Rosińska, Ross, Rowan, Rüdiger, Ruggi, Rutins, Ryan, Sachdev,
  Sadecki, Sadeghian, Sakellariadou, Salconi, Saleem, Salemi, Samajdar, Sammut,
  Sampson, Sanchez, Sanchez, Sanchis-Gual, Sandberg, Sanders, Sassolas,
  Sathyaprakash, Saulson, Sauter, Savage, Sawadsky, Schale, Scheel, Scheuer,
  Schmidt, Schmidt, Schnabel, Schofield, Schönbeck, Schreiber, Schuette,
  Schulte, Schutz, Schwalbe, Scott, Scott, Seidel, Sellers, Sengupta, Sentenac,
  Sequino, Sergeev, Shaddock, Shaffer, Shah, Shahriar, Shaner, Shao, Shapiro,
  Shawhan, Sheperd, Shoemaker, Shoemaker, Siellez, Siemens, Sieniawska, Sigg,
  Silva, Singer, Singh, Singhal, Sintes, Slagmolen, Smith, Smith, Smith,
  Somala, Son, Sonnenberg, Sorazu, Sorrentino, Souradeep, Spencer, Srivastava,
  Staats, Staley, Steinke, Steinlechner, Steinlechner, Steinmeyer, Stevenson,
  Stone, Stops, Strain, Stratta, Strigin, Strunk, Sturani, Stuver,
  Summerscales, Sun, Sunil, Suresh, Sutton, Swinkels, Szczepańczyk, Tacca,
  Tait, Talbot, Talukder, Tanner, Tápai, Taracchini, Tasson, Taylor, Taylor,
  Tewari, Theeg, Thies, Thomas, Thomas, Thomas, Thorne, Thorne, Thrane, Tiwari,
  Tiwari, Tokmakov, Toland, Tonelli, Tornasi, Torres-Forné, Torrie, Töyrä,
  Travasso, Traylor, Trinastic, Tringali, Trozzo, Tsang, Tse, Tso, Tsukada,
  Tsuna, Tuyenbayev, Ueno, Ugolini, Unnikrishnan, Urban, Usman, Vahlbruch,
  Vajente, Valdes, Bakel, Beuzekom, Brand, Broeck, Vander-Hyde, Schaaf,
  Heijningen, Veggel, Vardaro, Varma, Vass, Vasúth, Vecchio, Vedovato, Veitch,
  Veitch, Venkateswara, Venugopalan, Verkindt, Vetrano, Viceré, Viets,
  Vinciguerra, Vine, Vinet, Vitale, Vo, Vocca, Vorvick, Vyatchanin, Wade, Wade,
  Wade, Walet, Walker, Wallace, Walsh, Wang, Wang, Wang, Wang, Wang, Ward,
  Warner, Was, Watchi, Weaver, Wei, Weinert, Weinstein, Weiss, Wen, Wessel,
  Weßels, Westerweck, Westphal, Wette, Whelan, Whitcomb, Whiting, Whittle,
  Wilken, Williams, Williams, Williamson, Willis, Willke, Wimmer, Winkler,
  Wipf, Wittel, Woan, Woehler, Wofford, Wong, Worden, Wright, Wu, Wysocki,
  Xiao, Yamamoto, Yancey, Yang, Yap, Yazback, Yu, Yu, Yvert,
  Zadro{\textbackslash}.zny, Zanolin, Zelenova, Zendri, Zevin, Zhang, Zhang,
  Zhang, Zhang, Zhao, Zhou, Zhou, Zhu, Zhu, Zimmerman, Zucker, Zweizig, Burns,
  Veres, Kocevski, Racusin, Goldstein, Connaughton, Briggs, Blackburn, Hamburg,
  Hui, Kienlin, McEnery, Preece, Wilson-Hodge, Bissaldi, Cleveland, Gibby,
  Giles, Kippen, McBreen, Meegan, Paciesas, Poolakkil, Roberts, Stanbro,
  Savchenko, Ferrigno, Kuulkers, Bazzano, Bozzo, Brandt, Chenevez, Courvoisier,
  Diehl, Domingo, Hanlon, Jourdain, Laurent, Lebrun, Lutovinov, Mereghetti,
  Natalucci, Rodi, Roques, Sunyaev, Ubertini, \&
  and}]{abbott_gravitational_2017}
Abbott, B.~P., Abbott, R., Abbott, T.~D., {et~al.} 2017,
  \href{http://dx.doi.org/10.3847/2041-8213/aa920c}{\JournalTitle{The
  Astrophysical Journal}, 848, L13}, publisher: American Astronomical Society

\bibitem[{Abbott {et~al.}(2021{\natexlab{a}})Abbott, Abbott, Abraham, Acernese,
  Ackley, Adams, Adams, Adhikari, Adya, Affeldt, Agarwal, Agathos, Agatsuma,
  Aggarwal, Aguiar, Aiello, Ain, Ajith, Akutsu, Aleman, Allen, Allocca, Altin,
  Amato, Anand, Ananyeva, Anderson, Anderson, Ando, Angelova, Ansoldi, Antelis,
  Antier, Appert, Arai, Arai, Arai, Araki, Araya, Araya, Areeda, Arène,
  Aritomi, Arnaud, Aronson, Arun, Asada, Asali, Ashton, Aso, Aston, Astone,
  Aubin, Aufmuth, AultONeal, Austin, Babak, Badaracco, Bader, Bae, Bae, Baer,
  Bagnasco, Bai, Baiotti, Baird, Bajpai, Ball, Ballardin, Ballmer, Bals,
  Balsamo, Baltus, Banagiri, Bankar, Bankar, Barayoga, Barbieri, Barish,
  Barker, Barneo, Barone, Barr, Barsotti, Barsuglia, Barta, Bartlett, Barton,
  Bartos, Bassiri, Basti, Bawaj, Bayley, Baylor, Bazzan, Bécsy, Bedakihale,
  Bejger, Belahcene, Benedetto, Beniwal, Benjamin, Benkel, Bennett, Bentley,
  BenYaala, Bergamin, Berger, Bernuzzi, Berry, Bersanetti, Bertolini,
  Betzwieser, Bhandare, Bhandari, Bhattacharjee, Bhaumik, Bidler, Bilenko,
  Billingsley, Birney, Birnholtz, Biscans, Bischi, Biscoveanu, Bisht, Biswas,
  Bitossi, Bizouard, Blackburn, Blackman, Blair, Blair, Blair, Bobba, Bode,
  Boer, Bogaert, Boldrini, Bondu, Bonilla, Bonnand, Booker, Boom, Bork, Boschi,
  Bose, Bose, Bossilkov, Boudart, Bouffanais, Bozzi, Bradaschia, Brady,
  Bramley, Branch, Branchesi, Brau, Breschi, Briant, Briggs, Brillet,
  Brinkmann, Brockill, Brooks, Brooks, Brown, Brunett, Bruno, Bruntz, Bryant,
  Buikema, Bulik, Bulten, Buonanno, Buscicchio, Buskulic, Byer, Cadonati,
  Caesar, Cagnoli, Cahillane, III, Bustillo, Callaghan, Callister, Calloni,
  Camp, Canepa, Cannavacciuolo, Cannon, Cao, Cao, Cao, Capocasa, Capote,
  Carapella, Carbognani, Carlin, Carney, Carpinelli, Carullo, Carver, Diaz,
  Casentini, Castaldi, Caudill, Cavaglià, Cavalier, Cavalieri, Cella,
  Cerdá-Durán, Cesarini, Chaibi, Chakravarti, Champion, Chan, Chan, Chan,
  Chan, Chandra, Chanial, Chao, Charlton, Chase, Chassande-Mottin, Chatterjee,
  Chaturvedi, Chatziioannou, Chen, Chen, Chen, Chen, Chen, Chen, Chen, Chen,
  Chen, Cheng, Cheong, Cheung, Chia, Chiadini, Chiang, Chierici, Chincarini,
  Chiofalo, Chiummo, Cho, Cho, Choate, Choudhary, Choudhary, Christensen, Chu,
  Chu, Chu, Chua, Chung, Ciani, Ciecielag, Cieślar, Cifaldi, Ciobanu, Ciolfi,
  Cipriano, Cirone, Clara, Clark, Clark, Clarke, Clearwater, Clesse, Cleva,
  Coccia, Cohadon, Cohen, Cohen, Colleoni, Collette, Colpi, Compton, Jr, Conti,
  Cooper, Corban, Corbitt, Cordero-Carrión, Corezzi, Corley, Cornish, Corre,
  Corsi, Cortese, Costa, Cotesta, Coughlin, Coughlin, Coulon, Countryman,
  Cousins, Couvares, Covas, Coward, Cowart, Coyne, Coyne, Creighton, Creighton,
  Criswell, Croquette, Crowder, Cudell, Cullen, Cumming, Cummings, Cuoco,
  Cury{\textbackslash}lo, Canton, Dálya, Dana, DaneshgaranBajastani, D'Angelo,
  Danilishin, D'Antonio, Danzmann, Darsow-Fromm, Dasgupta, Datrier, Dattilo,
  Dave, Davier, Davies, Davis, Daw, Dean, DeBra, Deenadayalan, Degallaix,
  Laurentis, Deléglise, Favero, Lillo, Lillo, Pozzo, DeMarchi, Matteis,
  D'Emilio, Demos, Dent, Depasse, Pietri, Rosa, Rossi, DeSalvo, Simone,
  Dhurandhar, Díaz, Jr, Didio, Dietrich, Fiore, Fronzo, Giorgio, Giovanni,
  Girolamo, Lieto, Ding, Pace, Palma, Renzo, Divakarla, Dmitriev, Doctor,
  D'Onofrio, Donovan, Dooley, Doravari, Dorrington, Drago, Driggers, Drori, Du,
  Ducoin, Dupej, Durante, D'Urso, Duverne, Dwyer, Easter, Ebersold, Eddolls,
  Edelman, Edo, Edy, Effler, Eguchi, Eichholz, Eikenberry, Eisenmann,
  Eisenstein, Ejlli, Enomoto, Errico, Essick, Estellés, Estevez, Etienne,
  Etzel, Evans, Evans, Ewing, Fafone, Fair, Fairhurst, Fan, Farah, Farinon,
  Farr, Farr, Farrow, Fauchon-Jones, Favata, Fays, Fazio, Feicht, Fejer, Feng,
  Fenyvesi, Ferguson, Fernandez-Galiana, Ferrante, Ferreira, Fidecaro, Figura,
  Fiori, Fishbach, Fisher, Fittipaldi, Fiumara, Flaminio, Floden, Flynn, Fong,
  Font, Fornal, Forsyth, Franke, Frasca, Frasconi, Frederick, Frei, Freise,
  Frey, Fritschel, Frolov, Fronzé, Fujii, Fujikawa, Fukunaga, Fukushima,
  Fulda, Fyffe, Gabbard, Gadre, Gaebel, Gair, Gais, Galaudage, Gamba,
  Ganapathy, Ganguly, Gao, Gaonkar, Garaventa, García-Núñez,
  García-Quirós, Garufi, Gateley, Gaudio, Gayathri, Ge, Gemme, Gennai,
  George, Gergely, Gewecke, Ghonge, Ghosh, Ghosh, Ghosh, Ghosh, Ghosh,
  Giacomazzo, Giacoppo, Giaime, Giardina, Gibson, Gier, Giesler, Giri, Gissi,
  Glanzer, Gleckl, Godwin, Goetz, Goetz, Gohlke, Goncharov, González,
  Gopakumar, Gosselin, Gouaty, Grace, Grado, Granata, Granata, Grant, Gras,
  Grassia, Gray, Gray, Greco, Green, Green, Gretarsson, Gretarsson, Griffith,
  Griffiths, Griggs, Grignani, Grimaldi, Grimes, Grimm, Grote, Grunewald,
  Gruning, Guerrero, Guidi, Guimaraes, Guixé, Gulati, Guo, Guo, Gupta, Gupta,
  Gupta, Gustafson, Gustafson, Guzman, Ha, Haegel, Hagiwara, Haino, Halim,
  Hall, Hamilton, Hammond, Han, Haney, Hanks, Hanna, Hannam, Hannuksela,
  Hansen, Hansen, Hanson, Harder, Hardwick, Haris, Harms, Harry, Harry,
  Hartwig, Hasegawa, Haskell, Hasskew, Haster, Hattori, Haughian, Hayakawa,
  Hayama, Hayes, Healy, Heidmann, Heintze, Heinze, Heinzel, Heitmann, Hellman,
  Hello, Helmling-Cornell, Hemming, Hendry, Heng, Hennes, Hennig, Hennig,
  Vivanco, Heurs, Hild, Hill, Himemoto, Hinderer, Hines, Hiranuma, Hirata,
  Hirose, Ho, Hochheim, Hofman, Hohmann, Holgado, Holland, Hollows, Holmes,
  Holt, Holz, Hong, Hopkins, Hough, Howell, Hoy, Hoyland, Hreibi, Hsieh, Hsu,
  Huang, Huang, Huang, Huang, Huang, Huang, Hübner, Huddart, Huerta, Hughey,
  Hui, Hui, Husa, Huttner, Huxford, Huynh-Dinh, Ide, Idzkowski, Iess, Ikenoue,
  Imam, Inayoshi, Inchauspe, Ingram, Inoue, Intini, Ioka, Isi, Isleif, Ito,
  Itoh, Iyer, Izumi, JaberianHamedan, Jacqmin, Jadhav, Jadhav, James, Jan,
  Jani, Janssens, Janthalur, Jaranowski, Jariwala, Jaume, Jenkins, Jeon,
  Jeunon, Jia, Jiang, Jin, Johns, Jones, Jones, Jones, Jones, Jones, Jonker,
  Ju, Jung, Jung, Junker, Kaihotsu, Kajita, Kakizaki, Kalaghatgi, Kalogera,
  Kamai, Kamiizumi, Kanda, Kandhasamy, Kang, Kanner, Kao, Kapadia, Kapasi,
  Karat, Karathanasis, Karki, Kashyap, Kasprzack, Kastaun, Katsanevas,
  Katsavounidis, Katzman, Kaur, Kawabe, Kawaguchi, Kawai, Kawasaki, Kéfélian,
  Keitel, Key, Khadka, Khalili, Khan, Khan, Khazanov, Khetan, Khursheed,
  Kijbunchoo, Kim, Kim, Kim, Kim, Kim, Kim, Kimball, Kimura, King,
  Kinley-Hanlon, Kirchhoff, Kissel, Kita, Kitazawa, Kleybolte, Klimenko, Knee,
  Knowles, Knyazev, Koch, Koekoek, Kojima, Kokeyama, Koley, Kolitsidou,
  Kolstein, Komori, Kondrashov, Kong, Kontos, Koper, Korobko, Kotake, Kovalam,
  Kozak, Kozakai, Kozu, Kringel, Krishnendu, Królak, Kuehn, Kuei, Kumar,
  Kumar, Kumar, Kumar, Kume, Kuns, Kuo, Kuo, Kuromiya, Kuroyanagi, Kusayanagi,
  Kwak, Kwang, Laghi, Lalande, Lam, Lamberts, Landry, Landry, Lane, Lang,
  Lange, Lantz, Rosa, Lartaux-Vollard, Lasky, Laxen, Lazzarini, Lazzaro, Leaci,
  Leavey, Lecoeuche, Lee, Lee, Lee, Lee, Lee, Lee, Lehmann, Lemaître, Leon,
  Leonardi, Leroy, Letendre, Levin, Leviton, Li, Li, Li, Li, Li, Li, Lin, Lin,
  Lin, Lin, Lin, Linde, Linker, Linley, Littenberg, Liu, Liu, Liu, Liu,
  Llorens-Monteagudo, Lo, Lockwood, Lollie, London, Longo, Lopez, Lorenzini,
  Loriette, Lormand, Losurdo, Lough, Lousto, Lovelace, Lück, Lumaca, Lundgren,
  Luo, Macas, MacInnis, Macleod, MacMillan, Macquet, Hernandez,
  Magaña-Sandoval, Magazzù, Magee, Maggiore, Majorana, Makarem, Maksimovic,
  Maliakal, Malik, Man, Mandic, Mangano, Mango, Mansell, Manske, Mantovani,
  Mapelli, Marchesoni, Marchio, Marion, Mark, Márka, Márka, Markakis,
  Markosyan, Markowitz, Maros, Marquina, Marsat, Martelli, Martin, Martin,
  Martinez, Martinez, Martinovic, Martynov, Marx, Masalehdan, Mason, Massera,
  Masserot, Massinger, Masso-Reid, Mastrogiovanni, Matas, Mateu-Lucena,
  Matichard, Matiushechkina, Mavalvala, McCann, McCarthy, McClelland, McClincy,
  McCormick, McCuller, McGhee, McGuire, McIsaac, McIver, McManus, McRae,
  McWilliams, Meacher, Mehmet, Mehta, Melatos, Melchor, Mendell,
  Menendez-Vazquez, Menoni, Mercer, Mereni, Merfeld, Merilh, Merritt,
  Merzougui, Meshkov, Messenger, Messick, Meyers, Meylahn, Mhaske, Miani, Miao,
  Michaloliakos, Michel, Michimura, Middleton, Milano, Miller, Millhouse,
  Mills, Milotti, Milovich-Goff, Minazzoli, Minenkov, Mio, Mir, Mishkin,
  Mishra, Mishra, Mistry, Mitra, Mitrofanov, Mitselmakher, Mittleman, Miyakawa,
  Miyamoto, Miyazaki, Miyo, Miyoki, Mo, Mogushi, Mohapatra, Mohite, Molina,
  Molina-Ruiz, Mondin, Montani, Moore, Moraru, Morawski, More, Moreno, Moreno,
  Mori, Morisaki, Moriwaki, Mours, Mow-Lowry, Mozzon, Muciaccia, Mukherjee,
  Mukherjee, Mukherjee, Mukherjee, Mukund, Mullavey, Munch, Muñiz, Murray,
  Musenich, Nadji, Nagano, Nagano, Nagar, Nakamura, Nakano, Nakano, Nakashima,
  Nakayama, Nardecchia, Narikawa, Naticchioni, Nayak, Nayak, Negishi, Neil,
  Neilson, Nelemans, Nelson, Nery, Neunzert, Ng, Ng, Nguyen, Nguyen, Nguyen,
  Quynh, Ni, Nichols, Nishizawa, Nissanke, Nocera, Noh, Norman, North, Nozaki,
  Nuttall, Oberling, O'Brien, Obuchi, O'Dell, Ogaki, Oganesyan, Oh, Oh, Oh,
  Ohashi, Ohishi, Ohkawa, Ohme, Ohta, Okada, Okutani, Okutomi, Olivetto,
  Oohara, Ooi, Oram, O'Reilly, Ormiston, Ormsby, Ortega, O'Shaughnessy, O'Shea,
  Oshino, Ossokine, Osthelder, Otabe, Ottaway, Overmier, Pace, Pagano, Page,
  Pagliaroli, Pai, Pai, Palamos, Palashov, Palomba, Pan, Panda, Pang, Pang,
  Pankow, Pannarale, Pant, Paoletti, Paoli, Paolone, Parisi, Park, Parker,
  Pascucci, Pasqualetti, Passaquieti, Passuello, Patel, Patricelli, Payne,
  Pechsiri, Pedraza, Pegoraro, Pele, Arellano, Penn, Perego, Pereira, Pereira,
  Perez, Périgois, Perreca, Perriès, Petermann, Petterson, Pfeiffer, Pham,
  Phukon, Piccinni, Pichot, Piendibene, Piergiovanni, Pierini, Pierro, Pillant,
  Pilo, Pinard, Pinto, Piotrzkowski, Piotrzkowski, Pirello, Pitkin, Placidi,
  Plastino, Pluchar, Poggiani, Polini, Pong, Ponrathnam, Popolizio, Porter,
  Powell, Pracchia, Pradier, Prajapati, Prasai, Prasanna, Pratten, Prestegard,
  Principe, Prodi, Prokhorov, Prosposito, Prudenzi, Puecher, Punturo, Puosi,
  Puppo, Pürrer, Qi, Quetschke, Quinonez, Quitzow-James, Raab, Raaijmakers,
  Radkins, Radulesco, Raffai, Rail, Raja, Rajan, Ramirez, Ramirez,
  Ramos-Buades, Rana, Rapagnani, Rapol, Ratto, Ray, Raymond, Raza, Razzano,
  Read, Rees, Regimbau, Rei, Reid, Reitze, Relton, Rettegno, Ricci, Richardson,
  Richardson, Richardson, Ricker, Riemenschneider, Riles, Rizzo, Robertson,
  Robie, Robinet, Rocchi, Rocha, Rodriguez, Rodriguez-Soto, Rolland, Rollins,
  Roma, Romanelli, Romano, Romel, Romero, Romero-Shaw, Romie, Rose, Rosińska,
  Rosofsky, Ross, Rowan, Rowlinson, Roy, Roy, Rozza, Ruggi, Ryan, Sachdev,
  Sadecki, Sadiq, Sago, Saito, Saito, Sakai, Sakai, Sakellariadou, Sakuno,
  Salafia, Salconi, Saleem, Salemi, Samajdar, Sanchez, Sanchez, Sanchez,
  Sanchis-Gual, Sanders, Sanuy, Saravanan, Sarin, Sassolas, Satari,
  Sathyaprakash, Sato, Sato, Sauter, Savage, Savant, Sawada, Sawant, Sawant,
  Sayah, Schaetzl, Scheel, Scheuer, Schindler-Tyka, Schmidt, Schnabel,
  Schneewind, Schofield, Schönbeck, Schulte, Schutz, Schwartz, Scott, Scott,
  Seglar-Arroyo, Seidel, Sekiguchi, Sekiguchi, Sellers, Sengupta, Sennett,
  Sentenac, Seo, Sequino, Sergeev, Setyawati, Shaffer, Shahriar, Shams, Shao,
  Sharifi, Sharma, Sharma, Shawhan, Shcheblanov, Shen, Shibagaki, Shikauchi,
  Shimizu, Shimoda, Shimode, Shink, Shinkai, Shishido, Shoda, Shoemaker,
  Shoemaker, Shukla, ShyamSundar, Sieniawska, Sigg, Singer, Singh, Singh,
  Singha, Sintes, Sipala, Skliris, Slagmolen, Slaven-Blair, Smetana, Smith,
  Smith, Somala, Somiya, Son, Soni, Soni, Sorazu, Sordini, Sorrentino,
  Sorrentino, Sotani, Soulard, Souradeep, Sowell, Spagnuolo, Spencer, Spera,
  Srivastava, Srivastava, Staats, Stachie, Steer, Steinlechner, Steinlechner,
  Stops, Stevenson, Stover, Strain, Strang, Stratta, Strunk, Sturani, Stuver,
  Südbeck, Sudhagar, Sudhir, Sugimoto, Suh, Summerscales, Sun, Sun, Sunil,
  Sur, Suresh, Sutton, Suzuki, Suzuki, Swinkels, Szczepańczyk, Szewczyk,
  Tacca, Tagoshi, Tait, Takahashi, Takahashi, Takamori, Takano, Takeda, Takeda,
  Talbot, Tanaka, Tanaka, Tanaka, Tanaka, Tanaka, Tanasijczuk, Tanioka, Tanner,
  Tao, Tapia, Martin, Tasson, Telada, Tenorio, Terkowski, Test,
  Thirugnanasambandam, Thomas, Thomas, Thompson, Thondapu, Thorne, Thrane,
  Tiwari, Tiwari, Tiwari, Toland, Tolley, Tomaru, Tomigami, Tomura, Tonelli,
  Torres-Forné, Torrie, Melo, Töyrä, Trapananti, Travasso, Traylor,
  Tringali, Tripathee, Troiano, Trovato, Trozzo, Trudeau, Tsai, Tsai, Tsang,
  Tsang, Tsao, Tse, Tso, Tsubono, Tsuchida, Tsukada, Tsuna, Tsutsui, Tsuzuki,
  Turconi, Tuyenbayev, Ubhi, Uchikata, Uchiyama, Udall, Ueda, Uehara, Ueno,
  Ueshima, Ugolini, Unnikrishnan, Uraguchi, Urban, Ushiba, Usman, Utina,
  Vahlbruch, Vajente, Vajpeyi, Valdes, Valentini, Valsan, Bakel, Beuzekom,
  Brand, Broeck, Vander-Hyde, Schaaf, Heijningen, Vanosky, Putten, Vardaro,
  Vargas, Varma, Vasúth, Vecchio, Vedovato, Veitch, Veitch, Venkateswara,
  Venneberg, Venugopalan, Verkindt, Verma, Veske, Vetrano, Viceré, Viets,
  Villa-Ortega, Vinet, Vitale, Vo, Vocca, Reis, Wrangel, Vorvick, Vyatchanin,
  Wade, Wade, Wagner, Walet, Walker, Wallace, Wallace, Walsh, Wang, Wang, Wang,
  Ward, Warner, Was, Washimi, Washington, Watchi, Weaver, Wei, Weinert,
  Weinstein, Weiss, Weller, Wellmann, Wen, Weßels, Westhouse, Wette, Whelan,
  White, Whiting, Whittle, Wilken, Williams, Williams, Williamson, Willis,
  Willke, Wilson, Winkler, Wipf, Wlodarczyk, Woan, Woehler, Wofford, Wong, Wu,
  Wu, Wu, Wu, Wysocki, Xiao, Xu, Yamada, Yamamoto, Yamamoto, Yamamoto,
  Yamamoto, Yamashita, Yamazaki, Yang, Yang, Yang, Yang, Yang, Yap, Yeeles,
  Yelikar, Ying, Yokogawa, Yokoyama, Yokozawa, Yoon, Yoshioka, Yu, Yu,
  Yuzurihara, Zadro{\textbackslash}.zny, Zanolin, Zappa, Zeidler, Zelenova,
  Zendri, Zevin, Zhan, Zhang, Zhang, Zhang, Zhang, Zhang, Zhao, Zhao, Zhao,
  Zhao, Zhou, Zhu, Zhu, Zimmerman, Zlochower, Zucker, \&
  Zweizig}]{abbott_observation_2021}
Abbott, R., Abbott, T.~D., Abraham, S., {et~al.} 2021{\natexlab{a}},
  \href{http://dx.doi.org/10.3847/2041-8213/ac082e}{\JournalTitle{The
  Astrophysical Journal Letters}, 915, L5}, publisher: American Astronomical
  Society

\bibitem[{Abbott {et~al.}(2021{\natexlab{b}})Abbott, Abbott, Abraham, Acernese,
  Ackley, Adams, Adams, Adhikari, Adya, Affeldt, Agathos, Agatsuma, Aggarwal,
  Aguiar, Aiello, Ain, Ajith, Allen, Allocca, Altin, Amato, Anand, Ananyeva,
  Anderson, Anderson, Angelova, Ansoldi, Antelis, Antier, Appert, Arai, Araya,
  Areeda, Arène, Arnaud, Aronson, Arun, Asali, Ascenzi, Ashton, Aston, Astone,
  Aubin, Aufmuth, AultONeal, Austin, Avendano, Babak, Badaracco, Bader, Bae,
  Baer, Bagnasco, Baird, Ball, Ballardin, Ballmer, Bals, Balsamo, Baltus,
  Banagiri, Bankar, Bankar, Barayoga, Barbieri, Barish, Barker, Barneo, Barnum,
  Barone, Barr, Barsotti, Barsuglia, Barta, Bartlett, Bartos, Bassiri, Basti,
  Bawaj, Bayley, Bazzan, Becher, Bécsy, Bedakihale, Bejger, Belahcene,
  Beniwal, Benjamin, Bennett, Bentley, Bergamin, Berger, Bergmann, Bernuzzi,
  Berry, Bersanetti, Bertolini, Betzwieser, Bhandare, Bhandari, Bhattacharjee,
  Bidler, Bilenko, Billingsley, Birney, Birnholtz, Biscans, Bischi, Biscoveanu,
  Bisht, Bitossi, Bizouard, Blackburn, Blackman, Blair, Blair, Blair, Blanch,
  Bobba, Bode, Boer, Boetzel, Bogaert, Boldrini, Bondu, Bonilla, Bonnand,
  Booker, Boom, Bork, Boschi, Bose, Bossilkov, Boudart, Bouffanais, Bozzi,
  Bradaschia, Brady, Bramley, Branchesi, Brau, Breschi, Briant, Briggs,
  Brighenti, Brillet, Brinkmann, Brockill, Brooks, Brooks, Brown, Brunett,
  Bruno, Bruntz, Buikema, Bulik, Bulten, Buonanno, Buscicchio, Buskulic, Byer,
  Cabero, Cadonati, Caesar, Cagnoli, Cahillane, Bustillo, Callaghan, Callister,
  Calloni, Camp, Canepa, Cannon, Cao, Cao, Carapella, Carbognani, Carney,
  Carpinelli, Carullo, Carver, Diaz, Casentini, Caudill, Cavaglià, Cavalier,
  Cavalieri, Cella, Cerdá-Durán, Cesarini, Chaibi, Chakravarti, Chan, Chan,
  Chandra, Chanial, Chao, Charlton, Chase, Chassande-Mottin, Chatterjee,
  Chattopadhyay, Chaturvedi, Chatziioannou, Chen, Chen, Chen, Chen, Cheng,
  Cheong, Chia, Chiadini, Chierici, Chincarini, Chiummo, Cho, Cho, Cho, Choate,
  Christensen, Chu, Chua, Chung, Chung, Ciani, Ciecielag, Cieślar, Cifaldi,
  Ciobanu, Ciolfi, Cipriano, Cirone, Clara, Clark, Clark, Clarke, Clearwater,
  Clesse, Cleva, Coccia, Cohadon, Cohen, Colleoni, Collette, Collins, Colpi,
  Constancio, Conti, Cooper, Corban, Corbitt, Cordero-Carrión, Corezzi,
  Corley, Cornish, Corre, Corsi, Cortese, Costa, Cotesta, Coughlin, Coughlin,
  Coulon, Countryman, Couvares, Covas, Coward, Cowart, Coyne, Coyne, Creighton,
  Creighton, Croquette, Crowder, Cudell, Cullen, Cumming, Cummings, Cunningham,
  Cuoco, Curylo, Canton, Dálya, Dana, DaneshgaranBajastani, D'Angelo,
  Danilishin, D'Antonio, Danzmann, Darsow-Fromm, Dasgupta, Datrier, Dattilo,
  Dave, Davier, Davies, Davis, Daw, Dean, DeBra, Deenadayalan, Degallaix,
  Laurentis, Deléglise, Favero, Lillo, Lillo, Pozzo, DeMarchi, Matteis,
  D'Emilio, Demos, Denker, Dent, Depasse, Pietri, Rosa, Rossi, DeSalvo, Varona,
  Dhurandhar, Díaz, Diaz-Ortiz, Didio, Dietrich, Fiore, DiFronzo, Giorgio,
  Giovanni, Giovanni, Girolamo, Lieto, Ding, Pace, Palma, Renzo, Divakarla,
  Dmitriev, Doctor, D'Onofrio, Donovan, Dooley, Doravari, Dorrington, Downes,
  Drago, Driggers, Du, Ducoin, Dupej, Durante, D'Urso, Duverne, Dwyer, Easter,
  Eddolls, Edelman, Edo, Edy, Effler, Eichholz, Eikenberry, Eisenmann,
  Eisenstein, Ejlli, Errico, Essick, Estellés, Estevez, Etienne, Etzel, Evans,
  Evans, Ewing, Fafone, Fair, Fairhurst, Fan, Farah, Farinon, Farr, Farr,
  Fauchon-Jones, Favata, Fays, Fazio, Feicht, Fejer, Feng, Fenyvesi, Ferguson,
  Fernandez-Galiana, Ferrante, Ferreira, Fidecaro, Figura, Fiori, Fiorucci,
  Fishbach, Fisher, Fishner, Fittipaldi, Fitz-Axen, Fiumara, Flaminio, Floden,
  Flynn, Fong, Font, Forsyth, Fournier, Frasca, Frasconi, Frei, Freise, Frey,
  Frey, Fritschel, Frolov, Fronzé, Fulda, Fyffe, Gabbard, Gadre, Gaebel, Gair,
  Gais, Galaudage, Gamba, Ganapathy, Ganguly, Gaonkar, Garaventa,
  García-Quirós, Garufi, Gateley, Gaudio, Gayathri, Gemme, Gennai, George,
  George, Gergely, Ghonge, Ghosh, Ghosh, Ghosh, Giacomazzo, Giacoppo, Giaime,
  Giardina, Gibson, Gier, Gill, Giri, Glanzer, Gleckl, Godwin, Goetz, Goetz,
  Gohlke, Goncharov, González, Gopakumar, Gossan, Gosselin, Gouaty, Grace,
  Grado, Granata, Granata, Grant, Gras, Grassia, Gray, Gray, Greco, Green,
  Green, Gretarsson, Griggs, Grignani, Grimaldi, Grimes, Grimm, Grote,
  Grunewald, Gruning, Guerrero, Guidi, Guimaraes, Guixé, Gulati, Guo, Gupta,
  Gupta, Gupta, Gustafson, Gustafson, Guzman, Haegel, Halim, Hall, Hamilton,
  Hammond, Haney, Hanke, Hanks, Hanna, Hannuksela, Hannuksela, Hansen, Hansen,
  Hanson, Harder, Hardwick, Haris, Harms, Harry, Harry, Hartwig, Hasskew,
  Haster, Haughian, Hayes, Healy, Heidmann, Heintze, Heinze, Heinzel, Heitmann,
  Hellman, Hello, Helmling-Cornell, Hemming, Hendry, Heng, Hennes, Hennig,
  Hennig, Vivanco, Heurs, Hild, Hill, Hines, Hochheim, Hofgard, Hofman,
  Hohmann, Holgado, Holland, Hollows, Holmes, Holt, Holz, Hopkins, Horst,
  Hough, Howell, Hoy, Hoyland, Huang, Hübner, Huddart, Huerta, Hughey, Hui,
  Husa, Huttner, Hutzler, Huxford, Huynh-Dinh, Idzkowski, Iess, Imperato,
  Inchauspe, Ingram, Intini, Isi, Iyer, JaberianHamedan, Jacqmin, Jadhav,
  Jadhav, James, Jani, Janssens, Janthalur, Jaranowski, Jariwala, Jaume,
  Jenkins, Jeunon, Jiang, Johns, Jones, Jones, Jones, Jones, Jones, Jonker, Ju,
  Junker, Kalaghatgi, Kalogera, Kamai, Kandhasamy, Kang, Kanner, Kapadia,
  Kapasi, Karathanasis, Karki, Kashyap, Kasprzack, Kastaun, Katsanevas,
  Katsavounidis, Katzman, Kawabe, Kéfélian, Keitel, Key, Khadka, Khalili,
  Khan, Khan, Khazanov, Khetan, Khursheed, Kijbunchoo, Kim, Kim, Kim, Kim, Kim,
  Kim, Kimball, King, Kinley-Hanlon, Kirchhoff, Kissel, Kleybolte, Klimenko,
  Knowles, Knyazev, Koch, Koehlenbeck, Koekoek, Koley, Kolstein, Komori,
  Kondrashov, Kontos, Koper, Korobko, Korth, Kovalam, Kozak, Krämer, Kringel,
  Krishnendu, Królak, Kuehn, Kumar, Kumar, Kumar, Kumar, Kuns, Kwang, Lackey,
  Laghi, Lalande, Lam, Lamberts, Landry, Lane, Lang, Lange, Lantz, Lanza, Rosa,
  Lartaux-Vollard, Lasky, Laxen, Lazzarini, Lazzaro, Leaci, Leavey, Lecoeuche,
  Lee, Lee, Lee, Lee, Lehmann, Leon, Leroy, Letendre, Levin, Li, Li, Li, Li,
  Li, Linde, Linker, Linley, Littenberg, Liu, Liu, Llorens-Monteagudo, Lo,
  Lockwood, London, Longo, Lorenzini, Loriette, Lormand, Losurdo, Lough,
  Lousto, Lovelace, Lück, Lumaca, Lundgren, Ma, Macas, MacInnis, Macleod,
  MacMillan, Macquet, Hernandez, Magaña-Sandoval, Magazzù, Magee, Majorana,
  Maksimovic, Maliakal, Malik, Man, Mandic, Mangano, Mansell, Manske,
  Mantovani, Mapelli, Marchesoni, Marion, Márka, Márka, Markakis, Markosyan,
  Markowitz, Maros, Marquina, Marsat, Martelli, Martin, Martin, Martinez,
  Martinez, Martynov, Masalehdan, Mason, Massera, Masserot, Massinger,
  Masso-Reid, Mastrogiovanni, Matas, Mateu-Lucena, Matichard, Matiushechkina,
  Mavalvala, Maynard, McCann, McCarthy, McClelland, McCormick, McCuller,
  McGuire, McIsaac, McIver, McManus, McRae, McWilliams, Meacher, Meadors,
  Mehmet, Mehta, Melatos, Melchor, Mendell, Menendez-Vazquez, Mercer, Mereni,
  Merfeld, Merilh, Merritt, Merzougui, Meshkov, Messenger, Messick, Metzdorff,
  Meyers, Meylahn, Mhaske, Miani, Miao, Michaloliakos, Michel, Middleton,
  Milano, Miller, Miller, Millhouse, Mills, Milotti, Milovich-Goff, Minazzoli,
  Minenkov, Mir, Mishkin, Mishra, Mistry, Mitra, Mitrofanov, Mitselmakher,
  Mittleman, Mo, Mogushi, Mohapatra, Mohite, Molina, Molina-Ruiz, Mondin,
  Montani, Moore, Moraru, Morawski, Moreno, Morisaki, Mours, Mow-Lowry, Mozzon,
  Muciaccia, Mukherjee, Mukherjee, Mukherjee, Mukherjee, Mukund, Mullavey,
  Munch, Muñiz, Murray, Nadji, Nagar, Nardecchia, Naticchioni, Nayak, Neil,
  Neilson, Nelemans, Nelson, Nery, Neunzert, Ng, Ng, Nguyen, Nguyen, Nguyen,
  Nichols, Nissanke, Nocera, Noh, North, Nothard, Nuttall, Oberling, O'Brien,
  O'Dell, Oganesyan, Ogin, Oh, Oh, Ohme, Ohta, Okada, Olivetto, Oppermann,
  Oram, O'Reilly, Ormiston, Ormsby, Ortega, O'Shaughnessy, Ossokine, Osthelder,
  Ottaway, Overmier, Owen, Pace, Pagano, Page, Pagliaroli, Pai, Pai, Palamos,
  Palashov, Palomba, Pan, Panda, Pang, Pankow, Pannarale, Pant, Paoletti,
  Paoli, Paolone, Parker, Pascucci, Pasqualetti, Passaquieti, Passuello, Patel,
  Patricelli, Payne, Pechsiri, Pedraza, Pegoraro, Pele, Penn, Perego, Perez,
  Périgois, Perreca, Perriès, Petermann, Petterson, Pfeiffer, Pham, Phukon,
  Piccinni, Pichot, Piendibene, Piergiovanni, Pierini, Pierro, Pillant, Pilo,
  Pinard, Pinto, Piotrzkowski, Pirello, Pitkin, Placidi, Plastino, Pluchar,
  Poggiani, Polini, Pong, Ponrathnam, Popolizio, Porter, Poverman, Powell,
  Pracchia, Prajapati, Prasai, Prasanna, Pratten, Prestegard, Principe, Prodi,
  Prokhorov, Prosposito, Puecher, Punturo, Puosi, Puppo, Pürrer, Qi,
  Quetschke, Quinonez, Quitzow-James, Raab, Raaijmakers, Radkins, Radulesco,
  Raffai, Rafferty, Rail, Raja, Rajan, Rajbhandari, Rakhmanov, Ramirez,
  Ramirez, Ramos-Buades, Rana, Rao, Rapagnani, Rapol, Ratto, Raymond, Razzano,
  Read, Regimbau, Rei, Reid, Reitze, Rettegno, Ricci, Richardson, Richardson,
  Richardson, Ricker, Riemenschneider, Riles, Rizzo, Robertson, Robinet,
  Rocchi, Rocha, Rodriguez, Rodriguez-Soto, Rolland, Rollins, Roma, Romanelli,
  Romano, Romel, Romero, Romero-Shaw, Romie, Ronchini, Rose, Rose, Rose,
  Rosell, Rosińska, Rosofsky, Ross, Rowan, Rowlinson, Roy, Roy, Ruggi, Ryan,
  Sachdev, Sadecki, Sakellariadou, Salafia, Salconi, Saleem, Samajdar, Sanchez,
  Sanchez, Sanchez, Sanchis-Gual, Sanders, Santiago, Santos, Saravanan, Sarin,
  Sassolas, Sathyaprakash, Sauter, Savage, Savant, Sawant, Sayah, Schaetzl,
  Schale, Scheel, Scheuer, Schindler-Tyka, Schmidt, Schnabel, Schofield,
  Schönbeck, Schreiber, Schulte, Schutz, Schwarm, Schwartz, Scott, Scott,
  Seglar-Arroyo, Seidel, Sellers, Sengupta, Sennett, Sentenac, Sequino,
  Sergeev, Setyawati, Shaffer, Shahriar, Sharifi, Sharma, Sharma, Shawhan,
  Shen, Shikauchi, Shink, Shoemaker, Shoemaker, Shukla, ShyamSundar,
  Sieniawska, Sigg, Singer, Singh, Singh, Singha, Singhal, Sintes, Sipala,
  Skliris, Slagmolen, Slaven-Blair, Smetana, Smith, Smith, Somala, Son, Soni,
  Sorazu, Sordini, Sorrentino, Sorrentino, Soulard, Souradeep, Sowell, Spencer,
  Spera, Srivastava, Srivastava, Staats, Stachie, Steer, Steinke, Steinlechner,
  Steinlechner, Steinmeyer, Stevenson, Stolle-McAllister, Stops, Stover,
  Strain, Stratta, Strunk, Sturani, Stuver, Südbeck, Sudhagar, Sudhir, Suh,
  Summerscales, Sun, Sun, Sunil, Sur, Suresh, Sutton, Swinkels, Szczepańczyk,
  Tacca, Tait, Talbot, Tanasijczuk, Tanner, Tao, Tapia, Martin, Tasson, Taylor,
  Tenorio, Terkowski, Thirugnanasambandam, Thomas, Thomas, Thomas, Thompson,
  Thondapu, Thorne, Thrane, Tiwari, Tiwari, Tiwari, Toland, Tolley, Tonelli,
  Tornasi, Torres-Forné, Torrie, Melo, Töyrä, Tran, Trapananti, Travasso,
  Traylor, Tringali, Tripathee, Trovato, Trudeau, Tsai, Tsang, Tse, Tso,
  Tsukada, Tsuna, Tsutsui, Turconi, Ubhi, Udall, Ueno, Ugolini, Unnikrishnan,
  Urban, Usman, Utina, Vahlbruch, Vajente, Vajpeyi, Valdes, Valentini, Valsan,
  Bakel, Beuzekom, Brand, Broeck, Vander-Hyde, Schaaf, Heijningen, Vardaro,
  Vargas, Varma, Vass, Vasúth, Vecchio, Vedovato, Veitch, Veitch,
  Venkateswara, Venneberg, Venugopalan, Verkindt, Verma, Veske, Vetrano,
  Viceré, Viets, Villa-Ortega, Vinet, Vitale, Vo, Vocca, Vorvick, Vyatchanin,
  Wade, Wade, Wade, Walet, Walker, Wallace, Wallace, Walsh, Wang, Wang, Wang,
  Wang, Ward, Warner, Was, Washington, Watchi, Weaver, Wei, Weinert, Weinstein,
  Weiss, Wellmann, Wen, Weßels, Westhouse, Wette, Whelan, White, White,
  Whiting, Whittle, Wilken, Williams, Williams, Williamson, Willis, Willke,
  Wilson, Wimmer, Winkler, Wipf, Woan, Woehler, Wofford, Wong, Wrangel, Wright,
  Wu, Wysocki, Xiao, Yamamoto, Yang, Yang, Yang, Yap, Yeeles, Yoon, Yu, Yu,
  Yuen, Zadro{\textbackslash}.zny, Zanolin, Zelenova, Zendri, Zevin, Zhang,
  Zhang, Zhang, Zhang, Zhao, Zhao, Zhou, Zhou, Zhu, Zimmerman, Zucker, \&
  Zweizig}]{abbott_population_2021}
---. 2021{\natexlab{b}},
  \href{http://dx.doi.org/10.3847/2041-8213/abe949}{\JournalTitle{The
  Astrophysical Journal Letters}, 913, L7}, publisher: American Astronomical
  Society

\bibitem[{Abe {et~al.}(2021)Abe, Asami, Gando, Gando, Gima, Goto, Hachiya,
  Hata, Hayashida, Hosokawa, Ichimura, Ieki, Ikeda, Inoue, Ishidoshiro, Kamei,
  Kawada, Kishimoto, Kinoshita, Koga, Maemura, Mitsui, Miyake, Nakamura,
  Nakamura, Nakamura, Ozaki, Sakai, Sambonsugi, Shimizu, Shirai, Shiraishi,
  Suzuki, Suzuki, Takeuchi, Tamae, Ueshima, Wada, Watanabe, Yoshida, Obara,
  Kozlov, Chernyak, Takemoto, Yoshida, Umehara, Fushimi, Ichikawa, Nakamura,
  Yoshida, Berger, Fujikawa, Learned, Maricic, Axani, Winslow, Fu, Ouellet,
  Efremenko, Karwowski, Markoff, Tornow, Li, Detwiler, Enomoto, Decowski,
  Grant, O'Donnell, \& Dell'Oro}]{abe_search_2021}
Abe, S., Asami, S., Gando, A., {et~al.} 2021,
  \href{http://dx.doi.org/10.3847/1538-4357/abd5bc}{\JournalTitle{The
  Astrophysical Journal}, 909, 116}, publisher: American Astronomical Society

\bibitem[{Acernese {et~al.}(2014)Acernese, Agathos, Agatsuma, Aisa, Allemandou,
  Allocca, Amarni, Astone, Balestri, Ballardin, Barone, Baronick, Barsuglia,
  Basti, Basti, Bauer, Bavigadda, Bejger, Beker, Belczynski, Bersanetti,
  Bertolini, Bitossi, Bizouard, Bloemen, Blom, Boer, Bogaert, Bondi, Bondu,
  Bonelli, Bonnand, Boschi, Bosi, Bouedo, Bradaschia, Branchesi, Briant,
  Brillet, Brisson, Bulik, Bulten, Buskulic, Buy, Cagnoli, Calloni, Campeggi,
  Canuel, Carbognani, Cavalier, Cavalieri, Cella, Cesarini, Mottin, Chincarini,
  Chiummo, Chua, Cleva, Coccia, Cohadon, Colla, Colombini, Conte, Coulon,
  Cuoco, Dalmaz, D'Antonio, Dattilo, Davier, Day, Debreczeni, Degallaix,
  Deléglise, Pozzo, Dereli, Rosa, Fiore, Lieto, Virgilio, Doets, Dolique,
  Drago, Ducrot, Endr{\textbackslash}Hoczi, Fafone, Farinon, Ferrante, Ferrini,
  Fidecaro, Fiori, Flaminio, Fournier, Franco, Frasca, Frasconi, Gammaitoni,
  Garufi, Gaspard, Gatto, Gemme, Gendre, Genin, Gennai, Ghosh, Giacobone,
  Giazotto, Gouaty, Granata, Greco, Groot, Guidi, Harms, Heidmann, Heitmann,
  Hello, Hemming, Hennes, Hofman, Jaranowski, Jonker, Kasprzack, Kéfélian,
  Kowalska, Kraan, Królak, Kutynia, Lazzaro, Leonardi, Leroy, Letendre, Li,
  Lieunard, Lorenzini, Loriette, Losurdo, Magazzù, Majorana, Maksimovic,
  Malvezzi, Man, Mangano, Mantovani, Marchesoni, Marion, Marque, Martelli,
  Martellini, Masserot, Meacher, Meidam, Mezzani, Michel, Milano, Minenkov,
  Moggi, Mohan, Montani, Morgado, Mours, Mul, Nagy, Nardecchia, Naticchioni,
  Nelemans, Neri, Neri, Nocera, Pacaud, Palomba, Paoletti, Paoli, Pasqualetti,
  Passaquieti, Passuello, Perciballi, Petit, Pichot, Piergiovanni, Pillant,
  Piluso, Pinard, Poggiani, Prijatelj, Prodi, Punturo, Puppo, Rabeling, Rácz,
  Rapagnani, Razzano, Re, Regimbau, Ricci, Robinet, Rocchi, Rolland, Romano,
  Rosińska, Ruggi, Saracco, Sassolas, Schimmel, Sentenac, Sequino, Shah,
  Siellez, Straniero, Swinkels, Tacca, Tonelli, Travasso, Turconi, Vajente,
  Bakel, Beuzekom, Brand, Broeck, Sluys, Heijningen, Vasúth, Vedovato, Veitch,
  Verkindt, Vetrano, Viceré, Vinet, Visser, Vocca, Ward, Was, Wei, Yvert,
  Zadro~{\textbackslash}.zny, \& Zendri}]{acernese_advanced_2014}
Acernese, F., Agathos, M., Agatsuma, K., {et~al.} 2014,
  \href{http://dx.doi.org/10.1088/0264-9381/32/2/024001}{\JournalTitle{Classical
  and Quantum Gravity}, 32, 024001}, publisher: IOP Publishing

\bibitem[{Adhikari {et~al.}(2020)Adhikari, Aguiar, Arai, Barr, Bassiri,
  Billingsley, Birney, Blair, Briggs, Brooks, Brown, Cao, Constancio, Cooper,
  Corbitt, Coyne, Daw, Eichholz, Fejer, Freise, Frolov, Gras, Green, Grote,
  Gustafson, Hall, Hammond, Harms, Harry, Haughian, Hellman, Hennig, Hennig,
  Hild, Johnson, Kamai, Kapasi, Komori, Korobko, Kuns, Lantz, Leavey,
  Magana-Sandoval, Markosyan, Martin, Martin, Martynov, Mcclelland, Mcghee,
  Mills, Mitrofanov, Molina-Ruiz, Mow-Lowry, Murray, Ng, Prokhorov, Quetschke,
  Reid, Reitze, Richardson, Robie, Romero-Shaw, Rowan, Schnabel, Schneewind,
  Shapiro, Shoemaker, Slagmolen, Smith, Steinlechner, Tait, Tanner, Torrie,
  Vanheijningen, Veitch, Wallace, Wessels, Willke, Wipf, Yamamoto, Zhao,
  Barsotti, Ward, Bell, Byer, Wade, Korth, Seifert, Smith, Koptsov, Tornasi,
  Markowitz, Mansell, Mcrae, Altin, Yap, Van~Veggel, Eddolls, Bonilla,
  Ferreira, Silva, Okada, Taira, Heinert, Hough, Strain, Cumming, Route,
  Shaddock, Evans, \& Weiss}]{adhikari_cryogenic_2020}
Adhikari, R.~X., Aguiar, O., Arai, K., {et~al.} 2020,
  \href{http://dx.doi.org/10.1088/1361-6382/ab9143}{\JournalTitle{Classical and
  Quantum Gravity}, 37, 165003}, arXiv: 2001.11173

\bibitem[{Aksulu {et~al.}(2020)Aksulu, Wijers, van Eerten, \&
  van der Horst}]{aksulu_new_2020}
Aksulu, M.~D., Wijers, R. A. M.~J., van Eerten, H.~J., \& van der Horst,
  A.~J. 2020,
  \href{http://dx.doi.org/10.1093/mnras/staa2297}{\JournalTitle{Monthly Notices
  of the Royal Astronomical Society}, 497, 4672}

\bibitem[{Aksulu {et~al.}(2022)Aksulu, Wijers, van Eerten, \&
  van der Horst}]{aksulu_exploring_2022}
---. 2022, \href{http://dx.doi.org/10.1093/mnras/stac246}{\JournalTitle{Monthly
  Notices of the Royal Astronomical Society}, 511, 2848}

\bibitem[{Alexander {et~al.}(2017)Alexander, Berger, Fong, Williams, Guidorzi,
  Margutti, Metzger, Annis, Blanchard, Brout, Brown, Chen, Chornock,
  Cowperthwaite, Drout, Eftekhari, Frieman, Holz, Nicholl, Rest, Sako,
  Soares-Santos, \& Villar}]{alexander_electromagnetic_2017}
Alexander, K.~D., Berger, E., Fong, W., {et~al.} 2017,
  \href{http://dx.doi.org/10.3847/2041-8213/aa905d}{\JournalTitle{The
  Astrophysical Journal}, 848, L21}, publisher: American Astronomical Society

\bibitem[{Anand {et~al.}(2021)Anand, Coughlin, Kasliwal, Bulla, Ahumada,
  Sagués~Carracedo, Almualla, Andreoni, Stein, Foucart, Singer, Sollerman,
  Bellm, Bolin, Caballero-García, Castro-Tirado, Cenko, De, Dekany, Duev,
  Feeney, Fremling, Goldstein, Golkhou, Graham, Guessoum, Hankins, Hu, Kong,
  Kool, Kulkarni, Kumar, Laher, Masci, Mróz, Nissanke, Porter, Reusch, Riddle,
  Rosnet, Rusholme, Serabyn, Sánchez-Ramírez, Rigault, Shupe, Smith,
  Soumagnac, Walters, \& Valeev}]{anand_optical_2021}
Anand, S., Coughlin, M.~W., Kasliwal, M.~M., {et~al.} 2021,
  \href{http://dx.doi.org/10.1038/s41550-020-1183-3}{\JournalTitle{Nature
  Astronomy}, 5, 46}

\bibitem[{Antier {et~al.}(2020)Antier, Agayeva, Almualla, Awiphan, Baransky,
  Barynova, Beradze, Blažek, Boër, Burkhonov, Christensen, Coleiro, Corre,
  Coughlin, Crisp, Dietrich, Ducoin, Duverne, Marchal-Duval, Gendre, Gokuldass,
  Eggenstein, Eymar, Hello, Howell, Ismailov, Kann, Karpov, Klotz,
  Kochiashvili, Lachaud, Leroy, Lin, Li, Mašek, Mo, Menard, Morris, Noysena,
  Orange, Prouza, Rattanamala, Sadibekova, Saint-Gelais, Serrau, Simon,
  Stachie, Thöne, Tillayev, Turpin, Postigo, Vasylenko, Vidadi, Was, Wang,
  Zhang, Zhang, \& Zhang}]{antier_grandma_2020}
Antier, S., Agayeva, S., Almualla, M., {et~al.} 2020,
  \href{http://dx.doi.org/10.1093/mnras/staa1846}{\JournalTitle{Monthly Notices
  of the Royal Astronomical Society}, 497, 5518}

\bibitem[{Ascenzi {et~al.}(2019)Ascenzi, Lillo, Haster, Ohme, \&
  Pannarale}]{ascenzi_constraining_2019}
Ascenzi, S., Lillo, N.~D., Haster, C.-J., Ohme, F., \& Pannarale, F. 2019,
  \href{http://dx.doi.org/10.3847/1538-4357/ab1b15}{\JournalTitle{The
  Astrophysical Journal}, 877, 94}

\bibitem[{Ashkar {et~al.}(2021)Ashkar, Brun, Füßling, Hoischen, Ohm, Prokoph,
  Reichherzer, Schüssler, \& Seglar-Arroyo}]{ashkar_hess_2021}
Ashkar, H., Brun, F., Füßling, M., {et~al.} 2021,
  \href{http://dx.doi.org/10.1088/1475-7516/2021/03/045}{\JournalTitle{Journal
  of Cosmology and Astroparticle Physics}, 2021, 045}, publisher: IOP
  Publishing

\bibitem[{Ashton {et~al.}(2019)Ashton, Hübner, Lasky, Talbot, Ackley,
  Biscoveanu, Chu, Divakarla, Easter, Goncharov, Vivanco, Harms, Lower,
  Meadors, Melchor, Payne, Pitkin, Powell, Sarin, Smith, \&
  Thrane}]{ashton_bilby_2019}
Ashton, G., Hübner, M., Lasky, P.~D., {et~al.} 2019,
  \href{http://dx.doi.org/10.3847/1538-4365/ab06fc}{\JournalTitle{The
  Astrophysical Journal Supplement Series}, 241, 27}, publisher: American
  Astronomical Society

\bibitem[{Babak {et~al.}(2017)Babak, Taracchini, \&
  Buonanno}]{babak_validating_2017}
Babak, S., Taracchini, A., \& Buonanno, A. 2017,
  \href{http://dx.doi.org/10.1103/PhysRevD.95.024010}{\JournalTitle{Physical
  Review D}, 95, 024010}, publisher: American Physical Society

\bibitem[{Barbieri {et~al.}(2019)Barbieri, Salafia, Perego, Colpi, \&
  Ghirlanda}]{barbieri_light-curve_2019}
Barbieri, C., Salafia, O.~S., Perego, A., Colpi, M., \& Ghirlanda, G. 2019,
  \href{http://dx.doi.org/10.1051/0004-6361/201935443}{\JournalTitle{Astronomy
  \& Astrophysics}, 625, A152}, arXiv: 1903.04543

\bibitem[{Barrett {et~al.}(2018)Barrett, Gaebel, Neijssel, Vigna-Gómez,
  Stevenson, Berry, Farr, \& Mandel}]{barrett_accuracy_2018}
Barrett, J.~W., Gaebel, S.~M., Neijssel, C.~J., {et~al.} 2018,
  \href{http://dx.doi.org/10.1093/mnras/sty908}{\JournalTitle{Monthly Notices
  of the Royal Astronomical Society}, 477, 4685}

\bibitem[{Beniamini {et~al.}(2015)Beniamini, Nava, Duran, \&
  Piran}]{beniamini_energies_2015}
Beniamini, P., Nava, L., Duran, R.~B., \& Piran, T. 2015,
  \href{http://dx.doi.org/10.1093/mnras/stv2033}{\JournalTitle{Monthly Notices
  of the Royal Astronomical Society}, 454, 1073}

\bibitem[{Beniamini {et~al.}(2016)Beniamini, Nava, \&
  Piran}]{beniamini_revised_2016}
Beniamini, P., Nava, L., \& Piran, T. 2016,
  \href{http://dx.doi.org/10.1093/mnras/stw1331}{\JournalTitle{Monthly Notices
  of the Royal Astronomical Society}, 461, 51}

\bibitem[{Beniamini \& van~der Horst(2017)}]{beniamini_electrons_2017}
Beniamini, P., \& van~der Horst, A.~J. 2017,
  \href{http://dx.doi.org/10.1093/mnras/stx2203}{\JournalTitle{Monthly Notices
  of the Royal Astronomical Society}, 472, 3161}

\bibitem[{Blandford \& Znajek(1977)}]{blandford_electromagnetic_1977}
Blandford, R.~D., \& Znajek, R.~L. 1977,
  \href{http://dx.doi.org/10.1093/mnras/179.3.433}{\JournalTitle{Monthly
  Notices of the Royal Astronomical Society}, 179, 433}

\bibitem[{Boersma {et~al.}(2021)Boersma, Leeuwen, Adams, Adebahr, Kutkin,
  Oosterloo, Blok, Brink, Coolen, Connor, Damstra, Dénes, Hess, Hulst, Hut,
  Ivashina, Loose, Lucero, Maan, Mika, Moss, Mulder, Oostrum, Ruiter, Schuur,
  Smits, Vermaas, Vohl, \& Ziemke}]{boersma_search_2021}
Boersma, O.~M., Leeuwen, J.~v., Adams, E. a.~K., {et~al.} 2021,
  \href{http://dx.doi.org/10.1051/0004-6361/202140578}{\JournalTitle{Astronomy
  \& Astrophysics}, 650, A131}, publisher: EDP Sciences

\bibitem[{Braun {et~al.}(2019)Braun, Bonaldi, Bourke, Keane, \&
  Wagg}]{braun_anticipated_2019}
Braun, R., Bonaldi, A., Bourke, T., Keane, E., \& Wagg, J. 2019,
  \href{http://arxiv.org/abs/1912.12699}{\JournalTitle{arXiv:1912.12699
  [astro-ph]}}, arXiv: 1912.12699

\bibitem[{Breschi {et~al.}(2021{\natexlab{a}})Breschi, Gamba, \&
  Bernuzzi}]{breschi_bayesian_2021}
Breschi, M., Gamba, R., \& Bernuzzi, S. 2021{\natexlab{a}},
  \href{http://dx.doi.org/10.1103/PhysRevD.104.042001}{\JournalTitle{Physical
  Review D}, 104, 042001}

\bibitem[{Breschi {et~al.}(2021{\natexlab{b}})Breschi, Perego, Bernuzzi,
  Del~Pozzo, Nedora, Radice, \& Vescovi}]{breschi_at2017gfo_2021}
Breschi, M., Perego, A., Bernuzzi, S., {et~al.} 2021{\natexlab{b}},
  \href{http://dx.doi.org/10.1093/mnras/stab1287}{\JournalTitle{Monthly Notices
  of the Royal Astronomical Society}}

\bibitem[{Broekgaarden {et~al.}(2021)Broekgaarden, Berger, Neijssel,
  Vigna-Gómez, Chattopadhyay, Stevenson, Chruslinska, Justham, de Mink, \&
  Mandel}]{broekgaarden_impact_2021}
Broekgaarden, F.~S., Berger, E., Neijssel, C.~J., {et~al.} 2021,
  \href{http://dx.doi.org/10.1093/mnras/stab2716}{\JournalTitle{Monthly Notices
  of the Royal Astronomical Society}, 508, 5028}

\bibitem[{Burgay {et~al.}(2003)Burgay, D'Amico, Possenti, Manchester, Lyne,
  Joshi, McLaughlin, Kramer, Sarkissian, Camilo, Kalogera, Kim, \&
  Lorimer}]{burgay_increased_2003}
Burgay, M., D'Amico, N., Possenti, A., {et~al.} 2003,
  \href{http://dx.doi.org/10.1038/nature02124}{\JournalTitle{Nature}, 426, 531}

\bibitem[{Capano {et~al.}(2020)Capano, Tews, Brown, Margalit, De, Kumar, Brown,
  Krishnan, \& Reddy}]{capano_stringent_2020}
Capano, C.~D., Tews, I., Brown, S.~M., {et~al.} 2020,
  \href{http://dx.doi.org/10.1038/s41550-020-1014-6}{\JournalTitle{Nature
  Astronomy}, 4, 625}, arXiv: 1908.10352

\bibitem[{Carson {et~al.}(2019)Carson, Steiner, \&
  Yagi}]{carson_constraining_2019}
Carson, Z., Steiner, A.~W., \& Yagi, K. 2019,
  \href{http://dx.doi.org/10.1103/PhysRevD.99.043010}{\JournalTitle{Physical
  Review D}, 99, 043010}, arXiv: 1812.08910

\bibitem[{Chornock {et~al.}(2017)Chornock, Berger, Kasen, Cowperthwaite,
  Nicholl, Villar, Alexander, Blanchard, Eftekhari, Fong, Margutti, Williams,
  Annis, Brout, Brown, Chen, Drout, Farr, Foley, Frieman, Fryer, Herner, Holz,
  Kessler, Matheson, Metzger, Quataert, Rest, Sako, Scolnic, Smith, \&
  Soares-Santos}]{chornock_electromagnetic_2017}
Chornock, R., Berger, E., Kasen, D., {et~al.} 2017,
  \href{http://dx.doi.org/10.3847/2041-8213/aa905c}{\JournalTitle{The
  Astrophysical Journal}, 848, L19}, publisher: American Astronomical Society

\bibitem[{Coughlin {et~al.}(2019)Coughlin, Dietrich, Margalit, \&
  Metzger}]{coughlin_multimessenger_2019}
Coughlin, M.~W., Dietrich, T., Margalit, B., \& Metzger, B.~D. 2019,
  \href{http://dx.doi.org/10.1093/mnrasl/slz133}{\JournalTitle{Monthly Notices
  of the Royal Astronomical Society: Letters}, 489, L91}

\bibitem[{Coughlin {et~al.}(2018)Coughlin, Dietrich, Doctor, Kasen, Coughlin,
  Jerkstrand, Leloudas, McBrien, Metzger, O’Shaughnessy, \&
  Smartt}]{coughlin_constraints_2018}
Coughlin, M.~W., Dietrich, T., Doctor, Z., {et~al.} 2018,
  \href{http://dx.doi.org/10.1093/mnras/sty2174}{\JournalTitle{Monthly Notices
  of the Royal Astronomical Society}, 480, 3871}

\bibitem[{Coulter {et~al.}(2017)Coulter, Foley, Kilpatrick, Drout, Piro,
  Shappee, Siebert, Simon, Ulloa, Kasen, Madore, Murguia-Berthier, Pan,
  Prochaska, Ramirez-Ruiz, Rest, \& Rojas-Bravo}]{coulter_swope_2017}
Coulter, D.~A., Foley, R.~J., Kilpatrick, C.~D., {et~al.} 2017,
  \href{http://dx.doi.org/10.1126/science.aap9811}{\JournalTitle{Science}, 358,
  1556}, publisher: American Association for the Advancement of Science
  Section: Research Article

\bibitem[{Cutler \& Flanagan(1994)}]{cutler_gravitational_1994}
Cutler, C., \& Flanagan, E.~E. 1994,
  \href{http://dx.doi.org/10.1103/PhysRevD.49.2658}{\JournalTitle{Physical
  Review D}, 49, 2658}

\bibitem[{Dietrich {et~al.}(2020)Dietrich, Coughlin, Pang, Bulla, Heinzel,
  Issa, Tews, \& Antier}]{dietrich_multimessenger_2020}
Dietrich, T., Coughlin, M.~W., Pang, P. T.~H., {et~al.} 2020,
  \href{http://dx.doi.org/10.1126/science.abb4317}{\JournalTitle{Science}, 370,
  1450}, publisher: American Association for the Advancement of Science

\bibitem[{Dobie {et~al.}(2021)Dobie, Murphy, Kaplan, Hotokezaka,
  Bonilla Ataides, Mahony, \& Sadler}]{dobie_radio_2021}
Dobie, D., Murphy, T., Kaplan, D.~L., {et~al.} 2021,
  \href{http://dx.doi.org/10.1093/mnras/stab1468}{\JournalTitle{Monthly Notices
  of the Royal Astronomical Society}, 505, 2647}

\bibitem[{Duque {et~al.}(2019)Duque, Daigne, \& Mochkovitch}]{duque_radio_2019}
Duque, R., Daigne, F., \& Mochkovitch, R. 2019,
  \href{http://dx.doi.org/10.1051/0004-6361/201935926}{\JournalTitle{Astronomy
  \& Astrophysics}, 631, A39}

\bibitem[{D’Emilio {et~al.}(2021)D’Emilio, Green, \&
  Raymond}]{demilio_density_2021}
D’Emilio, V., Green, R., \& Raymond, V. 2021,
  \href{http://dx.doi.org/10.1093/mnras/stab2623}{\JournalTitle{Monthly Notices
  of the Royal Astronomical Society}, 508, 2090}

\bibitem[{Eerten {et~al.}(2012)Eerten, Horst, \&
  MacFadyen}]{eerten_gamma-ray_2012}
Eerten, H.~v., Horst, A. v.~d., \& MacFadyen, A. 2012,
  \href{http://dx.doi.org/10.1088/0004-637X/749/1/44}{\JournalTitle{The
  Astrophysical Journal}, 749, 44}, publisher: American Astronomical Society

\bibitem[{Eichler {et~al.}(1989)Eichler, Livio, Piran, \&
  Schramm}]{eichler_nucleosynthesis_1989}
Eichler, D., Livio, M., Piran, T., \& Schramm, D.~N. 1989,
  \href{http://dx.doi.org/10.1038/340126a0}{\JournalTitle{Nature}, 340, 126}

\bibitem[{Fernández {et~al.}(2020)Fernández, Foucart, \&
  Lippuner}]{fernandez_landscape_2020}
Fernández, R., Foucart, F., \& Lippuner, J. 2020,
  \href{http://dx.doi.org/10.1093/mnras/staa2209}{\JournalTitle{Monthly Notices
  of the Royal Astronomical Society}, 497, 3221}, publisher: Oxford Academic

\bibitem[{Feroz {et~al.}(2009)Feroz, Hobson, \& Bridges}]{feroz_multinest_2009}
Feroz, F., Hobson, M.~P., \& Bridges, M. 2009,
  \href{http://dx.doi.org/10.1111/j.1365-2966.2009.14548.x}{\JournalTitle{Monthly
  Notices of the Royal Astronomical Society}, 398, 1601}

\bibitem[{Finn \& Chernoff(1993)}]{finn_observing_1993}
Finn, L.~S., \& Chernoff, D.~F. 1993,
  \href{http://dx.doi.org/10.1103/PhysRevD.47.2198}{\JournalTitle{Physical
  review D: Particles and fields}, 47, 2198}

\bibitem[{Foucart {et~al.}(2018)Foucart, Hinderer, \&
  Nissanke}]{foucart_remnant_2018}
Foucart, F., Hinderer, T., \& Nissanke, S. 2018,
  \href{http://dx.doi.org/10.1103/PhysRevD.98.081501}{\JournalTitle{Physical
  Review D}, 98, 081501}, arXiv: 1807.00011

\bibitem[{Fragione(2021)}]{fragione_black-holeneutron-star_2021}
Fragione, G. 2021,
  \href{http://dx.doi.org/10.3847/2041-8213/ac3bcd}{\JournalTitle{The
  Astrophysical Journal Letters}, 923, L2}, publisher: American Astronomical
  Society

\bibitem[{Fragione \& Loeb(2021)}]{fragione_constraining_2021}
Fragione, G., \& Loeb, A. 2021,
  \href{http://dx.doi.org/10.1093/mnras/stab666}{\JournalTitle{Monthly Notices
  of the Royal Astronomical Society}, 503, 2861}

\bibitem[{{GCN archive for S200105ae.}(2020)}]{httpsgcn1:online}
{GCN archive for S200105ae.} 2020,
  \url{https://gcn.gsfc.nasa.gov/other/S200105ae.gcn3}

\bibitem[{{GCN archive for S200115j.}(2020)}]{httpsgcn2:online}
{GCN archive for S200115j.} 2020,
  \url{https://gcn.gsfc.nasa.gov/other/S200115j.gcn3}

\bibitem[{Gompertz {et~al.}(2020)Gompertz, Levan, \&
  Tanvir}]{gompertz_search_2020}
Gompertz, B.~P., Levan, A.~J., \& Tanvir, N.~R. 2020,
  \href{http://dx.doi.org/10.3847/1538-4357/ab8d24}{\JournalTitle{The
  Astrophysical Journal}, 895, 58}, publisher: American Astronomical Society

\bibitem[{Haggard {et~al.}(2017)Haggard, Nynka, Ruan, Kalogera, Cenko, Evans,
  \& Kennea}]{haggard_deep_2017}
Haggard, D., Nynka, M., Ruan, J.~J., {et~al.} 2017,
  \href{http://dx.doi.org/10.3847/2041-8213/aa8ede}{\JournalTitle{The
  Astrophysical Journal}, 848, L25}

\bibitem[{Hallinan {et~al.}(2017)Hallinan, Corsi, Mooley, Hotokezaka, Nakar,
  Kasliwal, Kaplan, Frail, Myers, Murphy, De, Dobie, Allison, Bannister,
  Bhalerao, Chandra, Clarke, Giacintucci, Ho, Horesh, Kassim, Kulkarni, Lenc,
  Lockman, Lynch, Nichols, Nissanke, Palliyaguru, Peters, Piran, Rana, Sadler,
  \& Singer}]{hallinan_radio_2017}
Hallinan, G., Corsi, A., Mooley, K.~P., {et~al.} 2017,
  \href{http://dx.doi.org/10.1126/science.aap9855}{\JournalTitle{Science}, 358,
  1579}, publisher: American Association for the Advancement of Science
  Section: Research Article

\bibitem[{Hannam {et~al.}(2014)Hannam, Schmidt, Bohé, Haegel, Husa, Ohme,
  Pratten, \& Pürrer}]{hannam_simple_2014}
Hannam, M., Schmidt, P., Bohé, A., {et~al.} 2014,
  \href{http://dx.doi.org/10.1103/PhysRevLett.113.151101}{\JournalTitle{Physical
  Review Letters}, 113, 151101}, publisher: American Physical Society

\bibitem[{Hild {et~al.}(2011)Hild, Abernathy, Acernese, Amaro-Seoane,
  Andersson, Arun, Barone, Barr, Barsuglia, Beker, Beveridge, Birindelli, Bose,
  Bosi, Braccini, Bradaschia, Bulik, Calloni, Cella, Mottin, Chelkowski,
  Chincarini, Clark, Coccia, Colacino, Colas, Cumming, Cunningham, Cuoco,
  Danilishin, Danzmann, Salvo, Dent, Rosa, Fiore, Virgilio, Doets, Fafone,
  Falferi, Flaminio, Franc, Frasconi, Freise, Friedrich, Fulda, Gair, Gemme,
  Genin, Gennai, Giazotto, Glampedakis, Gräf, Granata, Grote, Guidi,
  Gurkovsky, Hammond, Hannam, Harms, Heinert, Hendry, Heng, Hennes, Hough,
  Husa, Huttner, Jones, Khalili, Kokeyama, Kokkotas, Krishnan, Li, Lorenzini,
  Lück, Majorana, Mandel, Mandic, Mantovani, Martin, Michel, Minenkov,
  Morgado, Mosca, Mours, Müller–Ebhardt, Murray, Nawrodt, Nelson,
  Oshaughnessy, Ott, Palomba, Paoli, Parguez, Pasqualetti, Passaquieti,
  Passuello, Pinard, Plastino, Poggiani, Popolizio, Prato, Punturo, Puppo,
  Rabeling, Rapagnani, Read, Regimbau, Rehbein, Reid, Ricci, Richard, Rocchi,
  Rowan, Rüdiger, Santamaría, Sassolas, Sathyaprakash, Schnabel, Schwarz,
  Seidel, Sintes, Somiya, Speirits, Strain, Strigin, Sutton, Tarabrin,
  Thüring, Brand, Veggel, Broeck, Vecchio, Veitch, Vetrano, Vicere,
  Vyatchanin, Willke, Woan, \& Yamamoto}]{hild_sensitivity_2011}
Hild, S., Abernathy, M., Acernese, F., {et~al.} 2011,
  \href{http://dx.doi.org/10.1088/0264-9381/28/9/094013}{\JournalTitle{Classical
  and Quantum Gravity}, 28, 094013}, publisher: IOP Publishing

\bibitem[{Hinderer {et~al.}(2019)Hinderer, Nissanke, Foucart, Hotokezaka,
  Vincent, Kasliwal, Schmidt, Williamson, Nichols, Duez, Kidder, Pfeiffer, \&
  Scheel}]{hinderer_distinguishing_2019}
Hinderer, T., Nissanke, S., Foucart, F., {et~al.} 2019,
  \href{http://dx.doi.org/10.1103/PhysRevD.100.063021}{\JournalTitle{Physical
  Review D}, 100, 063021}, publisher: American Physical Society

\bibitem[{Hinshaw {et~al.}(2013)Hinshaw, Larson, Komatsu, Spergel, Bennett,
  Dunkley, Nolta, Halpern, Hill, Odegard, Page, Smith, Weiland, Gold, Jarosik,
  Kogut, Limon, Meyer, Tucker, Wollack, \& Wright}]{hinshaw_nine-year_2013}
Hinshaw, G., Larson, D., Komatsu, E., {et~al.} 2013,
  \href{http://dx.doi.org/10.1088/0067-0049/208/2/19}{\JournalTitle{The
  Astrophysical Journal Supplement Series}, 208, 19}

\bibitem[{Hotokezaka {et~al.}(2019)Hotokezaka, Nakar, Gottlieb, Nissanke,
  Masuda, Hallinan, Mooley, \& Deller}]{hotokezaka_hubble_2019}
Hotokezaka, K., Nakar, E., Gottlieb, O., {et~al.} 2019,
  \href{http://dx.doi.org/10.1038/s41550-019-0820-1}{\JournalTitle{Nature
  Astronomy}, 3, 940}

\bibitem[{Hotokezaka {et~al.}(2016)Hotokezaka, Nissanke, Hallinan, Lazio,
  Nakar, \& Piran}]{hotokezaka_radio_2016}
Hotokezaka, K., Nissanke, S., Hallinan, G., {et~al.} 2016,
  \href{http://dx.doi.org/10.3847/0004-637X/831/2/190}{\JournalTitle{The
  Astrophysical Journal}, 831, 190}, publisher: American Astronomical Society

\bibitem[{Husa {et~al.}(2016)Husa, Khan, Hannam, Pürrer, Ohme, Forteza, \&
  Bohé}]{husa_frequency-domain_2016}
Husa, S., Khan, S., Hannam, M., {et~al.} 2016,
  \href{http://dx.doi.org/10.1103/PhysRevD.93.044006}{\JournalTitle{Physical
  Review D}, 93, 044006}, publisher: American Physical Society

\bibitem[{Jiménez-Forteza {et~al.}(2017)Jiménez-Forteza, Keitel, Husa,
  Hannam, Khan, \& Pürrer}]{jimenez-forteza_hierarchical_2017}
Jiménez-Forteza, X., Keitel, D., Husa, S., {et~al.} 2017,
  \href{http://dx.doi.org/10.1103/PhysRevD.95.064024}{\JournalTitle{Physical
  Review D}, 95, 064024}, publisher: American Physical Society

\bibitem[{Just {et~al.}(2016)Just, Obergaulinger, Janka, Bauswein, \&
  Schwarz}]{just_neutron-star_2016}
Just, O., Obergaulinger, M., Janka, H.-T., Bauswein, A., \& Schwarz, N. 2016,
  \href{http://dx.doi.org/10.3847/2041-8205/816/2/L30}{\JournalTitle{The
  Astrophysical Journal}, 816, L30}, publisher: American Astronomical Society

\bibitem[{Kasliwal {et~al.}(2020)Kasliwal, Anand, Ahumada, Stein, Carracedo,
  Andreoni, Coughlin, Singer, Kool, De, Kumar, AlMualla, Yao, Bulla, Dobie,
  Reusch, Perley, Cenko, Bhalerao, Kaplan, Sollerman, Goobar, Copperwheat,
  Bellm, Anupama, Corsi, Nissanke, Agudo, Bagdasaryan, Barway, Belicki, Bloom,
  Bolin, Buckley, Burdge, Burruss, Caballero-García, Cannella, Castro-Tirado,
  Cook, Cooke, Cunningham, Dahiwale, Deshmukh, Dichiara, Duev, Dutta, Feeney,
  Franckowiak, Frederick, Fremling, Gal-Yam, Gatkine, Ghosh, Goldstein,
  Golkhou, Graham, Graham, Hankins, Helou, Hu, Ip, Jaodand, Karambelkar, Kong,
  Kowalski, Khandagale, Kulkarni, Kumar, Laher, Li, Mahabal, Masci, Miller,
  Mogotsi, Mohite, Mooley, Mroz, Newman, Ngeow, Oates, Patil, Pandey, Pavana,
  Pian, Riddle, Sánchez-Ramírez, Sharma, Singh, Smith, Soumagnac, Taggart,
  Tan, Tzanidakis, Troja, Valeev, Walters, Waratkar, Webb, Yu, Zhang, Zhou, \&
  Zolkower}]{kasliwal_kilonova_2020}
Kasliwal, M.~M., Anand, S., Ahumada, T., {et~al.} 2020,
  \href{http://dx.doi.org/10.3847/1538-4357/abc335}{\JournalTitle{The
  Astrophysical Journal}, 905, 145}, publisher: American Astronomical Society

\bibitem[{Kelley(2021)}]{kelley_kalepy_2021}
Kelley, L.~Z. 2021,
  \href{http://dx.doi.org/10.21105/joss.02784}{\JournalTitle{Journal of Open
  Source Software}, 6, 2784}

\bibitem[{Khan {et~al.}(2016)Khan, Husa, Hannam, Ohme, Pürrer, Forteza, \&
  Bohé}]{khan_frequency-domain_2016}
Khan, S., Husa, S., Hannam, M., {et~al.} 2016,
  \href{http://dx.doi.org/10.1103/PhysRevD.93.044007}{\JournalTitle{Physical
  Review D}, 93, 044007}, publisher: American Physical Society

\bibitem[{Krüger \& Foucart(2020)}]{kruger_estimates_2020}
Krüger, C.~J., \& Foucart, F. 2020,
  \href{http://dx.doi.org/10.1103/PhysRevD.101.103002}{\JournalTitle{Physical
  Review D}, 101, 103002}, arXiv: 2002.07728

\bibitem[{Kyutoku {et~al.}(2013)Kyutoku, Ioka, \&
  Shibata}]{kyutoku_anisotropic_2013}
Kyutoku, K., Ioka, K., \& Shibata, M. 2013,
  \href{http://dx.doi.org/10.1103/PhysRevD.88.041503}{\JournalTitle{Physical
  Review D}, 88, 041503}

\bibitem[{Kyutoku {et~al.}(2021)Kyutoku, Shibata, \&
  Taniguchi}]{kyutoku_coalescence_2021}
Kyutoku, K., Shibata, M., \& Taniguchi, K. 2021,
  \href{http://dx.doi.org/10.1007/s41114-021-00033-4}{\JournalTitle{Living
  Reviews in Relativity}, 24, 5}

\bibitem[{Lattimer \& Prakash(2001)}]{lattimer_neutron_2001}
Lattimer, J.~M., \& Prakash, M. 2001,
  \href{http://dx.doi.org/10.1086/319702}{\JournalTitle{The Astrophysical
  Journal}, 550, 426}

\bibitem[{Lazzati {et~al.}(2021)Lazzati, Perna, Ciolfi, Giacomazzo,
  López-Cámara, \& Morsony}]{lazzati_two_2021}
Lazzati, D., Perna, R., Ciolfi, R., {et~al.} 2021,
  \href{http://dx.doi.org/10.3847/2041-8213/ac1794}{\JournalTitle{The
  Astrophysical Journal Letters}, 918, L6}, publisher: American Astronomical
  Society

\bibitem[{Liebling \& Palenzuela(2016)}]{liebling_electromagnetic_2016}
Liebling, S.~L., \& Palenzuela, C. 2016,
  \href{http://dx.doi.org/10.1103/PhysRevD.94.064046}{\JournalTitle{Physical
  Review D}, 94, 064046}, publisher: American Physical Society

\bibitem[{{LIGO Scientific Collaboration and Virgo Collaboration}
  {et~al.}(2017){LIGO Scientific Collaboration and Virgo Collaboration},
  Abbott, Abbott, Abbott, Acernese, Ackley, Adams, Adams, Addesso, Adhikari,
  Adya, Affeldt, Afrough, Agarwal, Agathos, Agatsuma, Aggarwal, Aguiar, Aiello,
  Ain, Ajith, Allen, Allen, Allocca, Altin, Amato, Ananyeva, Anderson,
  Anderson, Angelova, Antier, Appert, Arai, Araya, Areeda, Arnaud, Arun,
  Ascenzi, Ashton, Ast, Aston, Astone, Atallah, Aufmuth, Aulbert, AultONeal,
  Austin, Avila-Alvarez, Babak, Bacon, Bader, Bae, Bailes, Baker, Baldaccini,
  Ballardin, Ballmer, Banagiri, Barayoga, Barclay, Barish, Barker, Barkett,
  Barone, Barr, Barsotti, Barsuglia, Barta, Barthelmy, Bartlett, Bartos,
  Bassiri, Basti, Batch, Bawaj, Bayley, Bazzan, Bécsy, Beer, Bejger,
  Belahcene, Bell, Berger, Bergmann, Bernuzzi, Bero, Berry, Bersanetti,
  Bertolini, Betzwieser, Bhagwat, Bhandare, Bilenko, Billingsley, Billman,
  Birch, Birney, Birnholtz, Biscans, Biscoveanu, Bisht, Bitossi, Biwer,
  Bizouard, Blackburn, Blackman, Blair, Blair, Blair, Bloemen, Bock, Bode,
  Boer, Bogaert, Bohe, Bondu, Bonilla, Bonnand, Boom, Bork, Boschi, Bose,
  Bossie, Bouffanais, Bozzi, Bradaschia, Brady, Branchesi, Brau, Briant,
  Brillet, Brinkmann, Brisson, Brockill, Broida, Brooks, Brown, Brown, Brunett,
  Buchanan, Buikema, Bulik, Bulten, Buonanno, Buskulic, Buy, Byer, Cabero,
  Cadonati, Cagnoli, Cahillane, Calderón~Bustillo, Callister, Calloni, Camp,
  Canepa, Canizares, Cannon, Cao, Cao, Capano, Capocasa, Carbognani, Caride,
  Carney, Carullo, Casanueva~Diaz, Casentini, Caudill, Cavaglià, Cavalier,
  Cavalieri, Cella, Cepeda, Cerdá-Durán, Cerretani, Cesarini, Chamberlin,
  Chan, Chao, Charlton, Chase, Chassande-Mottin, Chatterjee, Chatziioannou,
  Cheeseboro, Chen, Chen, Chen, Cheng, Chia, Chincarini, Chiummo, Chmiel, Cho,
  Cho, Chow, Christensen, Chu, Chua, Chua, Chung, Chung, Ciani, Ciolfi,
  Cirelli, Cirone, Clara, Clark, Clearwater, Cleva, Cocchieri, Coccia, Cohadon,
  Cohen, Colla, Collette, Cominsky, Constancio, Conti, Cooper, Corban, Corbitt,
  Cordero-Carrión, Corley, Cornish, Corsi, Cortese, Costa, Coughlin, Coughlin,
  Coulon, Countryman, Couvares, Covas, Cowan, Coward, Cowart, Coyne, Coyne,
  Creighton, Creighton, Cripe, Crowder, Cullen, Cumming, Cunningham, Cuoco,
  Dal~Canton, Dálya, Danilishin, D’Antonio, Danzmann, Dasgupta,
  Da~Silva~Costa, Dattilo, Dave, Davier, Davis, Daw, Day, De, DeBra, Degallaix,
  De~Laurentis, Deléglise, Del~Pozzo, Demos, Denker, Dent, De~Pietri,
  Dergachev, De~Rosa, DeRosa, De~Rossi, DeSalvo, de~Varona, Devenson,
  Dhurandhar, Díaz, Dietrich, Di~Fiore, Di~Giovanni, Di~Girolamo, Di~Lieto,
  Di~Pace, Di~Palma, Di~Renzo, Doctor, Dolique, Donovan, Dooley, Doravari,
  Dorrington, Douglas, Dovale~Álvarez, Downes, Drago, Dreissigacker, Driggers,
  Du, Ducrot, Dudi, Dupej, Dwyer, Edo, Edwards, Effler, Eggenstein, Ehrens,
  Eichholz, Eikenberry, Eisenstein, Essick, Estevez, Etienne, Etzel, Evans,
  Evans, Factourovich, Fafone, Fair, Fairhurst, Fan, Farinon, Farr, Farr,
  Fauchon-Jones, Favata, Fays, Fee, Fehrmann, Feicht, Fejer, Fernandez-Galiana,
  Ferrante, Ferreira, Ferrini, Fidecaro, Finstad, Fiori, Fiorucci, Fishbach,
  Fisher, Fitz-Axen, Flaminio, Fletcher, Fong, Font, Forsyth, Forsyth,
  Fournier, Frasca, Frasconi, Frei, Freise, Frey, Frey, Fries, Fritschel,
  Frolov, Fulda, Fyffe, Gabbard, Gadre, Gaebel, Gair, Gammaitoni, Ganija,
  Gaonkar, Garcia-Quiros, Garufi, Gateley, Gaudio, Gaur, Gayathri, Gehrels,
  Gemme, Genin, Gennai, George, George, Gergely, Germain, Ghonge, Ghosh, Ghosh,
  Ghosh, Giaime, Giardina, Giazotto, Gill, Glover, Goetz, Goetz, Gomes,
  Goncharov, González, Gonzalez~Castro, Gopakumar, Gorodetsky, Gossan,
  Gosselin, Gouaty, Grado, Graef, Granata, Grant, Gras, Gray, Greco, Green,
  Gretarsson, Groot, Grote, Grunewald, Gruning, Guidi, Guo, Gupta, Gupta,
  Gushwa, Gustafson, Gustafson, Halim, Hall, Hall, Hamilton, Hammond, Haney,
  Hanke, Hanks, Hanna, Hannam, Hannuksela, Hanson, Hardwick, Harms, Harry,
  Harry, Hart, Haster, Haughian, Healy, Heidmann, Heintze, Heitmann, Hello,
  Hemming, Hendry, Heng, Hennig, Heptonstall, Heurs, Hild, Hinderer, Ho, Hoak,
  Hofman, Holt, Holz, Hopkins, Horst, Hough, Houston, Howell, Hreibi, Hu,
  Huerta, Huet, Hughey, Husa, Huttner, Huynh-Dinh, Indik, Inta, Intini, Isa,
  Isac, Isi, Iyer, Izumi, Jacqmin, Jani, Jaranowski, Jawahar, Jiménez-Forteza,
  Johnson, Johnson-McDaniel, Jones, Jones, Jonker, Ju, Junker, Kalaghatgi,
  Kalogera, Kamai, Kandhasamy, Kang, Kanner, Kapadia, Karki, Karvinen,
  Kasprzack, Kastaun, Katolik, Katsavounidis, Katzman, Kaufer, Kawabe,
  Kéfélian, Keitel, Kemball, Kennedy, Kent, Key, Khalili, Khan, Khan, Khan,
  Khazanov, Kijbunchoo, Kim, Kim, Kim, Kim, Kim, Kim, Kimbrell, King, King,
  Kinley-Hanlon, Kirchhoff, Kissel, Kleybolte, Klimenko, Knowles, Koch,
  Koehlenbeck, Koley, Kondrashov, Kontos, Korobko, Korth, Kowalska, Kozak,
  Krämer, Kringel, Krishnan, Królak, Kuehn, Kumar, Kumar, Kumar, Kuo,
  Kutynia, Kwang, Lackey, Lai, Landry, Lang, Lange, Lantz, Lanza, Larson,
  Lartaux-Vollard, Lasky, Laxen, Lazzarini, Lazzaro, Leaci, Leavey, Lee, Lee,
  Lee, Lee, Lee, Lehmann, Lenon, Leon, Leonardi, Leroy, Letendre, Levin, Li,
  Linker, Littenberg, Liu, Liu, Lo, Lockerbie, London, Lord, Lorenzini,
  Loriette, Lormand, Losurdo, Lough, Lousto, Lovelace, Lück, Lumaca, Lundgren,
  Lynch, Ma, Macas, Macfoy, Machenschalk, MacInnis, Macleod, Magaña~Hernandez,
  Magaña-Sandoval, Magaña~Zertuche, Magee, Majorana, Maksimovic, Man, Mandic,
  Mangano, Mansell, Manske, Mantovani, Marchesoni, Marion, Márka, Márka,
  Markakis, Markosyan, Markowitz, Maros, Marquina, Marsh, Martelli, Martellini,
  Martin, Martin, Martynov, Marx, Mason, Massera, Masserot, Massinger,
  Masso-Reid, Mastrogiovanni, Matas, Matichard, Matone, Mavalvala, Mazumder,
  McCarthy, McClelland, McCormick, McCuller, McGuire, McIntyre, McIver,
  McManus, McNeill, McRae, McWilliams, Meacher, Meadors, Mehmet, Meidam,
  Mejuto-Villa, Melatos, Mendell, Mercer, Merilh, Merzougui, Meshkov,
  Messenger, Messick, Metzdorff, Meyers, Miao, Michel, Middleton, Mikhailov,
  Milano, Miller, Miller, Miller, Millhouse, Milovich-Goff, Minazzoli,
  Minenkov, Ming, Mishra, Mitra, Mitrofanov, Mitselmakher, Mittleman, Moffa,
  Moggi, Mogushi, Mohan, Mohapatra, Molina, Montani, Moore, Moraru, Moreno,
  Morisaki, Morriss, Mours, Mow-Lowry, Mueller, Muir, Mukherjee, Mukherjee,
  Mukherjee, Mukund, Mullavey, Munch, Muñiz, Muratore, Murray, Nagar, Napier,
  Nardecchia, Naticchioni, Nayak, Neilson, Nelemans, Nelson, Nery, Neunzert,
  Nevin, Newport, Newton, Ng, Nguyen, Nguyen, Nichols, Nielsen, Nissanke, Nitz,
  Noack, Nocera, Nolting, North, Nuttall, Oberling, O’Dea, Ogin, Oh, Oh,
  Ohme, Okada, Oliver, Oppermann, Oram, O’Reilly, Ormiston, Ortega,
  O’Shaughnessy, Ossokine, Ottaway, Overmier, Owen, Pace, Page, Page, Pai,
  Pai, Palamos, Palashov, Palomba, Pal-Singh, Pan, Pan, Pang, Pang, Pankow,
  Pannarale, Pant, Paoletti, Paoli, Papa, Parida, Parker, Pascucci,
  Pasqualetti, Passaquieti, Passuello, Patil, Patricelli, Pearlstone, Pedraza,
  Pedurand, Pekowsky, Pele, Penn, Perez, Perreca, Perri, Pfeiffer, Phelps,
  Piccinni, Pichot, Piergiovanni, Pierro, Pillant, Pinard, Pinto, Pirello,
  Pitkin, Poe, Poggiani, Popolizio, Porter, Post, Powell, Prasad, Pratt,
  Pratten, Predoi, Prestegard, Prijatelj, Principe, Privitera, Prix, Prodi,
  Prokhorov, Puncken, Punturo, Puppo, Pürrer, Qi, Quetschke, Quintero,
  Quitzow-James, Raab, Rabeling, Radkins, Raffai, Raja, Rajan, Rajbhandari,
  Rakhmanov, Ramirez, Ramos-Buades, Rapagnani, Raymond, Razzano, Read,
  Regimbau, Rei, Reid, Reitze, Ren, Reyes, Ricci, Ricker, Rieger, Riles, Rizzo,
  Robertson, Robie, Robinet, Rocchi, Rolland, Rollins, Roma, Romano, Romano,
  Romel, Romie, Rosińska, Ross, Rowan, Rüdiger, Ruggi, Rutins, Ryan, Sachdev,
  Sadecki, Sadeghian, Sakellariadou, Salconi, Saleem, Salemi, Samajdar, Sammut,
  Sampson, Sanchez, Sanchez, Sanchis-Gual, Sandberg, Sanders, Sassolas,
  Sathyaprakash, Saulson, Sauter, Savage, Sawadsky, Schale, Scheel, Scheuer,
  Schmidt, Schmidt, Schnabel, Schofield, Schönbeck, Schreiber, Schuette,
  Schulte, Schutz, Schwalbe, Scott, Scott, Seidel, Sellers, Sengupta, Sentenac,
  Sequino, Sergeev, Shaddock, Shaffer, Shah, Shahriar, Shaner, Shao, Shapiro,
  Shawhan, Sheperd, Shoemaker, Shoemaker, Siellez, Siemens, Sieniawska, Sigg,
  Silva, Singer, Singh, Singhal, Sintes, Slagmolen, Smith, Smith, Smith,
  Somala, Son, Sonnenberg, Sorazu, Sorrentino, Souradeep, Spencer, Srivastava,
  Staats, Staley, Steinke, Steinlechner, Steinlechner, Steinmeyer, Stevenson,
  Stone, Stops, Strain, Stratta, Strigin, Strunk, Sturani, Stuver,
  Summerscales, Sun, Sunil, Suresh, Sutton, Swinkels, Szczepańczyk, Tacca,
  Tait, Talbot, Talukder, Tanner, Tápai, Taracchini, Tasson, Taylor, Taylor,
  Tewari, Theeg, Thies, Thomas, Thomas, Thomas, Thorne, Thorne, Thrane, Tiwari,
  Tiwari, Tokmakov, Toland, Tonelli, Tornasi, Torres-Forné, Torrie, Töyrä,
  Travasso, Traylor, Trinastic, Tringali, Trozzo, Tsang, Tse, Tso, Tsukada,
  Tsuna, Tuyenbayev, Ueno, Ugolini, Unnikrishnan, Urban, Usman, Vahlbruch,
  Vajente, Valdes, Vallisneri, van Bakel, van Beuzekom, van~den Brand, Van
  Den~Broeck, Vander-Hyde, van~der Schaaf, van Heijningen, van Veggel, Vardaro,
  Varma, Vass, Vasúth, Vecchio, Vedovato, Veitch, Veitch, Venkateswara,
  Venugopalan, Verkindt, Vetrano, Viceré, Viets, Vinciguerra, Vine, Vinet,
  Vitale, Vo, Vocca, Vorvick, Vyatchanin, Wade, Wade, Wade, Walet, Walker,
  Wallace, Walsh, Wang, Wang, Wang, Wang, Wang, Ward, Warner, Was, Watchi,
  Weaver, Wei, Weinert, Weinstein, Weiss, Wen, Wessel, Weßels, Westerweck,
  Westphal, Wette, Whelan, Whitcomb, Whiting, Whittle, Wilken, Williams,
  Williams, Williamson, Willis, Willke, Wimmer, Winkler, Wipf, Wittel, Woan,
  Woehler, Wofford, Wong, Worden, Wright, Wu, Wysocki, Xiao, Yamamoto, Yancey,
  Yang, Yap, Yazback, Yu, Yu, Yvert, Zadrożny, Zanolin, Zelenova, Zendri,
  Zevin, Zhang, Zhang, Zhang, Zhang, Zhao, Zhou, Zhou, Zhu, Zhu, Zimmerman,
  Zucker, \&
  Zweizig}]{ligo_scientific_collaboration_and_virgo_collaboration_gw170817_2017}
{LIGO Scientific Collaboration and Virgo Collaboration}, Abbott, B., Abbott,
  R., {et~al.} 2017,
  \href{http://dx.doi.org/10.1103/PhysRevLett.119.161101}{\JournalTitle{Physical
  Review Letters}, 119, 161101}, publisher: American Physical Society

\bibitem[{Maggiore {et~al.}(2020)Maggiore, Broeck, Bartolo, Belgacem, Bertacca,
  Bizouard, Branchesi, Clesse, Foffa, García-Bellido, Grimm, Harms, Hinderer,
  Matarrese, Palomba, Peloso, Ricciardone, \&
  Sakellariadou}]{maggiore_science_2020}
Maggiore, M., Broeck, C. v.~d., Bartolo, N., {et~al.} 2020,
  \href{http://dx.doi.org/10.1088/1475-7516/2020/03/050}{\JournalTitle{Journal
  of Cosmology and Astroparticle Physics}, 2020, 050}, arXiv: 1912.02622

\bibitem[{Matas {et~al.}(2020)Matas, Dietrich, Buonanno, Hinderer, Pürrer,
  Foucart, Boyle, Duez, Kidder, Pfeiffer, \& Scheel}]{matas_aligned-spin_2020}
Matas, A., Dietrich, T., Buonanno, A., {et~al.} 2020,
  \href{http://dx.doi.org/10.1103/PhysRevD.102.043023}{\JournalTitle{Physical
  Review D}, 102, 043023}

\bibitem[{McKinney(2005)}]{mckinney_total_2005}
McKinney, J.~C. 2005, \href{http://dx.doi.org/10.1086/468184}{\JournalTitle{The
  Astrophysical Journal}, 630, L5}

\bibitem[{McKinney \& Gammie(2004)}]{mckinney_measurement_2004}
McKinney, J.~C., \& Gammie, C.~F. 2004,
  \href{http://dx.doi.org/10.1086/422244}{\JournalTitle{The Astrophysical
  Journal}, 611, 977}, publisher: IOP Publishing

\bibitem[{Metzger \& Berger(2012)}]{metzger_what_2012}
Metzger, B.~D., \& Berger, E. 2012,
  \href{http://dx.doi.org/10.1088/0004-637X/746/1/48}{\JournalTitle{The
  Astrophysical Journal}, 746, 48}, publisher: American Astronomical Society

\bibitem[{Miller {et~al.}(2019)Miller, Lamb, Dittmann, Bogdanov, Arzoumanian,
  Gendreau, Guillot, Harding, Ho, Lattimer, Ludlam, Mahmoodifar, Morsink, Ray,
  Strohmayer, Wood, Enoto, Foster, Okajima, Prigozhin, \&
  Soong}]{miller_psr_2019}
Miller, M.~C., Lamb, F.~K., Dittmann, A.~J., {et~al.} 2019,
  \href{http://dx.doi.org/10.3847/2041-8213/ab50c5}{\JournalTitle{The
  Astrophysical Journal}, 887, L24}, publisher: American Astronomical Society

\bibitem[{Mooley {et~al.}(2013)Mooley, Frail, Ofek, Miller, Kulkarni, \&
  Horesh}]{mooley_sensitive_2013}
Mooley, K.~P., Frail, D.~A., Ofek, E.~O., {et~al.} 2013,
  \href{http://dx.doi.org/10.1088/0004-637X/768/2/165}{\JournalTitle{The
  Astrophysical Journal}, 768, 165}, publisher: American Astronomical Society

\bibitem[{Mooley {et~al.}(2018)Mooley, Deller, Gottlieb, Nakar, Hallinan,
  Bourke, Frail, Horesh, Corsi, \& Hotokezaka}]{mooley_superluminal_2018}
Mooley, K.~P., Deller, A.~T., Gottlieb, O., {et~al.} 2018,
  \href{http://dx.doi.org/10.1038/s41586-018-0486-3}{\JournalTitle{Nature},
  561, 355}, number: 7723 Publisher: Nature Publishing Group

\bibitem[{Mészáros \& Rees(1992)}]{meszaros_high-entropy_1992}
Mészáros, P., \& Rees, M.~J. 1992,
  \href{http://dx.doi.org/10.1093/mnras/257.1.29P}{\JournalTitle{Monthly
  Notices of the Royal Astronomical Society}, 257, 29P}

\bibitem[{Nagakura {et~al.}(2014)Nagakura, Hotokezaka, Sekiguchi, Shibata, \&
  Ioka}]{nagakura_jet_2014}
Nagakura, H., Hotokezaka, K., Sekiguchi, Y., Shibata, M., \& Ioka, K. 2014,
  \href{http://dx.doi.org/10.1088/2041-8205/784/2/L28}{\JournalTitle{The
  Astrophysical Journal}, 784, L28}, publisher: American Astronomical Society

\bibitem[{Nakar \& Piran(2011)}]{nakar_detectable_2011}
Nakar, E., \& Piran, T. 2011,
  \href{http://dx.doi.org/10.1038/nature10365}{\JournalTitle{Nature}, 478, 82},
  number: 7367 Publisher: Nature Publishing Group

\bibitem[{Nakar \& Piran(2021)}]{nakar_afterglow_2021}
---. 2021, \href{http://dx.doi.org/10.3847/1538-4357/abd6cd}{\JournalTitle{The
  Astrophysical Journal}, 909, 114}

\bibitem[{Nakar {et~al.}(2002)Nakar, Piran, \&
  Granot}]{nakar_detectability_2002}
Nakar, E., Piran, T., \& Granot, J. 2002,
  \href{http://dx.doi.org/10.1086/342791}{\JournalTitle{The Astrophysical
  Journal}, 579, 699}

\bibitem[{Neijssel {et~al.}(2019)Neijssel, Vigna-Gómez, Stevenson, Barrett,
  Gaebel, Broekgaarden, de Mink, Szécsi, Vinciguerra, \&
  Mandel}]{neijssel_effect_2019}
Neijssel, C.~J., Vigna-Gómez, A., Stevenson, S., {et~al.} 2019,
  \href{http://dx.doi.org/10.1093/mnras/stz2840}{\JournalTitle{Monthly Notices
  of the Royal Astronomical Society}, 490, 3740}

\bibitem[{Nicholl {et~al.}(2021)Nicholl, Margalit, Schmidt, Smith, Ridley, \&
  Nuttall}]{nicholl_tight_2021}
Nicholl, M., Margalit, B., Schmidt, P., {et~al.} 2021,
  \href{http://dx.doi.org/10.1093/mnras/stab1523}{\JournalTitle{Monthly Notices
  of the Royal Astronomical Society}, 505, 3016}

\bibitem[{Nissanke {et~al.}(2010)Nissanke, Holz, Hughes, Dalal, \&
  Sievers}]{nissanke_exploring_2010}
Nissanke, S., Holz, D.~E., Hughes, S.~A., Dalal, N., \& Sievers, J.~L. 2010,
  \href{http://dx.doi.org/10.1088/0004-637X/725/1/496}{\JournalTitle{The
  Astrophysical Journal}, 725, 496}, publisher: American Astronomical Society

\bibitem[{Pan {et~al.}(2014)Pan, Buonanno, Taracchini, Kidder, Mroué,
  Pfeiffer, Scheel, \& Szilágyi}]{pan_inspiral-merger-ringdown_2014}
Pan, Y., Buonanno, A., Taracchini, A., {et~al.} 2014,
  \href{http://dx.doi.org/10.1103/PhysRevD.89.084006}{\JournalTitle{Physical
  Review D}, 89, 084006}, publisher: American Physical Society

\bibitem[{Pannarale \& Ohme(2014)}]{pannarale_prospects_2014}
Pannarale, F., \& Ohme, F. 2014,
  \href{http://dx.doi.org/10.1088/2041-8205/791/1/L7}{\JournalTitle{The
  Astrophysical Journal}, 791, L7}, publisher: American Astronomical Society

\bibitem[{Paschalidis {et~al.}(2015)Paschalidis, Ruiz, \&
  Shapiro}]{paschalidis_relativistic_2015}
Paschalidis, V., Ruiz, M., \& Shapiro, S.~L. 2015,
  \href{http://dx.doi.org/10.1088/2041-8205/806/1/L14}{\JournalTitle{The
  Astrophysical Journal}, 806, L14}, publisher: American Astronomical Society

\bibitem[{Paterson {et~al.}(2021)Paterson, Lundquist, Rastinejad, Fong, Sand,
  Andrews, Amaro, Eskandari, Wyatt, Daly, Bradley, Zhou-Wright, Valenti, Yang,
  Christensen, Gibbs, Shelly, Bilinski, Chomiuk, Corsi, Drout, Foley, Gabor,
  Garnavich, Grier, Hamden, Krantz, Olszewski, Paschalidis, Reichart, Rest,
  Smith, Strader, Trilling, Veillet, Wagner, Weiner, \&
  Zabludoff}]{paterson_searches_2021}
Paterson, K., Lundquist, M.~J., Rastinejad, J.~C., {et~al.} 2021,
  \href{http://dx.doi.org/10.3847/1538-4357/abeb71}{\JournalTitle{The
  Astrophysical Journal}, 912, 128}, publisher: American Astronomical Society

\bibitem[{Raaijmakers {et~al.}(2020)Raaijmakers, Greif, Riley, Hinderer,
  Hebeler, Schwenk, Watts, Nissanke, Guillot, Lattimer, \&
  Ludlam}]{raaijmakers_constraining_2020}
Raaijmakers, G., Greif, S.~K., Riley, T.~E., {et~al.} 2020,
  \href{http://dx.doi.org/10.3847/2041-8213/ab822f}{\JournalTitle{The
  Astrophysical Journal}, 893, L21}, publisher: American Astronomical Society

\bibitem[{Raaijmakers {et~al.}(2021{\natexlab{a}})Raaijmakers, Nissanke,
  Foucart, Kasliwal, Bulla, Fernández, Henkel, Hinderer, Hotokezaka,
  Lukošiūt{\textbackslash}.e, Venumadhav, Antier, Coughlin, Dietrich, \&
  Edwards}]{raaijmakers_challenges_2021}
Raaijmakers, G., Nissanke, S., Foucart, F., {et~al.} 2021{\natexlab{a}},
  \href{http://dx.doi.org/10.3847/1538-4357/ac222d}{\JournalTitle{The
  Astrophysical Journal}, 922, 269}, publisher: American Astronomical Society

\bibitem[{Raaijmakers {et~al.}(2021{\natexlab{b}})Raaijmakers, Greif, Hebeler,
  Hinderer, Nissanke, Schwenk, Riley, Watts, Lattimer, \&
  Ho}]{raaijmakers_constraints_2021}
Raaijmakers, G., Greif, S.~K., Hebeler, K., {et~al.} 2021{\natexlab{b}},
  \href{http://dx.doi.org/10.3847/2041-8213/ac089a}{\JournalTitle{The
  Astrophysical Journal Letters}, 918, L29}, publisher: American Astronomical
  Society

\bibitem[{Radice \& Dai(2019)}]{radice_multimessenger_2019}
Radice, D., \& Dai, L. 2019,
  \href{http://dx.doi.org/10.1140/epja/i2019-12716-4}{\JournalTitle{The
  European Physical Journal A}, 55, 50}

\bibitem[{Radice {et~al.}(2018)Radice, Perego, Zappa, \&
  Bernuzzi}]{radice_gw170817_2018}
Radice, D., Perego, A., Zappa, F., \& Bernuzzi, S. 2018,
  \href{http://dx.doi.org/10.3847/2041-8213/aaa402}{\JournalTitle{The
  Astrophysical Journal}, 852, L29}, publisher: American Astronomical Society

\bibitem[{Ridnaia {et~al.}(2020)Ridnaia, Svinkin, \&
  Frederiks}]{ridnaia_search_2020}
Ridnaia, A., Svinkin, D., \& Frederiks, D. 2020,
  \href{http://dx.doi.org/10.1088/1742-6596/1697/1/012030}{\JournalTitle{Journal
  of Physics: Conference Series}, 1697, 012030}, publisher: IOP Publishing

\bibitem[{Riley {et~al.}(2021)Riley, Watts, Ray, Bogdanov, Guillot, Morsink,
  Bilous, Arzoumanian, Choudhury, Deneva, Gendreau, Harding, Ho, Lattimer,
  Loewenstein, Ludlam, Markwardt, Okajima, Prescod-Weinstein, Remillard, Wolff,
  Fonseca, Cromartie, Kerr, Pennucci, Parthasarathy, Ransom, Stairs, Guillemot,
  \& Cognard}]{riley_nicer_2021}
Riley, T.~E., Watts, A.~L., Ray, P.~S., {et~al.} 2021,
  \href{http://dx.doi.org/10.3847/2041-8213/ac0a81}{\JournalTitle{The
  Astrophysical Journal Letters}, 918, L27}, publisher: American Astronomical
  Society

\bibitem[{Romero-Shaw {et~al.}(2020)Romero-Shaw, Talbot, Biscoveanu,
  D’Emilio, Ashton, Berry, Coughlin, Galaudage, Hoy, Hübner, Phukon, Pitkin,
  Rizzo, Sarin, Smith, Stevenson, Vajpeyi, Arène, Athar, Banagiri, Bose,
  Carney, Chatziioannou, Clark, Colleoni, Cotesta, Edelman, Estellés,
  García-Quirós, Ghosh, Green, Haster, Husa, Keitel, Kim, Hernandez-Vivanco,
  Magaña Hernandez, Karathanasis, Lasky, De Lillo, Lower, Macleod,
  Mateu-Lucena, Miller, Millhouse, Morisaki, Oh, Ossokine, Payne, Powell,
  Pratten, Pürrer, Ramos-Buades, Raymond, Thrane, Veitch, Williams, Williams,
  \& Xiao}]{romero-shaw_bayesian_2020}
Romero-Shaw, I.~M., Talbot, C., Biscoveanu, S., {et~al.} 2020,
  \href{http://dx.doi.org/10.1093/mnras/staa2850}{\JournalTitle{Monthly Notices
  of the Royal Astronomical Society}, 499, 3295}

\bibitem[{Ryan {et~al.}(2020)Ryan, van Eerten, Piro, \&
  Troja}]{ryan_gamma-ray_2020}
Ryan, G., van Eerten, H., Piro, L., \& Troja, E. 2020,
  \href{http://dx.doi.org/10.3847/1538-4357/ab93cf}{\JournalTitle{The
  Astrophysical Journal}, 896, 166}, arXiv: 1909.11691

\bibitem[{Salafia \& Giacomazzo(2021)}]{salafia_accretion--jet_2021}
Salafia, O.~S., \& Giacomazzo, B. 2021,
  \href{http://dx.doi.org/10.1051/0004-6361/202038590}{\JournalTitle{Astronomy
  \& Astrophysics}, 645, A93}, arXiv: 2006.07376

\bibitem[{Salafia \& Giacomazzo(2022)}]{salafia_accretion--jet_2022}
---. 2022,
  \href{http://dx.doi.org/10.1051/0004-6361/202038590e}{\JournalTitle{Astronomy
  \& Astrophysics}, 660, C1}, publisher: EDP Sciences

\bibitem[{Somiya(2012)}]{somiya_detector_2012}
Somiya, K. 2012,
  \href{http://dx.doi.org/10.1088/0264-9381/29/12/124007}{\JournalTitle{Classical
  and Quantum Gravity}, 29, 124007}, publisher: IOP Publishing

\bibitem[{Stone {et~al.}(2013)Stone, Loeb, \& Berger}]{stone_pulsations_2013}
Stone, N., Loeb, A., \& Berger, E. 2013,
  \href{http://dx.doi.org/10.1103/PhysRevD.87.084053}{\JournalTitle{Physical
  Review D}, 87, 084053}, publisher: American Physical Society

\bibitem[{{The KAGRA Collaboration} {et~al.}(2013){The KAGRA Collaboration},
  Aso, Michimura, Somiya, Ando, Miyakawa, Sekiguchi, Tatsumi, \&
  Yamamoto}]{the_kagra_collaboration_interferometer_2013}
{The KAGRA Collaboration}, Aso, Y., Michimura, Y., {et~al.} 2013,
  \href{http://dx.doi.org/10.1103/PhysRevD.88.043007}{\JournalTitle{Physical
  Review D}, 88, 043007}, publisher: American Physical Society

\bibitem[{{The LIGO Scientific Collaboration and The Virgo Collaboration}
  {et~al.}(2017){The LIGO Scientific Collaboration and The Virgo
  Collaboration}, Abbott, Abbott, Abbott, Acernese, Ackley, Adams, Adams,
  Addesso, Adhikari, Adya, Aggarwal, Aguiar, Aiello, Ain, Dreissigacker,
  Driggers, Du, Ducrot, Dupej, \&
  Dwyer}]{the_ligo_scientific_collaboration_and_the_virgo_collaboration_gravitational-wave_2017}
{The LIGO Scientific Collaboration and The Virgo Collaboration}, Abbott, B.~P.,
  Abbott, R., {et~al.} 2017,
  \href{http://dx.doi.org/10.1038/nature24471}{\JournalTitle{Nature}, 551, 85}

\bibitem[{{The LIGO Scientific Collaboration and The Virgo Collaboration}
  {et~al.}(2018){The LIGO Scientific Collaboration and The Virgo
  Collaboration}, Abbott, Abbott, Abbott, Acernese, Ackley, Adams, Adams,
  Addesso, Adhikari, Adya, Affeldt, Agarwal, Agathos, Agatsuma, Aggarwal,
  Aguiar, Aiello, Ain, Ajith, Allen, Allen, Allocca, Aloy, Altin, Amato,
  Ananyeva, Anderson, Anderson, Angelova, Antier, Appert, Arai, Araya, Areeda,
  Arène, Arnaud, Arun, Ascenzi, Ashton, Ast, Aston, Astone, Atallah, Aubin,
  Aufmuth, Aulbert, AultONeal, Austin, Avila-Alvarez, Babak, Bacon, Badaracco,
  Bader, Bae, Baker, Baldaccini, Ballardin, Ballmer, Banagiri, Barayoga,
  Barclay, Barish, Barker, Barkett, Barnum, Barone, Barr, Barsotti, Barsuglia,
  Barta, Bartlett, Bartos, Bassiri, Basti, Batch, Bawaj, Bayley, Bazzan,
  Bécsy, Beer, Bejger, Belahcene, Bell, Beniwal, Bensch, Berger, Bergmann,
  Bernuzzi, Bero, Berry, Bersanetti, Bertolini, Betzwieser, Bhandare, Bilenko,
  Bilgili, Billingsley, Billman, Birch, Birney, Birnholtz, Biscans, Biscoveanu,
  Bisht, Bitossi, Bizouard, Blackburn, Blackman, Blair, Blair, Blair, Bloemen,
  Bock, Bode, Boer, Boetzel, Bogaert, Bohe, Bondu, Bonilla, Bonnand, Booker,
  Boom, Booth, Bork, Boschi, Bose, Bossie, Bossilkov, Bosveld, Bouffanais,
  Bozzi, Bradaschia, Brady, Bramley, Branchesi, Brau, Briant, Brighenti,
  Brillet, Brinkmann, Brisson, Brockill, Brooks, Brown, Brunett, Buchanan,
  Buikema, Bulik, Bulten, Buonanno, Buskulic, Buy, Byer, Cabero, Cadonati,
  Cagnoli, Cahillane, Calderón~Bustillo, Callister, Calloni, Camp, Canepa,
  Canizares, Cannon, Cao, Cao, Capano, Capocasa, Carbognani, Caride, Carney,
  Carullo, Casanueva~Diaz, Casentini, Caudill, Cavaglià, Cavalier, Cavalieri,
  Cella, Cepeda, Cerdá-Durán, Cerretani, Cesarini, Chaibi, Chamberlin, Chan,
  Chao, Charlton, Chase, Chassande-Mottin, Chatterjee, Chatziioannou,
  Cheeseboro, Chen, Chen, Chen, Cheng, Chia, Chincarini, Chiummo, Chmiel, Cho,
  Cho, Chow, Christensen, Chu, Chua, Chua, Chung, Chung, Ciani, Ciobanu,
  Ciolfi, Cipriano, Cirelli, Cirone, Clara, Clark, Clearwater, Cleva,
  Cocchieri, Coccia, Cohadon, Cohen, Colla, Collette, Collins, Cominsky,
  Constancio, Conti, Cooper, Corban, Corbitt, Cordero-Carrión, Corley,
  Cornish, Corsi, Cortese, Costa, Cotesta, Coughlin, Coughlin, Coulon,
  Countryman, Couvares, Covas, Cowan, Coward, Cowart, Coyne, Coyne, Creighton,
  Creighton, Cripe, Crowder, Cullen, Cumming, Cunningham, Cuoco, Canton,
  Dálya, Danilishin, D’Antonio, Danzmann, Dasgupta, Da~Silva~Costa, Dattilo,
  Dave, Davier, Davis, Daw, Day, DeBra, Deenadayalan, Degallaix, De~Laurentis,
  Deléglise, Del~Pozzo, Demos, Denker, Dent, De~Pietri, Derby, Dergachev,
  De~Rosa, De~Rossi, DeSalvo, de~Varona, Dhurandhar, Díaz, Dietrich, Di~Fiore,
  Di~Giovanni, Di~Girolamo, Di~Lieto, Ding, Di~Pace, Di~Palma, Di~Renzo,
  Dmitriev, Doctor, Dolique, Donovan, Dooley, Doravari, Dorrington,
  Dovale~Álvarez, Downes, Drago, Dreissigacker, Driggers, Du, Dupej, Dwyer,
  Easter, Edo, Edwards, Effler, Eggenstein, Ehrens, Eichholz, Eikenberry,
  Eisenmann, Eisenstein, Essick, Estelles, Estevez, Etienne, Etzel, Evans,
  Evans, Fafone, Fair, Fairhurst, Fan, Farinon, Farr, Farr, Fauchon-Jones,
  Favata, Fays, Fee, Fehrmann, Feicht, Fejer, Feng, Fernandez-Galiana,
  Ferrante, Ferreira, Ferrini, Fidecaro, Fiori, Fiorucci, Fishbach, Fisher,
  Fishner, Fitz-Axen, Flaminio, Fletcher, Fong, Font, Forsyth, Forsyth,
  Fournier, Frasca, Frasconi, Frei, Freise, Frey, Frey, Fritschel, Frolov,
  Fulda, Fyffe, Gabbard, Gadre, Gaebel, Gair, Gammaitoni, Ganija, Gaonkar,
  Garcia, García-Quirós, Garufi, Gateley, Gaudio, Gaur, Gayathri, Gemme,
  Genin, Gennai, George, George, Gergely, Germain, Ghonge, Ghosh, Ghosh, Ghosh,
  Giacomazzo, Giaime, Giardina, Giazotto, Gill, Giordano, Glover, Goetz, Goetz,
  Goncharov, González, Gonzalez~Castro, Gopakumar, Gorodetsky, Gossan,
  Gosselin, Gouaty, Grado, Graef, Granata, Grant, Gras, Gray, Greco, Green,
  Green, Gretarsson, Groot, Grote, Grunewald, Gruning, Guidi, Gulati, Guo,
  Gupta, Gupta, Gushwa, Gustafson, Gustafson, Halim, Hall, Hall, Hamilton,
  Hamilton, Hammond, Haney, Hanke, Hanks, Hanna, Hannam, Hannuksela, Hanson,
  Hardwick, Harms, Harry, Harry, Hart, Haster, Haughian, Healy, Heidmann,
  Heintze, Heitmann, Hello, Hemming, Hendry, Heng, Hennig, Heptonstall,
  Hernandez, Heurs, Hild, Hinderer, Ho, Hoak, Hochheim, Hofman, Holland, Holt,
  Holz, Hopkins, Horst, Hough, Houston, Howell, Hreibi, Huerta, Huet, Hughey,
  Hulko, Husa, Huttner, Huynh-Dinh, Iess, Indik, Ingram, Inta, Intini, Irwin,
  Isa, Isac, Isi, Iyer, Izumi, Jacqmin, Jani, Jaranowski, Johnson, Johnson,
  Jones, Jones, Jonker, Ju, Junker, Kalaghatgi, Kalogera, Kamai, Kandhasamy,
  Kang, Kanner, Kapadia, Karki, Karvinen, Kasprzack, Katolik, Katsanevas,
  Katsavounidis, Katzman, Kaufer, Kawabe, Keerthana, Kéfélian, Keitel,
  Kemball, Kennedy, Key, Khalili, Khamesra, Khan, Khan, Khan, Khan, Khazanov,
  Kijbunchoo, Kim, Kim, Kim, Kim, Kim, Kim, King, King, Kinley-Hanlon,
  Kirchhoff, Kissel, Kleybolte, Klimenko, Knowles, Koch, Koehlenbeck, Koley,
  Kondrashov, Kontos, Korobko, Korth, Kowalska, Kozak, Krämer, Kringel,
  Krishnan, Królak, Kuehn, Kumar, Kumar, Kumar, Kuo, Kutynia, Kwang, Lackey,
  Lai, Landry, Landry, Lang, Lange, Lantz, Lanza, Lartaux-Vollard, Lasky,
  Laxen, Lazzarini, Lazzaro, Leaci, Leavey, Lee, Lee, Lee, Lee, Lee, Lehmann,
  Lenon, Leonardi, Leroy, Letendre, Levin, Li, Li, Li, Linker, Littenberg, Liu,
  Liu, Lo, Lockerbie, London, Longo, Lorenzini, Loriette, Lormand, Losurdo,
  Lough, Lousto, Lovelace, Lück, Lumaca, Lundgren, Lynch, Ma, Macas, Macfoy,
  Machenschalk, MacInnis, Macleod, Magaña~Hernandez, Magaña-Sandoval,
  Magaña~Zertuche, Magee, Majorana, Maksimovic, Man, Mandic, Mangano, Mansell,
  Manske, Mantovani, Marchesoni, Marion, Márka, Márka, Markakis, Markosyan,
  Markowitz, Maros, Marquina, Martelli, Martellini, Martin, Martin, Martynov,
  Mason, Massera, Masserot, Massinger, Masso-Reid, Mastrogiovanni, Matas,
  Matichard, Matone, Mavalvala, Mazumder, McCann, McCarthy, McClelland,
  McCormick, McCuller, McGuire, McIver, McManus, McRae, McWilliams, Meacher,
  Meadors, Mehmet, Meidam, Mejuto-Villa, Melatos, Mendell, Mendoza-Gandara,
  Mercer, Mereni, Merilh, Merzougui, Meshkov, Messenger, Messick, Metzdorff,
  Meyers, Miao, Michel, Middleton, Mikhailov, Milano, Miller, Miller, Miller,
  Miller, Millhouse, Mills, Milovich-Goff, Minazzoli, Minenkov, Ming, Mishra,
  Mitra, Mitrofanov, Mitselmakher, Mittleman, Moffa, Mogushi, Mohan, Mohapatra,
  Montani, Moore, Moraru, Moreno, Morisaki, Mours, Mow-Lowry, Mueller, Muir,
  Mukherjee, Mukherjee, Mukherjee, Mukund, Mullavey, Munch, Muñiz, Muratore,
  Murray, Nagar, Napier, Nardecchia, Naticchioni, Nayak, Neilson, Nelemans,
  Nelson, Nery, Neunzert, Nevin, Newport, Ng, Ng, Nguyen, Nguyen, Nichols,
  Nielsen, Nissanke, Nitz, Nocera, Nolting, North, Nuttall, Obergaulinger,
  Oberling, O’Brien, O’Dea, Ogin, Oh, Oh, Ohme, Ohta, Okada, Oliver,
  Oppermann, Oram, O’Reilly, Ormiston, Ortega, O’Shaughnessy, Ossokine,
  Ottaway, Overmier, Owen, Pace, Pagano, Page, Page, Pai, Pai, Palamos,
  Palashov, Palomba, Pal-Singh, Pan, Pan, Pang, Pang, Pankow, Pannarale, Pant,
  Paoletti, Paoli, Papa, Parida, Parker, Pascucci, Pasqualetti, Passaquieti,
  Passuello, Patil, Patricelli, Pearlstone, Pedersen, Pedraza, Pedurand,
  Pekowsky, Pele, Penn, Perego, Perez, Perreca, Perri, Pfeiffer, Phelps,
  Phukon, Piccinni, Pichot, Piergiovanni, Pierro, Pillant, Pinard, Pinto,
  Pirello, Pitkin, Poggiani, Popolizio, Porter, Possenti, Post, Powell, Prasad,
  Pratt, Pratten, Predoi, Prestegard, Principe, Privitera, Prodi, Prokhorov,
  Puncken, Punturo, Puppo, Pürrer, Qi, Quetschke, Quintero, Quitzow-James,
  Raab, Rabeling, Radkins, Raffai, Raja, Rajan, Rajbhandari, Rakhmanov,
  Ramirez, Ramos-Buades, Rana, Rapagnani, Raymond, Razzano, Read, Regimbau,
  Rei, Reid, Reitze, Ren, Ricci, Ricker, Riemenschneider, Riles, Rizzo,
  Robertson, Robie, Robinet, Robson, Rocchi, Rolland, Rollins, Roma, Romano,
  Romel, Romie, Rosińska, Ross, Rowan, Rüdiger, Ruggi, Rutins, Ryan, Sachdev,
  Sadecki, Sakellariadou, Salconi, Saleem, Salemi, Samajdar, Sammut, Sampson,
  Sanchez, Sanchez, Sanchis-Gual, Sandberg, Sanders, Sarin, Sassolas,
  Sathyaprakash, Saulson, Sauter, Savage, Sawadsky, Schale, Scheel, Scheuer,
  Schmidt, Schnabel, Schofield, Schönbeck, Schreiber, Schuette, Schulte,
  Schutz, Schwalbe, Scott, Scott, Seidel, Sellers, Sengupta, Sentenac, Sequino,
  Sergeev, Setyawati, Shaddock, Shaffer, Shah, Shahriar, Shaner, Shao, Shapiro,
  Shawhan, Shen, Shoemaker, Shoemaker, Siellez, Siemens, Sieniawska, Sigg,
  Silva, Singer, Singh, Singhal, Sintes, Slagmolen, Slaven-Blair, Smith, Smith,
  Smith, Somala, Son, Sorazu, Sorrentino, Souradeep, Spencer, Srivastava,
  Staats, Steinke, Steinlechner, Steinlechner, Steinmeyer, Steltner, Stevenson,
  Stocks, Stone, Stops, Strain, Stratta, Strigin, Strunk, Sturani, Stuver,
  Summerscales, Sun, Sunil, Suresh, Sutton, Swinkels, Szczepańczyk, Tacca,
  Tait, Talbot, Talukder, Tanner, Tápai, Taracchini, Tasson, Taylor, Taylor,
  Tewari, Theeg, Thies, Thomas, Thomas, Thomas, Thorne, Thrane, Tiwari, Tiwari,
  Tokmakov, Toland, Tonelli, Tornasi, Torres-Forné, Torrie, Töyrä, Travasso,
  Traylor, Trinastic, Tringali, Trovato, Trozzo, Tsang, Tse, Tso, Tsuna,
  Tsukada, Tuyenbayev, Ueno, Ugolini, Urban, Usman, Vahlbruch, Vajente, Valdes,
  van Bakel, van Beuzekom, van~den Brand, Van Den~Broeck, Vander-Hyde, van~der
  Schaaf, van Heijningen, van Veggel, Vardaro, Varma, Vass, Vasúth, Vecchio,
  Vedovato, Veitch, Veitch, Venkateswara, Venugopalan, Verkindt, Vetrano,
  Viceré, Viets, Vinciguerra, Vine, Vinet, Vitale, Vo, Vocca, Vorvick,
  Vyatchanin, Wade, Wade, Wade, Walet, Walker, Wallace, Walsh, Wang, Wang,
  Wang, Wang, Wang, Ward, Warner, Was, Watchi, Weaver, Wei, Weinert, Weinstein,
  Weiss, Wellmann, Wen, Wessel, Weßels, Westerweck, Wette, Whelan, Whiting,
  Whittle, Wilken, Williams, Williams, Williamson, Willis, Willke, Wimmer,
  Winkler, Wipf, Wittel, Woan, Woehler, Wofford, Wong, Worden, Wright, Wu,
  Wysocki, Xiao, Yam, Yamamoto, Yancey, Yang, Yap, Yazback, Yu, Yu, Yvert,
  Zadrożny, Zanolin, Zelenova, Zendri, Zevin, Zhang, Zhang, Zhang, Zhang,
  Zhang, Zhao, Zhou, Zhou, Zhu, Zhu, Zimmerman, Zlochower, Zucker, \&
  Zweizig}]{the_ligo_scientific_collaboration_and_the_virgo_collaboration_gw170817_2018}
{The LIGO Scientific Collaboration and The Virgo Collaboration}, Abbott, B.,
  Abbott, R., {et~al.} 2018,
  \href{http://dx.doi.org/10.1103/PhysRevLett.121.161101}{\JournalTitle{Physical
  Review Letters}, 121, 161101}, publisher: American Physical Society

\bibitem[{{The LIGO Scientific Collaboration, the Virgo Collaboration, the
  KAGRA Collaboration} {et~al.}(2021{\natexlab{a}}){The LIGO Scientific
  Collaboration, the Virgo Collaboration, the KAGRA Collaboration}, Abbott,
  Abe, Acernese, Ackley, Adhikari, \&
  Adhikari}]{the_ligo_scientific_collaboration_the_virgo_collaboration_the_kagra_collaboration_constraints_2021}
{The LIGO Scientific Collaboration, the Virgo Collaboration, the KAGRA
  Collaboration}, Abbott, R., Abe, H., {et~al.} 2021{\natexlab{a}},
  \href{https://arxiv.org/abs/2111.03604}{\JournalTitle{arXiv:2111.03604
  [astro-ph, physics:gr-qc]}}, arXiv: 2111.03604

\bibitem[{{The LIGO Scientific Collaboration, the Virgo Collaboration, the
  KAGRA Collaboration} {et~al.}(2021{\natexlab{b}}){The LIGO Scientific
  Collaboration, the Virgo Collaboration, the KAGRA Collaboration}, Abbott,
  Abbott, Acernese, Ackley, Adams, Adhikari, Adhikari, Adya, Affeldt, Agarwal,
  Agathos, Agatsuma, Aggarwal, Aguiar, Aiello, Ain, Ajith, Akcay, Akutsu,
  Albanesi, Allocca, Altin, Amato, Anand, Anand, Ananyeva, Anderson, Anderson,
  Ando, Andrade, Andres, Andrić, Angelova, Ansoldi, Antelis, Antier, Appert,
  Arai, Arai, Arai, Araki, Araya, Araya, Areeda, Arène, Aritomi, Arnaud,
  Arogeti, Aronson, Arun, Asada, Asali, Ashton, Aso, Assiduo, Aston, Astone,
  Aubin, Austin, Babak, Badaracco, Bader, Badger, Bae, Bae, Baer, Bagnasco,
  Bai, Baiotti, Baird, Bajpai, Ball, Ballardin, Ballmer, Balsamo, Baltus,
  Banagiri, Bankar, Barayoga, Barbieri, Barish, Barker, Barneo, Barone, Barr,
  Barsotti, Barsuglia, Barta, Bartlett, Barton, Bartos, Bassiri, Basti, Bawaj,
  Bayley, Baylor, Bazzan, Bécsy, Bedakihale, Bejger, Belahcene, Benedetto,
  Beniwal, Bennett, Bentley, BenYaala, Bergamin, Berger, Bernuzzi, Berry,
  Bersanetti, Bertolini, Betzwieser, Beveridge, Bhandare, Bhardwaj,
  Bhattacharjee, Bhaumik, Bilenko, Billingsley, Bini, Birney, Birnholtz,
  Biscans, Bischi, Biscoveanu, Bisht, Biswas, Bitossi, Bizouard, Blackburn,
  Blair, Blair, Blair, Bobba, Bode, Boer, Bogaert, Boldrini, Bonavena, Bondu,
  Bonilla, Bonnand, Booker, Boom, Bork, Boschi, Bose, Bose, Bossilkov, Boudart,
  Bouffanais, Bozzi, Bradaschia, Brady, Bramley, Branch, Branchesi, Brandt,
  Brau, Breschi, Briant, Briggs, Brillet, Brinkmann, Brockill, Brooks, Brooks,
  Brown, Brunett, Bruno, Bruntz, Bryant, Bulik, Bulten, Buonanno, Buscicchio,
  Buskulic, Buy, Byer, Davies, Cadonati, Cagnoli, Cahillane, Bustillo,
  Callaghan, Callister, Calloni, Cameron, Camp, Canepa, Canevarolo,
  Cannavacciuolo, Cannon, Cao, Cao, Capocasa, Capote, Carapella, Carbognani,
  Carlin, Carney, Carpinelli, Carrillo, Carullo, Carver, Diaz, Casentini,
  Castaldi, Caudill, Cavaglià, Cavalier, Cavalieri, Ceasar, Cella,
  Cerdá-Durán, Cesarini, Chaibi, Chakravarti, Subrahmanya, Champion, Chan,
  Chan, Chan, Chan, Chan, Chandra, Chanial, Chao, Chapman-Bird, Charlton,
  Chase, Chassande-Mottin, Chatterjee, Chatterjee, Chatterjee, Chaturvedi,
  Chaty, Chatziioannou, Chen, Chen, Chen, Chen, Chen, Chen, Chen, Chen, Cheng,
  Cheong, Cheung, Chia, Chiadini, Chiang, Chiarini, Chierici, Chincarini,
  Chiofalo, Chiummo, Cho, Cho, Choudhary, Choudhary, Christensen, Chu, Chu,
  Chu, Chua, Chung, Ciani, Ciecielag, Cieślar, Cifaldi, Ciobanu, Ciolfi,
  Cipriano, Cirone, Clara, Clark, Clark, Clarke, Clearwater, Clesse, Cleva,
  Coccia, Codazzo, Cohadon, Cohen, Cohen, Colleoni, Collette, Colombo, Colpi,
  Compton, Constancio~Jr., Conti, Cooper, Corban, Corbitt, Cordero-Carrión,
  Corezzi, Corley, Cornish, Corre, Corsi, Cortese, Costa, Cotesta, Coughlin,
  Coulon, Countryman, Cousins, Couvares, Coward, Cowart, Coyne, Coyne,
  Creighton, Creighton, Criswell, Croquette, Crowder, Cudell, Cullen, Cumming,
  Cummings, Cunningham, Cuoco, Curyło, Dabadie, Canton, Dall'Osso, Dálya,
  Dana, DaneshgaranBajastani, D'Angelo, Danila, Danilishin, D'Antonio,
  Danzmann, Darsow-Fromm, Dasgupta, Datrier, Datta, Dattilo, Dave, Davier,
  Davis, Davis, Daw, de~Alarcón, Dean, DeBra, Deenadayalan, Degallaix,
  De~Laurentis, Deléglise, Del~Favero, De~Lillo, De~Lillo, Del~Pozzo,
  DeMarchi, De~Matteis, D'Emilio, Demos, Dent, Depasse, De~Pietri, De~Rosa,
  De~Rossi, DeSalvo, De~Simone, Dhurandhar, Díaz, Diaz-Ortiz~Jr., Didio,
  Dietrich, Di~Fiore, Di~Fronzo, Di~Giorgio, Di~Giovanni, Di~Giovanni,
  Di~Girolamo, Di~Lieto, Ding, Di~Pace, Di~Palma, Di~Renzo, Divakarla,
  Dmitriev, Doctor, D'Onofrio, Donovan, Dooley, Doravari, Dorrington, Drago,
  Driggers, Drori, Ducoin, Dupej, Durante, D'Urso, Duverne, Dwyer, Eassa,
  Easter, Ebersold, Eckhardt, Eddolls, Edelman, Edo, Edy, Effler, Eguchi,
  Eichholz, Eikenberry, Eisenmann, Eisenstein, Ejlli, Engelby, Enomoto, Errico,
  Essick, Estellés, Estevez, Etienne, Etzel, Evans, Evans, Ewing, Fafone,
  Fair, Fairhurst, Farah, Farinon, Farr, Farr, Farrow, Fauchon-Jones, Favaro,
  Favata, Fays, Fazio, Feicht, Fejer, Fenyvesi, Ferguson, Fernandez-Galiana,
  Ferrante, Ferreira, Fidecaro, Figura, Fiori, Fishbach, Fisher, Fittipaldi,
  Fiumara, Flaminio, Floden, Fong, Font, Fornal, Forsyth, Franke, Frasca,
  Frasconi, Frederick, Freed, Frei, Freise, Frey, Fritschel, Frolov, Fronzé,
  Fujii, Fujikawa, Fukunaga, Fukushima, Fulda, Fyffe, Gabbard, Gabella, Gadre,
  Gair, Gais, Galaudage, Gamba, Ganapathy, Ganguly, Gao, Gaonkar, Garaventa,
  García, García-Núñez, García-Quirós, Garufi, Gateley, Gaudio, Gayathri,
  Ge, Gemme, Gennai, George, George, Gerberding, Gergely, Gewecke, Ghonge,
  Ghosh, Ghosh, Ghosh, Ghosh, Giacomazzo, Giacoppo, Giaime, Giardina, Gibson,
  Gier, Giesler, Giri, Gissi, Glanzer, Gleckl, Godwin, Goetz, Goetz, Gohlke,
  Golomb, Goncharov, González, Gopakumar, Gosselin, Gouaty, Gould, Grace,
  Grado, Granata, Granata, Grant, Gras, Grassia, Gray, Gray, Greco, Green,
  Green, Gretarsson, Gretarsson, Griffith, Griffiths, Griggs, Grignani,
  Grimaldi, Grimm, Grote, Grunewald, Gruning, Guerra, Guidi, Guimaraes, Guixé,
  Gulati, Guo, Guo, Gupta, Gupta, Gupta, Gustafson, Gustafson, Guzman, Ha,
  Haegel, Hagiwara, Haino, Halim, Hall, Hamilton, Hammond, Han, Haney, Hanks,
  Hanna, Hannam, Hannuksela, Hansen, Hansen, Hanson, Harder, Hardwick, Haris,
  Harms, Harry, Harry, Hartwig, Hasegawa, Haskell, Hasskew, Haster, Hattori,
  Haughian, Hayakawa, Hayama, Hayes, Healy, Heidmann, Heidt, Heintze, Heinze,
  Heinzel, Heitmann, Hellman, Hello, Helmling-Cornell, Hemming, Hendry, Heng,
  Hennes, Hennig, Hennig, Hernandez, Vivanco, Heurs, Hild, Hill, Himemoto,
  Hines, Hiranuma, Hirata, Hirose, Hochheim, Hofman, Hohmann, Holcomb, Holland,
  Holley-Bockelmann, Hollows, Holmes, Holt, Holz, Hong, Hopkins, Hough,
  Hourihane, Howell, Hoy, Hoyland, Hreibi, Hsieh, Hsu, Huang, Huang, Huang,
  Huang, Huang, Huang, Hübner, Huddart, Hughey, Hui, Hui, Husa, Huttner,
  Huxford, Huynh-Dinh, Ide, Idzkowski, Iess, Ikenoue, Imam, Inayoshi, Ingram,
  Inoue, Ioka, Isi, Isleif, Ito, Itoh, Iyer, Izumi, JaberianHamedan, Jacqmin,
  Jadhav, Jadhav, James, Jan, Jani, Janquart, Janssens, Janthalur, Jaranowski,
  Jariwala, Jaume, Jenkins, Jenner, Jeon, Jeunon, Jia, Jin, Johns,
  Johnson-McDaniel, Jones, Jones, Jones, Jones, Jones, Jonker, Ju, Jung, Jung,
  Junker, Juste, Kaihotsu, Kajita, Kakizaki, Kalaghatgi, Kalogera, Kamai,
  Kamiizumi, Kanda, Kandhasamy, Kang, Kanner, Kao, Kapadia, Kapasi, Karat,
  Karathanasis, Karki, Kashyap, Kasprzack, Kastaun, Katsanevas, Katsavounidis,
  Katzman, Kaur, Kawabe, Kawaguchi, Kawai, Kawasaki, Kéfélian, Keitel, Key,
  Khadka, Khalili, Khan, Khazanov, Khetan, Khursheed, Kijbunchoo, Kim, Kim,
  Kim, Kim, Kim, Kim, Kimball, Kimura, Kinley-Hanlon, Kirchhoff, Kissel, Kita,
  Kitazawa, Kleybolte, Klimenko, Knee, Knowles, Knyazev, Koch, Koekoek, Kojima,
  Kokeyama, Koley, Kolitsidou, Kolstein, Komori, Kondrashov, Kong, Kontos,
  Koper, Korobko, Kotake, Kovalam, Kozak, Kozakai, Kozu, Kringel, Krishnendu,
  Królak, Kuehn, Kuei, Kuijer, Kulkarni, Kumar, Kumar, Kumar, Kumar, Kume,
  Kuns, Kuo, Kuo, Kuromiya, Kuroyanagi, Kusayanagi, Kuwahara, Kwak, Lagabbe,
  Laghi, Lalande, Lam, Lamberts, Landry, Lane, Lang, Lange, Lantz, La~Rosa,
  Lartaux-Vollard, Lasky, Laxen, Lazzarini, Lazzaro, Leaci, Leavey, Lecoeuche,
  Lee, Lee, Lee, Lee, Lee, Lee, Lehmann, Lemaître, Leonardi, Leroy, Letendre,
  Levesque, Levin, Leviton, Leyde, Li, Li, Li, Li, Li, Li, Lin, Lin, Lin, Lin,
  Lin, Linde, Linker, Linley, Littenberg, Liu, Liu, Liu, Liu, Llamas,
  Llorens-Monteagudo, Lo, Lockwood, Loh, London, Longo, Lopez, Portilla,
  Lorenzini, Loriette, Lormand, Losurdo, Lott, Lough, Lousto, Lovelace,
  Lucaccioni, Lück, Lumaca, Lundgren, Luo, Lynam, Macas, MacInnis, Macleod,
  MacMillan, Macquet, Hernandez, Magazzù, Magee, Maggiore, Magnozzi, Mahesh,
  Majorana, Makarem, Maksimovic, Maliakal, Malik, Man, Mandic, Mangano, Mango,
  Mansell, Manske, Mantovani, Mapelli, Marchesoni, Marchio, Marion, Mark,
  Márka, Márka, Markakis, Markosyan, Markowitz, Maros, Marquina, Marsat,
  Martelli, Martin, Martin, Martinez, Martinez, Martinez, Martinovic, Martynov,
  Marx, Masalehdan, Mason, Massera, Masserot, Massinger, Masso-Reid,
  Mastrogiovanni, Matas, Mateu-Lucena, Matichard, Matiushechkina, Mavalvala,
  McCann, McCarthy, McClelland, McClincy, McCormick, McCuller, McGhee, McGuire,
  McIsaac, McIver, McRae, McWilliams, Meacher, Mehmet, Mehta, Meijer, Melatos,
  Melchor, Mendell, Menendez-Vazquez, Menoni, Mercer, Mereni, Merfeld, Merilh,
  Merritt, Merzougui, Meshkov, Messenger, Messick, Meyers, Meylahn, Mhaske,
  Miani, Miao, Michaloliakos, Michel, Michimura, Middleton, Milano, Miller,
  Miller, Miller, Millhouse, Mills, Milotti, Minazzoli, Minenkov, Mio, Mir,
  Miravet-Tenés, Mishra, Mishra, Mistry, Mitra, Mitrofanov, Mitselmakher,
  Mittleman, Miyakawa, Miyamoto, Miyazaki, Miyo, Miyoki, Mo, Modafferi, Moguel,
  Mogushi, Mohapatra, Mohite, Molina, Molina-Ruiz, Mondin, Montani, Moore,
  Moraru, Morawski, More, Moreno, Moreno, Mori, Morisaki, Moriwaki, Morrás,
  Mours, Mow-Lowry, Mozzon, Muciaccia, Mukherjee, Mukherjee, Mukherjee,
  Mukherjee, Mukherjee, Mukund, Mullavey, Munch, Muñiz, Murray, Musenich,
  Muusse, Nadji, Nagano, Nagano, Nagar, Nakamura, Nakano, Nakano, Nakashima,
  Nakayama, Napolano, Nardecchia, Narikawa, Naticchioni, Nayak, Nayak, Negishi,
  Neil, Neilson, Nelemans, Nelson, Nery, Neubauer, Neunzert, Ng, Ng, Nguyen,
  Nguyen, Nguyen, Quynh, Ni, Nichols, Nishizawa, Nissanke, Nitoglia, Nocera,
  Norman, North, Nozaki, Siles, Nuttall, Oberling, O'Brien, Obuchi, O'Dell,
  Oelker, Ogaki, Oganesyan, Oh, Oh, Oh, Ohashi, Ohishi, Ohkawa, Ohme, Ohta,
  Okada, Okutani, Okutomi, Olivetto, Oohara, Ooi, Oram, O'Reilly, Ormiston,
  Ormsby, Ortega, O'Shaughnessy, O'Shea, Oshino, Ossokine, Osthelder, Otabe,
  Ottaway, Overmier, Pace, Pagano, Page, Pagliaroli, Pai, Pai, Palamos,
  Palashov, Palomba, Pan, Pan, Panda, Pang, Pang, Pankow, Pannarale, Pant,
  Panther, Paoletti, Paoli, Paolone, Parisi, Park, Park, Parker, Pascucci,
  Pasqualetti, Passaquieti, Passuello, Patel, Pathak, Patricelli, Patron, Paul,
  Payne, Pedraza, Pegoraro, Pele, Arellano, Penn, Perego, Pereira, Pereira,
  Perez, Périgois, Perkins, Perreca, Perriès, Petermann, Petterson, Pfeiffer,
  Pham, Phukon, Piccinni, Pichot, Piendibene, Piergiovanni, Pierini, Pierro,
  Pillant, Pillas, Pilo, Pinard, Pinto, Pinto, Piotrzkowski, Piotrzkowski,
  Pirello, Pitkin, Placidi, Planas, Plastino, Pluchar, Poggiani, Polini, Pong,
  Ponrathnam, Popolizio, Porter, Poulton, Powell, Pracchia, Pradier, Prajapati,
  Prasai, Prasanna, Pratten, Principe, Prodi, Prokhorov, Prosposito, Prudenzi,
  Puecher, Punturo, Puosi, Puppo, Pürrer, Qi, Quetschke, Quitzow-James, Qutob,
  Raab, Raaijmakers, Radkins, Radulesco, Raffai, Rail, Raja, Rajan, Ramirez,
  Ramirez, Ramos-Buades, Rana, Rapagnani, Rapol, Ray, Raymond, Raza, Razzano,
  Read, Rees, Regimbau, Rei, Reid, Reid, Reitze, Relton, Renzini, Rettegno,
  Reza, Rezac, Ricci, Richards, Richardson, Richardson, Riemenschneider, Riles,
  Rinaldi, Rink, Rizzo, Robertson, Robie, Robinet, Rocchi, Rodriguez, Rolland,
  Rollins, Romanelli, Romano, Romel, Romero-Rodríguez, Romero-Shaw, Romie,
  Ronchini, Rosa, Rose, Rosińska, Ross, Rowan, Rowlinson, Roy, Roy, Roy,
  Rozza, Ruggi, Ruiz-Rocha, Ryan, Sachdev, Sadecki, Sadiq, Sago, Saito, Saito,
  Sakai, Sakai, Sakellariadou, Sakuno, Salafia, Salconi, Saleem, Salemi,
  Samajdar, Sanchez, Sanchez, Sanchez, Sanchis-Gual, Sanders, Sanuy, Saravanan,
  Sarin, Sassolas, Satari, Sathyaprakash, Sato, Sato, Sauter, Savage, Sawada,
  Sawant, Sawant, Sayah, Schaetzl, Scheel, Scheuer, Schiworski, Schmidt,
  Schmidt, Schnabel, Schneewind, Schofield, Schönbeck, Schulte, Schutz,
  Schwartz, Scott, Scott, Seglar-Arroyo, Sekiguchi, Sekiguchi, Sellers,
  Sengupta, Sentenac, Seo, Sequino, Sergeev, Setyawati, Shaffer, Shahriar,
  Shams, Shao, Sharma, Sharma, Shawhan, Shcheblanov, Shibagaki, Shikauchi,
  Shimizu, Shimoda, Shimode, Shinkai, Shishido, Shoda, Shoemaker, Shoemaker,
  ShyamSundar, Sieniawska, Sigg, Singer, Singh, Singh, Singha, Sintes, Sipala,
  Skliris, Slagmolen, Slaven-Blair, Smetana, Smith, Smith, Soldateschi, Somala,
  Somiya, Son, Soni, Soni, Sordini, Sorrentino, Sorrentino, Sotani, Soulard,
  Souradeep, Sowell, Spagnuolo, Spencer, Spera, Srinivasan, Srivastava,
  Srivastava, Staats, Stachie, Steer, Steinhoff, Steinlechner, Steinlechner,
  Stevenson, Stops, Stover, Strain, Strang, Stratta, Strunk, Sturani, Stuver,
  Sudhagar, Sudhir, Sugimoto, Suh, Sullivan, Sullivan, Summerscales, Sun, Sun,
  Sunil, Sur, Suresh, Sutton, Suzuki, Suzuki, Swinkels, Szczepańczyk,
  Szewczyk, Tacca, Tagoshi, Tait, Takahashi, Takahashi, Takamori, Takano,
  Takeda, Takeda, Talbot, Talbot, Tanaka, Tanaka, Tanaka, Tanaka, Tanaka,
  Tanasijczuk, Tanioka, Tanner, Tao, Tao, Martín, Taranto, Tasson, Telada,
  Tenorio, Terhune, Terkowski, Thirugnanasambandam, Thomas, Thomas, Thomas,
  Thompson, Thondapu, Thorne, Thrane, Tiwari, Tiwari, Tiwari, Toivonen, Toland,
  Tolley, Tomaru, Tomigami, Tomura, Tonelli, Torres-Forné, Torrie, Melo,
  Töyrä, Trapananti, Travasso, Traylor, Trevor, Tringali, Tripathee, Troiano,
  Trovato, Trozzo, Trudeau, Tsai, Tsai, Tsang, Tsang, Tsao, Tse, Tso, Tsubono,
  Tsuchida, Tsukada, Tsuna, Tsutsui, Tsuzuki, Turbang, Turconi, Tuyenbayev,
  Ubhi, Uchikata, Uchiyama, Udall, Ueda, Uehara, Ueno, Ueshima, Unnikrishnan,
  Uraguchi, Urban, Ushiba, Utina, Vahlbruch, Vajente, Vajpeyi, Valdes,
  Valentini, Valsan, van Bakel, van Beuzekom, Brand, Broeck, Vander-Hyde,
  van~der Schaaf, van Heijningen, Vanosky, van Putten, van Remortel, Vardaro,
  Vargas, Varma, Vasúth, Vecchio, Vedovato, Veitch, Veitch, Venneberg,
  Venugopalan, Verkindt, Verma, Verma, Veske, Vetrano, Viceré, Vidyant, Viets,
  Vijaykumar, Villa-Ortega, Vinet, Virtuoso, Vitale, Vo, Vocca, von Reis, von
  Wrangel, Vorvick, Vyatchanin, Wade, Wade, Wagner, Walet, Walker, Wallace,
  Wallace, Walsh, Wang, Wang, Wang, Ward, Warner, Was, Washimi, Washington,
  Watchi, Weaver, Webster, Weinert, Weinstein, Weiss, Weller, Weller, Wellmann,
  Wen, Weßels, Wette, Whelan, White, Whiting, Whittle, Wilken, Williams,
  Williams, Williams, Williamson, Willis, Willke, Wilson, Winkler, Wipf,
  Wlodarczyk, Woan, Woehler, Wofford, Wong, Wu, Wu, Wu, Wu, Wysocki, Xiao, Xu,
  Yamada, Yamamoto, Yamamoto, Yamamoto, Yamamoto, Yamashita, Yamazaki, Yang,
  Yang, Yang, Yang, Yang, Yap, Yeeles, Yelikar, Ying, Yokogawa, Yokoyama,
  Yokozawa, Yoo, Yoshioka, Yu, Yu, Yuzurihara, Zadrożny, Zanolin, Zeidler,
  Zelenova, Zendri, Zevin, Zhan, Zhang, Zhang, Zhang, Zhang, Zhang, Zhao, Zhao,
  Zhao, Zhao, Zheng, Zhou, Zhou, Zhu, Zhu, Zimmerman, Zlochower, Zucker, \&
  Zweizig}]{the_ligo_scientific_collaboration_the_virgo_collaboration_the_kagra_collaboration_gwtc-3_2021}
{The LIGO Scientific Collaboration, the Virgo Collaboration, the KAGRA
  Collaboration}, Abbott, R., Abbott, T.~D., {et~al.} 2021{\natexlab{b}},
  \href{http://arxiv.org/abs/2111.03606}{\JournalTitle{arXiv:2111.03606
  [astro-ph, physics:gr-qc]}}, arXiv: 2111.03606

\bibitem[{{The LIGO Scientific Collaboration, the Virgo Collaboration, the
  KAGRA Collaboration} {et~al.}(2021{\natexlab{c}}){The LIGO Scientific
  Collaboration, the Virgo Collaboration, the KAGRA Collaboration}, Abbott,
  Abe, Acernese, Ackley, Adhikari, Adhikari, Adkins, Adya, Affeldt, Agarwal,
  Agathos, Agatsuma, Aggarwal, Aguiar, Aiello, Ain, Ajith, Akutsu, de~Alarcón,
  Albanesi, Alfaidi, Allocca, Altin, Amato, Anand, Anand, Ananyeva, Anderson,
  Anderson, Ando, Andrade, Andres, Andrés-Carcasona, Andrić, Angelova,
  Ansoldi, Antelis, Antier, Apostolatos, Appavuravther, Appert, Apple, Arai,
  Araya, Araya, Areeda, Arène, Aritomi, Arnaud, Arogeti, Aronson, Arun, Asada,
  Asali, Ashton, Aso, Assiduo, Melo, Aston, Astone, Aubin, AultONeal, Austin,
  Babak, Badaracco, Bader, Badger, Bae, Bae, Baer, Bagnasco, Bai, Baird,
  Bajpai, Baka, Ball, Ballardin, Ballmer, Balsamo, Baltus, Banagiri, Banerjee,
  Bankar, Barayoga, Barbieri, Barish, Barker, Barneo, Barone, Barr, Barsotti,
  Barsuglia, Barta, Bartlett, Barton, Bartos, Basak, Bassiri, Basti, Bawaj,
  Bayley, Bazzan, Becher, Bécsy, Bedakihale, Beirnaert, Bejger, Belahcene,
  Benedetto, Beniwal, Benjamin, Bennett, Bentley, BenYaala, Bera, Berbel,
  Bergamin, Berger, Bernuzzi, Berry, Bersanetti, Bertolini, Betzwieser,
  Beveridge, Bhandare, Bhandari, Bhardwaj, Bhatt, Bhattacharjee, Bhaumik,
  Bianchi, Bilenko, Billingsley, Bini, Birney, Birnholtz, Biscans, Bischi,
  Biscoveanu, Bisht, Biswas, Bitossi, Bizouard, Blackburn, Blair, Blair, Blair,
  Bobba, Bode, Boër, Bogaert, Boldrini, Bolingbroke, Bonavena, Bondu, Bonilla,
  Bonnand, Booker, Boom, Bork, Boschi, Bose, Bose, Bossilkov, Boudart,
  Bouffanais, Bozzi, Bradaschia, Brady, Bramley, Branch, Branchesi, Brau,
  Breschi, Briant, Briggs, Brillet, Brinkmann, Brockill, Brooks, Brooks, Brown,
  Brunett, Bruno, Bruntz, Bryant, Bucci, Bulik, Bulten, Buonanno, Burtnyk,
  Buscicchio, Buskulic, Buy, Byer, Davies, Cabras, Cabrita, Cadonati, Caesar,
  Cagnoli, Cahillane, Bustillo, Callaghan, Callister, Calloni, Cameron, Camp,
  Canepa, Canevarolo, Cannavacciuolo, Cannon, Cao, Cao, Capocasa, Capote,
  Carapella, Carbognani, Carlassara, Carlin, Carney, Carpinelli, Carrillo,
  Carullo, Carver, Diaz, Casentini, Castaldi, Caudill, Cavaglià, Cavalier,
  Cavalieri, Cella, Cerdá-Durán, Cesarini, Chaibi, Subrahmanya, Champion,
  Chan, Chan, Chan, Chan, Chan, Chandra, Chang, Chanial, Chao, Chapman-Bird,
  Charlton, Chase, Chassande-Mottin, Chatterjee, Chatterjee, Chatterjee,
  Chaturvedi, Chaty, Chatziioannou, Chen, Chen, Chen, Chen, Chen, Chen, Chen,
  Chen, Chen, Cheng, Cheong, Cheung, Chia, Chiadini, Chiang, Chiarini,
  Chierici, Chincarini, Chiofalo, Chiummo, Choudhary, Choudhary, Christensen,
  Chu, Chu, Chua, Chung, Ciani, Ciecielag, Cieślar, Cifaldi, Ciobanu, Ciolfi,
  Cipriano, Clara, Clark, Clearwater, Clesse, Cleva, Coccia, Codazzo, Cohadon,
  Cohen, Colleoni, Collette, Colombo, Colpi, Compton, Constancio~Jr., Conti,
  Cooper, Corban, Corbitt, Cordero-Carrión, Corezzi, Corley, Cornish, Corre,
  Corsi, Cortese, Costa, Cotesta, Cottingham, Coughlin, Coulon, Countryman,
  Cousins, Couvares, Coward, Cowart, Coyne, Coyne, Creighton, Creighton,
  Criswell, Croquette, Crowder, Cudell, Cullen, Cumming, Cummings, Cunningham,
  Cuoco, Curyło, Dabadie, Canton, Dall'Osso, Dálya, Dana, D'Angelo,
  Danilishin, D'Antonio, Danzmann, Darsow-Fromm, Dasgupta, Datrier, Datta,
  Datta, Dattilo, Dave, Davier, Davis, Davis, Daw, Dean, DeBra, Deenadayalan,
  Degallaix, De~Laurentis, Deléglise, Del~Favero, De~Lillo, De~Lillo,
  Dell'Aquila, Del~Pozzo, DeMarchi, De~Matteis, D'Emilio, Demos, Dent, Depasse,
  De~Pietri, De~Rosa, De~Rossi, DeSalvo, De~Simone, Dhurandhar, Díaz, Didio,
  Dietrich, Di~Fiore, Di~Fronzo, Di~Giorgio, Di~Giovanni, Di~Giovanni,
  Di~Girolamo, Di~Lieto, Di~Michele, Ding, Di~Pace, Di~Palma, Di~Renzo,
  Divakarla, Divyajyoti, Dmitriev, Doctor, Donahue, D'Onofrio, Donovan, Dooley,
  Doravari, Drago, Driggers, Drori, Ducoin, Dupej, Dupletsa, Durante, D'Urso,
  Duverne, Dwyer, Eassa, Easter, Ebersold, Eckhardt, Eddolls, Edelman, Edo,
  Edy, Effler, Eguchi, Eichholz, Eikenberry, Eisenmann, Eisenstein, Ejlli,
  Engelby, Enomoto, Errico, Essick, Estellés, Estevez, Etienne, Etzel, Evans,
  Evans, Evstafyeva, Ewing, Fabrizi, Faedi, Fafone, Fair, Fairhurst, Fan,
  Farah, Farinon, Farr, Farr, Fauchon-Jones, Favaro, Favata, Fays, Fazio,
  Feicht, Fejer, Fenyvesi, Ferguson, Fernandez-Galiana, Ferrante, Ferreira,
  Fidecaro, Figura, Fiori, Fiori, Fishbach, Fisher, Fittipaldi, Fiumara,
  Flaminio, Floden, Fong, Font, Fornal, Forsyth, Franke, Frasca, Frasconi,
  Freed, Frei, Freise, Freitas, Frey, Fritschel, Frolov, Fronzé, Fujii,
  Fujikawa, Fujimoto, Fulda, Fyffe, Gabbard, Gabella, Gadre, Gair, Gais,
  Galaudage, Gamba, Ganapathy, Ganguly, Gao, Gaonkar, Garaventa, Núñez,
  García-Quirós, Garufi, Gateley, Gayathri, Ge, Gemme, Gennai, George,
  Gerberding, Gergely, Gewecke, Ghonge, Ghosh, Ghosh, Ghosh, Ghosh, Ghosh,
  Giacomazzo, Giacoppo, Giaime, Giardina, Gibson, Gier, Giesler, Giri, Gissi,
  Gkaitatzis, Glanzer, Gleckl, Godwin, Goetz, Goetz, Gohlke, Golomb, Goncharov,
  González, Gosselin, Gouaty, Gould, Goyal, Grace, Grado, Graham, Granata,
  Granata, Grant, Gras, Grassia, Gray, Gray, Greco, Green, Green, Gretarsson,
  Gretarsson, Griffith, Griffiths, Griggs, Grignani, Grimaldi, Grimes, Grimm,
  Grote, Grunewald, Gruning, Gruson, Guerra, Guidi, Guimaraes, Guixé, Gulati,
  Gunny, Guo, Guo, Gupta, Gupta, Gupta, Gupta, Gupta, Gustafson, Guzman, Ha,
  Hadiputrawan, Haegel, Haino, Halim, Hall, Hamilton, Hammond, Han, Haney,
  Hanks, Hanna, Hannam, Hannuksela, Hansen, Hansen, Hanson, Harder, Haris,
  Harms, Harry, Harry, Hartwig, Hasegawa, Haskell, Haster, Hathaway, Hattori,
  Haughian, Hayakawa, Hayama, Hayes, Healy, Heidmann, Heidt, Heintze, Heinze,
  Heinzel, Heitmann, Hellman, Hello, Helmling-Cornell, Hemming, Hendry, Heng,
  Hennes, Hennig, Hennig, Henshaw, Hernandez, Vivanco, Heurs, Hewitt,
  Higginbotham, Hild, Hill, Himemoto, Hines, Hirata, Hirose, Ho, Hochheim,
  Hofman, Hohmann, Holcomb, Holland, Hollows, Holmes, Holt, Holz, Hong, Hough,
  Hourihane, Howell, Hoy, Hoyland, Hreibi, Hsieh, Hsieh, Hsiung, Hsu, Huang,
  Huang, Huang, Huang, Huang, Huang, Hübner, Huddart, Hughey, Hui, Hui, Husa,
  Huttner, Huxford, Huynh-Dinh, Ide, Idzkowski, Iess, Inayoshi, Inoue, Iosif,
  Isi, Isleif, Ito, Itoh, Iyer, JaberianHamedan, Jacqmin, Jacquet, Jadhav,
  Jadhav, Jain, James, Jan, Jani, Janquart, Janssens, Janthalur, Jaranowski,
  Jariwala, Jaume, Jenkins, Jenner, Jeon, Jia, Jiang, Jin, Johns,
  Johnson-McDaniel, Johnston, Jones, Jones, Jones, Jones, Joshi, Ju, Jue, Jung,
  Jung, Junker, Juste, Kaihotsu, Kajita, Kakizaki, Kalaghatgi, Kalogera, Kamai,
  Kamiizumi, Kanda, Kandhasamy, Kang, Kanner, Kao, Kapadia, Kapasi,
  Karathanasis, Karki, Kashyap, Kasprzack, Kastaun, Kato, Katsanevas,
  Katsavounidis, Katzman, Kaur, Kawabe, Kawaguchi, Kéfélian, Keitel, Key,
  Khadka, Khalili, Khan, Khanam, Khazanov, Khetan, Khursheed, Kijbunchoo, Kim,
  Kim, Kim, Kim, Kim, Kim, Kim, Kimball, Kimura, Kinley-Hanlon, Kirchhoff,
  Kissel, Klimenko, Klinger, Knee, Knowles, Knust, Knyazev, Kobayashi, Koch,
  Koekoek, Kohri, Kokeyama, Koley, Kolitsidou, Kolstein, Komori, Kondrashov,
  Kong, Kontos, Koper, Korobko, Kovalam, Koyama, Kozak, Kozakai, Kringel,
  Krishnendu, Królak, Kuehn, Kuei, Kuijer, Kulkarni, Kumar, Kumar, Kumar,
  Kumar, Kume, Kuns, Kuromiya, Kuroyanagi, Kwak, Lacaille, Lagabbe, Laghi,
  Lalande, Lalleman, Lam, Lamberts, Landry, Lane, Lang, Lange, Lantz, La~Rosa,
  Lartaux-Vollard, Lasky, Laxen, Lazzarini, Lazzaro, Leaci, Leavey, LeBohec,
  Lecoeuche, Lee, Lee, Lee, Lee, Lee, Legred, Lehmann, Lemaître, Lenti,
  Leonardi, Leonova, Leroy, Letendre, Levesque, Levin, Leviton, Leyde, Li, Li,
  Li, Li, Li, Li, Li, Lin, Lin, Lin, Lin, Lin, Lin, Linde, Linker, Linley,
  Littenberg, Liu, Liu, Liu, Liu, Llamas, Lo, Lo, London, Longo, Lopez,
  Portilla, Lorenzini, Loriette, Lormand, Losurdo, Lott, Lough, Lousto,
  Lovelace, Lucaccioni, Lück, Lumaca, Lundgren, Luo, Lynam, Ma'arif, Macas,
  Machtinger, MacInnis, Macleod, MacMillan, Macquet, Hernandez, Magazzù,
  Magee, Maggiore, Magnozzi, Mahesh, Majorana, Maksimovic, Maliakal, Malik,
  Man, Mandic, Mangano, Mansell, Manske, Mantovani, Mapelli, Marchesoni, Pina,
  Marion, Mark, Márka, Márka, Markakis, Markosyan, Markowitz, Maros,
  Marquina, Marsat, Martelli, Martin, Martin, Martinez, Martinez, Martinez,
  Martinovic, Martynov, Marx, Masalehdan, Mason, Massera, Masserot, Masso-Reid,
  Mastrogiovanni, Matas, Mateu-Lucena, Matichard, Matiushechkina, Mavalvala,
  McCann, McCarthy, McClelland, McClincy, McCormick, McCuller, McGhee, McGuire,
  McIsaac, McIver, McRae, McWilliams, Meacher, Mehmet, Mehta, Meijer, Melatos,
  Melchor, Mendell, Menendez-Vazquez, Menoni, Mercer, Mereni, Merfeld, Merilh,
  Merritt, Merzougui, Meshkov, Messenger, Messick, Meyers, Meylahn, Mhaske,
  Miani, Miao, Michaloliakos, Michel, Michimura, Middleton, Mihaylov, Milano,
  Miller, Miller, Miller, Millhouse, Mills, Milotti, Minenkov, Mio, Mir,
  Miravet-Tenés, Mishkin, Mishra, Mishra, Mistry, Mitra, Mitrofanov,
  Mitselmakher, Mittleman, Miyakawa, Miyo, Miyoki, Mo, Modafferi, Moguel,
  Mogushi, Mohapatra, Mohite, Molina, Molina-Ruiz, Mondin, Montani, Moore,
  Moragues, Moraru, Morawski, More, Moreno, Moreno, Mori, Morisaki, Morisue,
  Moriwaki, Mours, Mow-Lowry, Mozzon, Muciaccia, Mukherjee, Mukherjee,
  Mukherjee, Mukherjee, Mukherjee, Mukund, Mullavey, Munch, Muñiz, Murray,
  Musenich, Muusse, Nadji, Nagano, Nagar, Nakamura, Nakano, Nakano, Nakayama,
  Napolano, Nardecchia, Narikawa, Narola, Naticchioni, Nayak, Nayak, Neil,
  Neilson, Nelson, Nelson, Nery, Neubauer, Neunzert, Ng, Ng, Nguyen, Nguyen,
  Nguyen, Quynh, Ni, Ni, Nichols, Nishimoto, Nishizawa, Nissanke, Nitoglia,
  Nocera, Norman, North, Nozaki, Nurbek, Nuttall, Obayashi, Oberling, O'Brien,
  O'Dell, Oelker, Ogaki, Oganesyan, Oh, Oh, Oh, Ohashi, Ohashi, Ohkawa, Ohme,
  Ohta, Okada, Okutani, Olivetto, Oohara, Oram, O'Reilly, Ormiston, Ormsby,
  O'Shaughnessy, O'Shea, Oshino, Ossokine, Osthelder, Otabe, Ottaway, Overmier,
  Pace, Pagano, Pagano, Page, Pagliaroli, Pai, Pai, Pal, Palamos, Palashov,
  Palomba, Pan, Pan, Panda, Pang, Pankow, Pannarale, Pant, Panther, Paoletti,
  Paoli, Paolone, Pappas, Parisi, Park, Park, Parker, Pascucci, Pasqualetti,
  Passaquieti, Passuello, Patel, Pathak, Patricelli, Patron, Paul, Payne,
  Pedraza, Pedurand, Pegoraro, Pele, Arellano, Penano, Penn, Perego, Pereira,
  Pereira, Perez, Périgois, Perkins, Perreca, Perriès, Pesios, Petermann,
  Petterson, Pfeiffer, Pham, Pham, Phukon, Phurailatpam, Piccinni, Pichot,
  Piendibene, Piergiovanni, Pierini, Pierro, Pillant, Pillas, Pilo, Pinard,
  Pineda-Bosque, Pinto, Pinto, Piotrzkowski, Piotrzkowski, Pirello, Pitkin,
  Placidi, Placidi, Planas, Plastino, Pluchar, Poggiani, Polini, Pong,
  Ponrathnam, Porter, Poulton, Poverman, Powell, Pracchia, Pradier, Prajapati,
  Prasai, Prasanna, Pratten, Principe, Prodi, Prokhorov, Prosposito, Prudenzi,
  Puecher, Punturo, Puosi, Puppo, Pürrer, Qi, Quartey, Quetschke, Quinonez,
  Quitzow-James, Qutob, Raab, Raaijmakers, Radkins, Radulesco, Raffai, Rail,
  Raja, Rajan, Ramirez, Ramirez, Ramos-Buades, Rana, Rapagnani, Ray, Raymond,
  Raza, Razzano, Read, Rees, Regimbau, Rei, Reid, Reid, Reitze, Relton,
  Renzini, Rettegno, Revenu, Reza, Rezac, Ricci, Richards, Richardson,
  Richardson, Riemenschneider, Riles, Rinaldi, Rink, Robertson, Robie, Robinet,
  Rocchi, Rodriguez, Rolland, Rollins, Romanelli, Romano, Romel, Romero,
  Romero-Shaw, Romie, Ronchini, Rosa, Rose, Rosińska, Ross, Rowan, Rowlinson,
  Roy, Roy, Roy, Rozza, Ruggi, Ruiz-Rocha, Ryan, Sachdev, Sadecki, Sadiq, Saha,
  Saito, Sakai, Sakellariadou, Sakon, Salafia, Salces-Carcoba, Salconi, Saleem,
  Salemi, Samajdar, Sanchez, Sanchez, Sanchez, Sanchis-Gual, Sanders, Sanuy,
  Saravanan, Sarin, Sassolas, Satari, Sathyaprakash, Sauter, Savage, Savant,
  Sawada, Sawant, Sayah, Schaetzl, Scheel, Scheuer, Schiworski, Schmidt,
  Schmidt, Schnabel, Schneewind, Schofield, Schönbeck, Schulte, Schutz,
  Schwartz, Scott, Scott, Seglar-Arroyo, Sekiguchi, Sellers, Sengupta,
  Sentenac, Seo, Sequino, Sergeev, Setyawati, Shaffer, Shahriar, Shaikh, Shams,
  Shao, Sharma, Sharma, Shawhan, Shcheblanov, Sheela, Shikano, Shikauchi,
  Shimizu, Shimode, Shinkai, Shishido, Shoda, Shoemaker, Shoemaker,
  ShyamSundar, Sieniawska, Sigg, Silenzi, Singer, Singh, Singh, Singh, Singha,
  Sintes, Sipala, Skliris, Slagmolen, Slaven-Blair, Smetana, Smith, Smith,
  Smith, Soldateschi, Somala, Somiya, Song, Soni, Soni, Sordini, Sorrentino,
  Sorrentino, Soulard, Souradeep, Sowell, Spagnuolo, Spencer, Spera,
  Spinicelli, Srivastava, Srivastava, Staats, Stachie, Stachurski, Steer,
  Steinhoff, Steinlechner, Steinlechner, Stergioulas, Stops, Stover, Strain,
  Strang, Stratta, Strong, Strunk, Sturani, Stuver, Suchenek, Sudhagar, Sudhir,
  Sugimoto, Suh, Sullivan, Sullivan, Summerscales, Sun, Sunil, Sur, Suresh,
  Sutton, Suzuki, Suzuki, Suzuki, Swinkels, Szczepańczyk, Szewczyk, Tacca,
  Tagoshi, Tait, Takahashi, Takahashi, Takano, Takeda, Takeda, Talbot, Talbot,
  Tanaka, Tanaka, Tanaka, Tanasijczuk, Tanioka, Tanner, Tao, Tao, Tapia,
  Martín, Taranto, Taruya, Tasson, Tenorio, Terhune, Terkowski,
  Thirugnanasambandam, Thomas, Thomas, Thompson, Thompson, Thondapu, Thorne,
  Thrane, Tiwari, Tiwari, Tiwari, Toivonen, Tolley, Tomaru, Tomura, Tonelli,
  Tornasi, Torres-Forné, Torrie, Melo, Töyrä, Trapananti, Travasso, Traylor,
  Trevor, Tringali, Tripathee, Troiano, Trovato, Trozzo, Trudeau, Tsai, Tsang,
  Tsang, Tsao, Tse, Tso, Tsuchida, Tsukada, Tsuna, Tsutsui, Turbang, Turconi,
  Tuyenbayev, Ubhi, Uchikata, Uchiyama, Udall, Ueda, Uehara, Ueno, Ueshima,
  Unnikrishnan, Urban, Ushiba, Utina, Vajente, Vajpeyi, Valdes, Valentini,
  Valsan, van Bakel, van Beuzekom, van Dael, Brand, Broeck, Vander-Hyde, van
  Haevermaet, van Heijningen, van Putten, van Remortel, Vardaro, Vargas, Varma,
  Vasúth, Vecchio, Vedovato, Veitch, Veitch, Venneberg, Venugopalan, Verkindt,
  Verma, Verma, Vermeulen, Veske, Vetrano, Viceré, Vidyant, Viets, Vijaykumar,
  Villa-Ortega, Vinet, Virtuoso, Vitale, Vocca, von Reis, von Wrangel, Vorvick,
  Vyatchanin, Wade, Wade, Wagner, Wald, Walet, Walker, Wallace, Wallace, Wang,
  Wang, Wang, Ward, Warner, Was, Washimi, Washington, Watchi, Weaver, Weaving,
  Webster, Weinert, Weinstein, Weiss, Weller, Weller, Wellmann, Wen, Weßels,
  Wette, Whelan, White, Whiting, Whittle, Wilken, Williams, Williams,
  Williamson, Willis, Willke, Wilson, Wipf, Wlodarczyk, Woan, Woehler, Wofford,
  Wong, Wong, Wright, Wu, Wu, Wu, Wysocki, Xiao, Yamada, Yamamoto, Yamamoto,
  Yamamoto, Yamashita, Yamazaki, Yang, Yang, Yang, Yang, Yang, Yang, Yap,
  Yeeles, Yeh, Yelikar, Ying, Yokoyama, Yokozawa, Yoo, Yoshioka, Yu, Yu,
  Yuzurihara, Zadrożny, Zanolin, Zeidler, Zelenova, Zendri, Zevin, Zhan,
  Zhang, Zhang, Zhang, Zhang, Zhang, Zhang, Zhao, Zhao, Zhao, Zhao, Zhou, Zhou,
  Zhu, Zhu, Zimmerman, Zucker, \&
  Zweizig}]{the_ligo_scientific_collaboration_the_virgo_collaboration_the_kagra_collaboration_tests_2021}
{The LIGO Scientific Collaboration, the Virgo Collaboration, the KAGRA
  Collaboration}, Abbott, R., Abe, H., {et~al.} 2021{\natexlab{c}},
  \href{http://arxiv.org/abs/2112.06861}{\JournalTitle{arXiv:2112.06861
  [astro-ph, physics:gr-qc, physics:hep-th]}}, arXiv: 2112.06861

\bibitem[{Yagi \& Yunes(2017)}]{yagi_approximate_2017}
Yagi, K., \& Yunes, N. 2017,
  \href{http://dx.doi.org/10.1016/j.physrep.2017.03.002}{\JournalTitle{Physics
  Reports}, 681, 1}

\bibitem[{Zappa {et~al.}(2019)Zappa, Bernuzzi, Pannarale, Mapelli, \&
  Giacobbo}]{zappa_black-hole_2019}
Zappa, F., Bernuzzi, S., Pannarale, F., Mapelli, M., \& Giacobbo, N. 2019,
  \href{http://dx.doi.org/10.1103/PhysRevLett.123.041102}{\JournalTitle{Physical
  Review Letters}, 123, 041102}

\end{thebibliography}
\begin{appendix}
\section{Additional BHNS population figures}
\label{app:bhnspop}
\begin{figure}[!htb]
    \centering
    \includegraphics[width=0.5\textwidth]{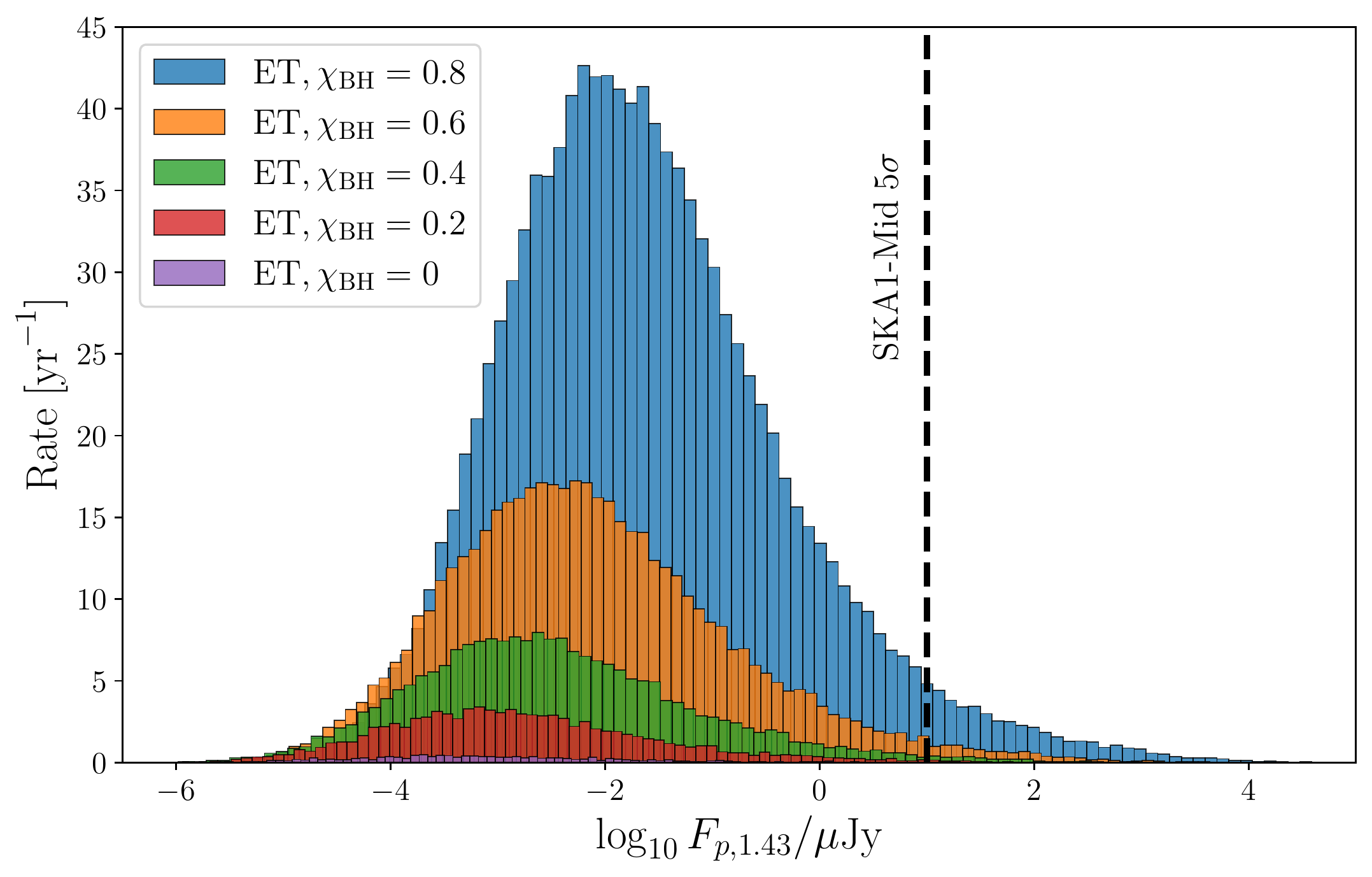}
    \caption{Flux distribution in $\log_{10}$ of sGRB afterglows associated with BHNS mergers detected in GWs with a 3G detector network that includes ET. The distributions are shown for different average BH spins: $\chi_\mathrm{BH} = 0$ (purple), $0.2$ (red), $0.4$ (green), $0.6$ (orange), and $0.8$ (blue). The vertical dashed line shows the SKA1-Mid 5$\sigma_\mathrm{RMS}$ detection limit.}
    \label{fig:fluxpopET}
\end{figure}

In Fig.~\ref{fig:fluxpopET}, we show the flux distribution of sGRB afterglows associated with BHNS mergers detected in GWs with a 3G detector network including ET. The different distributions correspond to different values for the average $\chi_\mathrm{BH}$ used in the simulations. Also displayed is the SKA1-Mid 5$\sigma_\mathrm{RMS}$ detection limit. Clearly, most of the sGRB afterglows are below the detection limit even with a radio telescope as sensitive as SKA1-Mid. As mentioned in Sect.~\ref{sec:aglowdetrates}, any change in the sensitivity of SKA1-Mid will have a big influence on the detection rates. This is especially evident for the distribution with $\chi_\mathrm{BH}=0.8$, which has a very steep slope below the 5$\sigma_\mathrm{RMS}$ detection limit.

\begin{figure}[!h]
    \centering
    \includegraphics[width=0.5\textwidth]{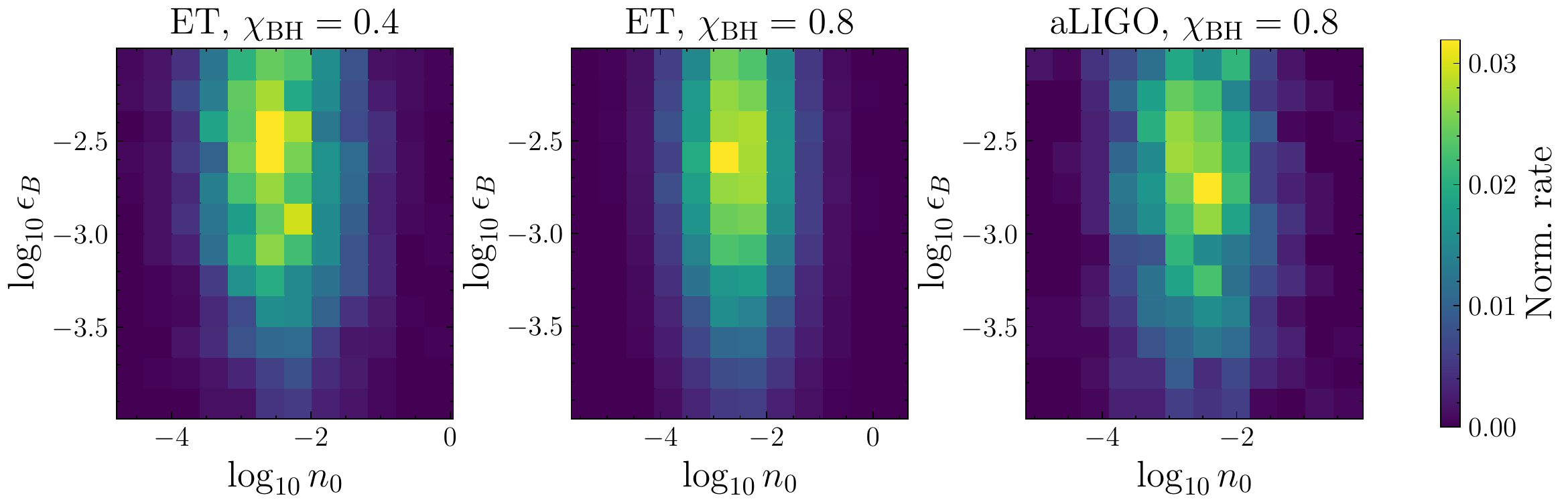}
    \caption{Normalised GW and radio detection rates for the ET and $\chi_\mathrm{BH} = \{0.4,0.8\}$ combinations (left and middle panel) and for the aLIGO and $\chi_\mathrm{BH} = 0.8$ combination (right panel). The rates are normalised to the combined detection rates of Fig.~\ref{fig:det_rates}, given by (from left to right): 3.3 $\mathrm{yr}^{-1}$, 44 $\mathrm{yr}^{-1}$, and 1.7 $\mathrm{yr}^{-1}$. A hard NS EOS ($R_\mathrm{NS} = 13.0$ km) is used in all cases. The rates are shown as a function of $n_0$ and $\epsilon_B$ of the detected sources.}
    \label{fig:eb_n_pop}
\end{figure}

In Fig.~\ref{fig:eb_n_pop}, we show the combined GW and radio detection rates as a function of $n_0$ and $\epsilon_B$. All rates are normalised to their respective total rate given in Fig.~\ref{fig:det_rates}. We used the normalised rates of the ET and $\chi_\mathrm{BH}=\{0.4,0.8\}$ combination and the aLIGO and $\chi_\mathrm{BH}=0.8$ combination. We assumed a hard NS EOS. Compared to the intrinsic distribution for $n_0$, the means of the detected distributions are a factor of $2.5-3.5$ higher as larger circumburst densities lead to larger peak fluxes (Eq.~\ref{eq:peakflux}). For $\epsilon_B$, this selection effect is less prominent, with means $\sim 1.5$ higher in the detected distributions. While the intrinsic distributions for $n_0$ and $\epsilon_B$ are similar, it is truncated at $10^{-2}$ for $\epsilon_B$ leading to lower means in the detected distributions.
\FloatBarrier
\section{Additional posterior of GW and EM data inference}
\label{app:PE}
\begin{figure}
    %\centering
    \includegraphics[width=0.48\textwidth]{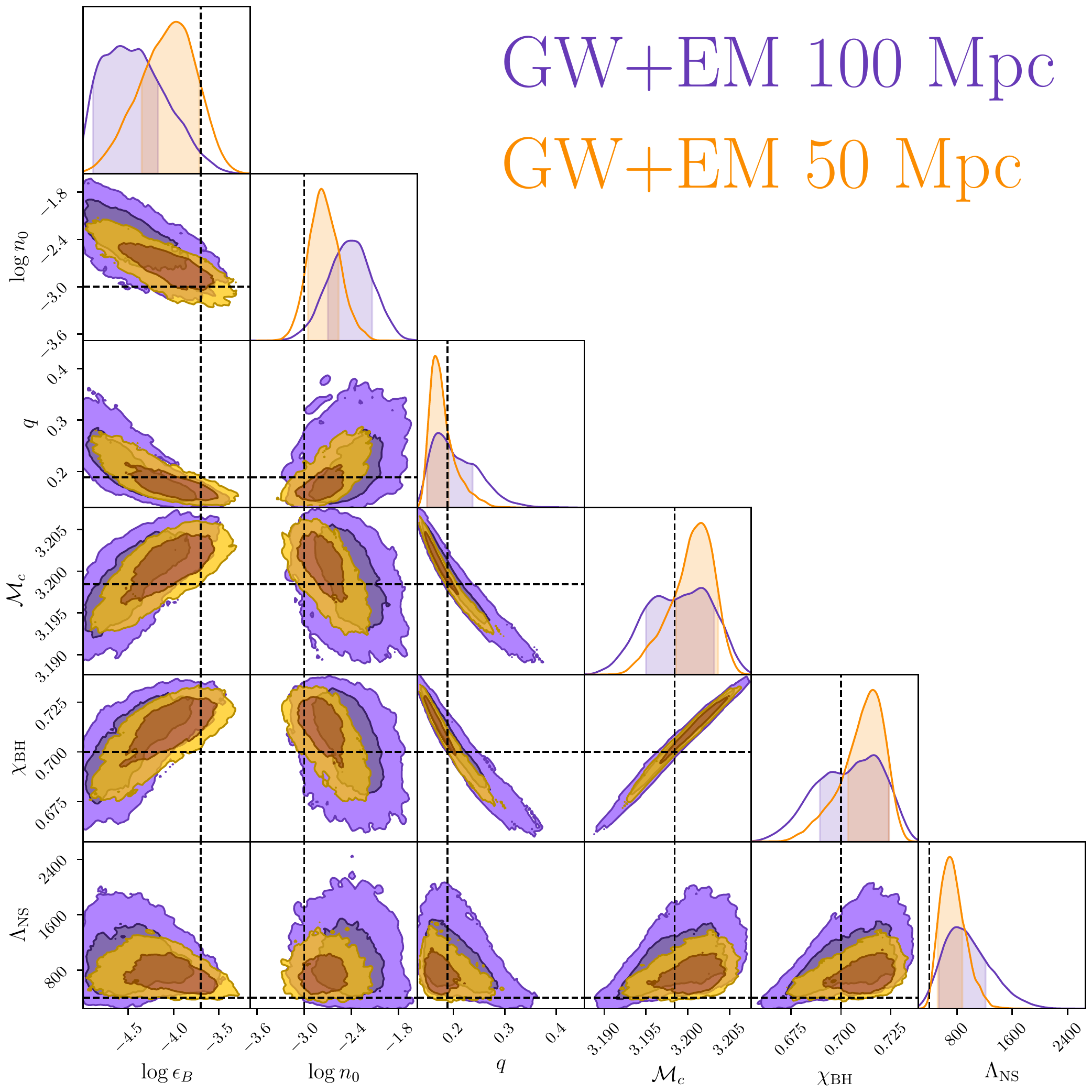}
    \caption{Posterior distributions of the binary source parameters together with $n_0$ and $\epsilon_B$. Purple contours give the inferred posterior distribution from the combination of the GW and radio data when the binary is placed at $d_L=100$ Mpc. Orange contours give the inferred posterior distribution from the combination of the GW and radio data when the binary is placed at $d_L=50$ Mpc. The fiducial injected values are indicated by the dashed black lines.}
    \label{fig:eBcorner}
\end{figure}

In Fig.~\ref{fig:eBcorner}, we show the combined posterior distribution of the binary source parameters together with $n_0$ and $\epsilon_B$ at distances $d_L=50, \ 100$ Mpc. Both $n_0$ and $\epsilon_B$ have varying degrees of correlation with the binary source parameters. We hypothesise that in the setup used in this work, improved estimates on $n_0$ and $\epsilon_B$, through different sources of EM emission for example, can help improve the estimates on the binary source parameters further. If we are instead able to improve our estimates on the binary source parameters, this will, in turn, also help with inferring $n_0$ and $\epsilon_B$.
\end{appendix}
\end{document}